\documentclass[journal]{IEEEtran}
\ifCLASSINFOpdf
\else
\fi

\usepackage{xcolor,colortbl}

\definecolor{Gray}{gray}{0.85}
\definecolor{LightCyan}{rgb}{0.88,1,1}
\newcolumntype{a}{>{\columncolor{Gray}}c}
\newcolumntype{b}{>{\columncolor{white}}c}

\usepackage{amssymb}
\usepackage{hyperref}
\usepackage{epstopdf}

\usepackage{graphicx}
\usepackage[caption=false]{subfig}

\usepackage{enumerate}
\usepackage{slashbox}
\usepackage{amsmath}
\usepackage{verbatim}

\usepackage{microtype}
\DisableLigatures{encoding = *, family = * }

\usepackage[nolist]{acronym}
\begin{acronym}
    \acro{HSM}{Hardware Security Module}
    \acro{RA}{Resolution Authority}
    \acro{CMIX}{Cryptographic Mix-Zone}
    \acro{CAM}{Cooperative Awareness Message}
    \acro{V2V}{vehicle-to-vehicle}
    \acro{V2I}{vehicle-to-infrastructure}
    \acro{VANET}{Vehicular Ad-hoc Network}
    \acro{OBU}{On-Board Unit}
    \acro{RSU}{Road-Side Unit}
    \acro{V2X}{V2V/V2I}
    \acro{VPKI}{Vehicular Public-Key Infrastructure}
    \acro{CA}{Certificate Authority}
    \acrodefplural{CA}{Certificate Authorities}
    \acro{REP}{random encryption periods}
    \acro{LTCA}{Long Term CA}
    \acro{PCA}{Pseudonym CA}
    \acro{RCA}{Root CA}
    \acro{LTC}{Long Term Certificate}
    \acro{TLS}{Transport Layer Security}
    \acro{CF}{Cuckoo Filter}
	\acro{CDF}{Cumulative Distribution Function}
   	\acro{ECDSA}{Elliptic Curve Digital Signature Algorithm}
	\acro{AAA}{Authentication, Authorization and Accounting}
	\acro{ACL}{Access Control List}
	\acro{AKI}{Accountable Key Infrastructure}
	\acro{API}{Application Programming Interface}
	\acro{BSM}{Basic Safety Message}
	\acro{BYOD}{Bring Your Own Device}
	\acro{BF}{Bloom Filter}
	\acro{C2C-CC}{Car2Car Communication Consortium}
	\acro{C2I}{Car-to-Infrastructure}
	\acro{C$^2$RL}{Compressed \ac{CRL}}
	\acro{CA}{Certification Authority}
	\acro{CN}{Common Name}
	\acro{CAMP VSC3}{Crash Avoidance Metrics Partnership Vehicle Safety Consortium}
	\acro{CIA}{Confidentiality, Integrity and Availability}
	\acro{CRL}{Certificate Revocation List}
	\acro{CDN}{Content Delivery Network}
	\acro{CP}{Chaff Pseudonym}
	\acro{COCA}{Cornell OnLine Certification Authority}
	\acro{CSR}{Certificate Signing Request}
	\acro{DAA}{Direct Anonymous Attestation}
	\acro{DDoS}{Distributed DoS}
	\acro{DDH}{Decisional Diffie-Helman}
	\acro{DENM}{Decentralized Environmental Notification Message}
	\acro{DHT}{Distributed Hash Table}
	\acro{DL/ECIES}{Discrete Logarithm and Elliptic Curve Integrated Encryption Scheme}
	\acro{DoS}{Denial of Service}
	\acro{DoT}{Department of Transportation}
	\acro{DPA}{Data Protection Agency}
	\acro{DSRC}{Dedicated Short Range Communication}
	\acro{DSS}{Digital Signature Standard}
	\acro{DTLS}{Datagram \ac{TLS}}
	\acro{ECU}{Electronic Control Unit}
	\acro{EDR}{Event Data Recorder}
	\acro{ETSI}{European Telecommunications Standards Institute}
	\acro{ECDSA}{Elliptic Curve Digital Signature Algorithm}
	\acro{ECC}{Elliptic Curve Cryptography}
	\acro{EVITA}{E-safety Vehicle Intrusion protected Applications}
	\acro{FOT}{Field Operational Testing}
	\acro{FPGA}{Field-Programmable Gate Array}
	\acro{GCP}{Google Cloud Platform}
	\acro{GKE}{Google Kubernetes Engine}
	\acro{GPA}{Global Passive Adversary}
	\acro{GN}{GeoNetworking}
	\acro{GS-VLR}{Group Signatures with Verifier Local Revocation}
	\acro{GS}{Group Signatures}
	\acro{GM}{Group Manager}
	\acro{GBA}{Generic Bootstrapping Architecture}
	\acro{GUI}{Graphic User Interface}
	\acro{KF}{Kalman Filter}
	\acro{HSM}{Hardware Security Module}
	\acro{HTTP}{Hypertext Transfer Protocol}
	\acro{IEEE}{Institute of Electrical and Electronics Engineers}
	\acro{IETF}{Internet Engineering Task Force}
	\acro{IoT}{Internet of Things}
	\acro{ITS}{Intelligent Transport System}
	\acro{IT}{Information Technologies}
	\acro{IMSI}{International Mobile Subscriber Identity}
	\acro{IMEI}{International Mobile Station Equipment Identity}
	\acro{IdP}{Identity Provider}
	\acro{IDS}{Intrusion Detection System}
	\acro{ISP}{Internet Service Provider}
	\acro{LEA}{Law Enforcement Agency}
	\acro{LCPP}{Lightweight Conditional Privacy Preservation}
	\acro{LTC}{Long Term Certificate}
	\acro{LTCA}{Long Term CA}
	\acro{H-LTCA}{Home-LTCA}
	\acro{F-LTCA}{Foreign-LTCA}
	\acro{LDAP}{Lightweight Directory Access Protocol}
	\acro{LBS}{Location Based Service}
	\acro{LTE}{Long Term Evolution}
	\acro{LuST}{Luxembourg SUMO Traffic}
	\acro{MCA}{Message \ac{CA}}
	\acro{MEA}{Misbehavior Evaluation Authority}
	\acro{MANET}{Mobile Ad-hoc Network}
	\acro{NIC}{Network Interface Card}
	\acro{MPB}{Most Pieces Broadcast}
	\acro{NoW}{Network on Wheel}
	\acro{OBU}{On-Board Unit}
	\acro{OEM}{Original Equipment Manufacturer}
	\acro{OCSP}{Online Certificate Status Protocol}
	\acro{PCA}{Pseudonym CA}
	\acro{PDP}{Policy Decision Point}
	\acro{PEP}{Policy Enforcement Point}
	\acro{PIR}{Private Information Retrieval}
	\acro{PKC}{Public Key Cryptography}
	\acro{PKCS}{Public Key Cryptosystem}
	\acro{PKI}{Public-Key Infrastructure}
	\acro{PRECIOSA}{Privacy Enabled Capability in Co-operative Systems and Safety Applications}
	\acro{PRESERVE}{Preparing Secure Vehicle-to-X Communication Systems}
	\acro{P2P}{peer-to-peer}
	\acro{PS}{Participatory Sensing}
	\acro{RA}{Resolution Authority}
	\acro{REST}{Representational State Transfer}
	\acro{RBAC}{Role Based Access Control}
	\acro{RCA}{Root \acs{CA}}
	\acro{RSSI}{Received Signal Strength Indication}
	\acro{RSU}{Road-Side Unit}
	\acro{SAML}{Security Assertion Markup Language}
	\acro{SAS}{Sample Aggregation Service}
	\acro{SCMS}{Security Credential Management System}
	\acro{SCORE@F}{Système COopératif Routier Expérimental Français}
	\acro{SDSI}{Simple Distributed Security Infrastructure}
	\acro{SRAAC}{Secure Revocable Anonymous Authenticated Inter-Vehicle Communication}
	\acro{SeVeCom}{Secure Vehicle Communication}
	\acro{SIT}{Sichere Informationstechnologie}
	\acro{SLC}{Short-Lived Certificate}
	\acro{SoA}{Service-oriented-Approach}
	\acro{SIFS}{Short Inter Frame Space}
	\acro{SSO}{Single-Sign-On}
	\acro{SSL}{Secure Sockets Layer}
	\acro{SOAP}{Simple Object Access Protocol}
	\acro{TACK}{Temporary Anonymous Certified Key}
	\acro{TS}{Task Service}
	\acro{TLS}{Transport Layer Security}
	\acro{TPM}{Trusted Platform Module}
	\acro{TTP}{Trusted Third Party}
	\acro{TVR}{Ticket Validation Repository}
	\acro{URI}{Uniform Resource Identifier}
	\acro{VANET}{Vehicular Ad-hoc Network}
	\acro{V2I}{Vehicle-to-Infrastructure}
	\acro{V2V}{Vehicle-to-Vehicle}
	\acro{V2X}{\ac{V2V}/\ac{V2I}}
	\acro{VC}{Vehicular Communication}
	\acro{VM}{Virtual Machine}
	\acro{VSN}{Vehicular Social Network}
	\acro{VSS}{\ac{VC} Security Subsystem}
	\acro{WAVE}{Wireless Access in Vehicular Environments}
	\acro{WSDL}{Web Services Discovery Language}
	\acro{W3C}{World Wide Web Consortium}
	\acro{V}{Vehicle}
	\acro{VANET}{Vehicular Ad-hoc Network}
	\acro{VLR}{Verifier-Local Revocation}
	\acro{VPKI}{Vehicular Public-Key Infrastructure}
	\acro{VM}{Virtual Machine}
	\acro{WS}{Web Service}
	\acro{WoT}{Web of Trust}
	\acro{WSACA}{\ac{WAVE} Service Advertisement \ac{CA}}
	\acro{XML}{Extensible Markup Language}
	\acro{XACML}{eXtensible Access Control Markup Language}
	\acro{3G}{3rd Generation}
	\acro{GNSS}{Global Navigation Satellite System}
	\acro{NTP}{Network Time Protocol}
\end{acronym}

\usepackage{makecell}

\usepackage{arydshln}

\usepackage{xcolor}

\usepackage{algpseudocode}

\usepackage{setspace}

\errorcontextlines\maxdimen

\makeatletter
\newcommand*{\algrule}[1][\algorithmicindent]{\makebox[#1][l]{\hspace*{.5em}\thealgruleextra\vrule height \thealgruleheight depth \thealgruledepth}}%
\newcommand*{\thealgruleextra}{}
\newcommand*{\thealgruleheight}{.75\baselineskip}
\newcommand*{\thealgruledepth}{.25\baselineskip}

\newcount\ALG@printindent@tempcnta
\def\ALG@printindent{%
	\ifnum \theALG@nested>0
	\ifx\ALG@text\ALG@x@notext
	\else
	\unskip
	\addvspace{-1pt}
	\ALG@printindent@tempcnta=1
	\loop
	\algrule[\csname ALG@ind@\the\ALG@printindent@tempcnta\endcsname]%
	\advance \ALG@printindent@tempcnta 1
	\ifnum \ALG@printindent@tempcnta<\numexpr\theALG@nested+1\relax
	\repeat
	\fi
	\fi
}%
\usepackage{etoolbox}
\patchcmd{\ALG@doentity}{\noindent\hskip\ALG@tlm}{\ALG@printindent}{}{\errmessage{failed to patch}}
\makeatother

\newbox\statebox
\newcommand{\myState}[1]{%
	\setbox\statebox=\vbox{#1}%
	\edef\thealgruleheight{\dimexpr \the\ht\statebox+1pt\relax}%
	\edef\thealgruledepth{\dimexpr \the\dp\statebox+1pt\relax}%
	\ifdim\thealgruleheight<.75\baselineskip
	\def\thealgruleheight{\dimexpr .75\baselineskip+1pt\relax}%
	\fi
	\ifdim\thealgruledepth<.25\baselineskip
	\def\thealgruledepth{\dimexpr .25\baselineskip+1pt\relax}%
	\fi
	\State #1%
	\def\thealgruleheight{\dimexpr .75\baselineskip+1pt\relax}%
	\def\thealgruledepth{\dimexpr .25\baselineskip+1pt\relax}%
}

\usepackage{footnote}

\usepackage{rotating}

\usepackage{soul}

\usepackage{algorithm}

\usepackage{tikz}
\usetikzlibrary{shadows,positioning,calc}
\tikzset{multiple/.style = {double copy shadow={shadow xshift=1ex,shadow
			yshift=-1.5ex,draw=black!30},fill=white,draw=black,thick,minimum height = 1cm,minimum
		width=2cm},
	ordinary/.style = {rectangle,draw,thick,minimum height = 1cm,minimum width=2cm}}

\usetikzlibrary{decorations.pathreplacing}

\usepackage{booktabs}
\usepackage{multirow}
\usepackage{siunitx}

\usepackage{algpseudocode}

\makeatletter
\renewcommand{\ALG@beginalgorithmic}{\footnotesize} 
\makeatother

\usepackage{afterpage}
\newlength{\oldtextfloatsep}

\usepackage[utf8]{inputenc}

\usepackage[normalem]{ulem}

\usepackage{booktabs} 

\usepackage{amsmath,amssymb,amsfonts}
\usepackage{graphicx}
\usepackage{textcomp}
\usepackage{makecell}
\usepackage{comment}

\usepackage{amssymb}
\usepackage{epstopdf}
\usepackage{amsmath}

\usepackage{setspace}

\usepackage{hyperref}
\usepackage{epstopdf}

\usepackage{algpseudocode}

\usepackage{setspace}

\errorcontextlines\maxdimen

\usepackage{algorithm}

\makeatletter
\renewcommand{\ALG@beginalgorithmic}{\footnotesize} 
\makeatother

\usepackage{booktabs}
\usepackage{multirow}
\usepackage{siunitx}

\usepackage{wrapfig}

\usepackage{blindtext}

\begin{document}

\title{Cooperative Location Privacy in Vehicular Networks: Why Simple Mix-zones are not Enough}
%
%
%

\author{Mohammad~Khodaei,~\IEEEmembership{Member,~IEEE,}
	and~Panos~Papadimitratos,~\IEEEmembership{Fellow,~IEEE}
	\thanks{The authors are with the Networked Systems Security group at KTH Royal Institute of Technology, Stockholm, Sweden. E-mail: \{khodaei, papadim\}@kth.se.}
}

\maketitle

\begin{abstract}
\acresetall
Vehicular communications disclose rich information about the vehicles and their whereabouts. Pseudonymous authentication secures communication while enhancing user privacy. To enhance location privacy, cryptographic mix-zones were proposed to facilitate vehicles covertly transition to new ephemeral credentials. The resilience to (\emph{syntactic} and \emph{semantic}) pseudonym linking (attacks) highly depends on the geometry of the mix-zones, mobility patterns, vehicle density, and arrival rates. We introduce a tracking algorithm for linking pseudonyms before and after a cryptographically protected mix-zone. Our experimental results show that an eavesdropper, leveraging standardized vehicular communication messages and road layout, could successfully link $\approx$73\% of pseudonyms during non-rush hours and $\approx$62\% of pseudonyms during rush hours after vehicles change their pseudonyms in a mix-zone. To mitigate such inference attacks, we present a novel \emph{cooperative mix-zone} scheme that enhances user privacy regardless of the vehicle mobility patterns, vehicle density, and arrival rate to the mix-zone. A subset of vehicles, termed \emph{relaying vehicles}, are selected to be responsible for emulating non-existing vehicles. Such vehicles cooperatively disseminate decoy traffic without affecting safety-critical operations: with 50\% of vehicles as relaying vehicles, the probability of linking pseudonyms (for the entire interval) drops from $\approx$68\% to $\approx$18\%. On average, this imposes 28~ms extra computation overhead, per second, on the \acp{RSU} and 4.67~ms extra computation overhead, per second, on the (relaying) vehicle side; it also introduces 1.46~KB/sec extra communication overhead by (relaying) vehicles and 45~KB/sec by \acp{RSU} for the dissemination of decoy traffic. Thus, user privacy is enhanced at the cost of low computation and communication overhead.
\end{abstract}

\begin{IEEEkeywords}
Privacy, Anonymity, Pseudonymity, Location Privacy, Mix Networks, \acl{VC}, \acsp{VANET}. 
\end{IEEEkeywords}

%
\IEEEpeerreviewmaketitle

\section{Introduction}
\label{sec:ieee-iot-tracking-introduction}

\acresetall

\ac{V2V} and \ac{V2I} communications seek to enhance transportation safety and efficiency. It has been well-understood that \ac{VC} systems are vulnerable to attacks and that the privacy of their users is at stake. As a result, security and privacy solutions have been developed by standardization bodies (IEEE 1609.2 WG~\cite{1609-2016} and \acs{ETSI}~\cite{ETSI-TS-102-940, EuropeanCommissionITS}), harmonization efforts (\ac{C2C-CC}~\cite{c2cPKIMemo}), and projects (\acs{SeVeCom}~\cite{papadimitratos2008secure}, \acs{PRESERVE}~\cite{feiri2015}, and CAMP~\cite{whyte2013security}). In \ac{VC} systems, vehicles disseminate \acp{CAM} and \acp{DENM} periodically at a high rate. To secure \ac{VC} systems, a consensus towards using \ac{PKC} to protect \ac{V2X} communication is reached: a set of Certification Authorities (CAs) constitutes the \ac{VPKI}, e.g.,~\cite{whyte2013security, khodaei2018Secmace}, providing multiple anonymous credentials, termed \emph{pseudonyms}, to legitimate vehicles. Vehicles switch from one pseudonym to a non-previously used one towards unlinkability of digitally signed messages, and improved sender privacy for \ac{V2V}/\ac{V2I} messages. Pseudonymity is conditional, in the sense that the corresponding long-term vehicle identity (\ac{LTC}) can be retrieved by the \ac{VPKI} entities if needed, e.g., for eviction of a faulty, misbehaving vehicle.

Due to the openness of wireless communication and dissemination of basic safety messages in plaintext (as confidentiality is not needed in \ac{VC} systems~\cite{1609-2016, papadimitratos2006securing}), an external entity could eavesdrop communications, towards inferring vehicle-sensitive information. Although pseudonymous authentication is a promising approach to protect user privacy, an adversary, eavesdropping all traffic in an area, could link successive pseudonymously authenticated messages. An adversary might observe an isolated pseudonym change, and associate the old and the new pseudonymous identifier through \emph{syntactic linking}, e.g.,~\cite{gerlach2007privacy, buttyan2009slow, tijinkPositionPaperC2CCC, khodaei2018PrivacyUniformity}. Alternatively, an adversary could leverage the physical constraints of the road layout~\cite{wiedersheim2010privacy}, together with data in message payloads, e.g., location, velocity, time, acceleration, the length and width\footnote{Length and width of vehicles are specified with a precision of 10 centimeters~\cite{ETSI-302-637-2, ullmann2016technical, US-EU-V2V-V2I-2013}.} of a victim's vehicle, to predict its trajectory towards linking messages \emph{semantically}, e.g.,~\cite{buttyan2009slow, wiedersheim2010privacy, Emara2013, Emara2017}. Such information could be unique, or one of few (locally rare), and thus, it could be easily linked. While appropriate pseudonym provisioning policies alleviate syntactic linking, by issuing time-aligned pseudonyms~\cite{khodaei2018Secmace, khodaei2014ScalableRobustVPKI, khodaei2016evaluating}, compromising user privacy by conducting semantic linking attacks is still feasible\footnote{Connecting such anonymous location profiles to real identities of vehicle owners is the final step, e.g., tracing their commutes and identifying home/work locations~\cite{gruteser2004protecting, golle2009anonymity, ma2013privacy}, the information obtained from \acp{VSN}~\cite{jin2016security}, or full de-anonymization of vehicles by \emph{honest-but-curious} \ac{VPKI} entities~\cite{khodaei2018Secmace}.}.

Different schemes were proposed, leveraging pseudonymous authentication primitives to mitigate inference, by an adversary, e.g., silent periods~\cite{Huang2005, sampigethaya2007amoeba, buttyan2007effectiveness, Emara2015}, silent cascades~\cite{huang2006silent}, SLOW~\cite{buttyan2009slow}, and random encryption periods~\cite{Wasef2010}. The common denominator among all \emph{silent period}-based schemes is that vehicles refrain from transmitting \acp{CAM} in certain intervals. This would result in diminished situational awareness, e.g., \emph{collision avoidance}~\cite{Lefevre2013}, thus increased probability of an accident, notably near intersections with congested traffic conditions~\cite{Corser2016}; thus, the practicality of such schemes is questionable.

Alternatively, vehicles could change their pseudonyms when approaching designated areas, termed mix-zones~\cite{beresford2003location, beresford2004mix, Freudiger2009Optimal}. The \ac{CMIX} was initially proposed~\cite{freudiger2007mix} in the \ac{VC} systems to establish a cryptographically protected region at appropriate times and places, e.g., at road intersections. All legitimate vehicles within the mix-zone receive a symmetric session key from a \ac{RSU}, responsible for the initiation of the pseudonym transition process and the symmetric key updates~\cite{freudiger2007mix}. Vehicles encrypt \acp{CAM} and opt in to change their pseudonyms while crossing these regions, towards pseudonym unlinkability (by an external eavesdropper). The achieved privacy protection level for such statically constructed mix-zones highly depends on the geometry of mix-zones, mobility patterns, and vehicle arrival rates. For example, based on the mix-zone geometries~\cite{Palanisamy2015}, or the traffic mobility pattern~\cite{codeca2015lust} and vehicle speed~\cite{wiedersheim2010privacy, freudiger2007mix, krumm2007inference, tomandl2012simulation}, an adversary can link successive pseudonyms of a given vehicle by observing the mix-zone entry and exit points. Such schemes are mostly effective when vehicle density and arrival/exit rates of vehicles in/from the mix-zones are uniformly distributed. Moreover, a fraction of non-cooperative vehicles within the mix-zone could affect anonymity by simply not changing their pseudonyms in a mix-zone; this yields a smaller anonymity set size for the mix-zone, compared to the one when all vehicles switch their pseudonyms.

Recently, a \emph{chaff-based} \ac{CMIX} scheme~\cite{vaas2018nowhere} was proposed: \acp{RSU} pre-generate and broadcast chaff \acp{CAM} to \emph{relaying} vehicles, responsible to periodically disseminate to emulate a non-existing (chaff) vehicle. This imposes significant computation overhead on \acp{RSU} and communication overhead: each \ac{RSU} needs to sign all chaff \acp{CAM} and distribute them to each relaying vehicle. Moreover, the \ac{RSU} needs to precisely predict trajectories of all vehicles in the neighborhood, to properly construct chaff \acp{CAM} for all chaff vehicles; the challenge lies in that traffic conditions are volatile, thus changes vehicles trajectories would invalidate (make relatively easily distinguishable) chaff \acp{CAM}. All these issues become clear in our performance evaluation in Sec.~\ref{subsec:ieee-iot-tracking-performance-comparison}.

\emph{Contributions:} In this paper, we fundamentally re-design the well-known approach (in different domains, e.g.,~\cite{chen2018taranet}) of introducing decoy traffic and re-design the \ac{VC}-specific scheme~\cite{vaas2018nowhere}, proposing a fully decentralized system (Sec.~\ref{sec:ieee-iot-tracking-cmix-with-decoy-traffic}). We show how to enhance user privacy, notably in low-density areas and non-rush hour periods, and how to mitigate syntactic and semantic linking attacks without affecting the operation of safety applications (Sec.~\ref{sec:ieee-iot-tracking-cmix-with-decoy-traffic}). Our scheme (i) enhances user privacy regardless of the vehicle mobility patterns, density, and arrival rate (to the mix-zone(s)); at the same time, (ii) it balances user privacy protection and (communication and computation) overhead based on vehicle density and mobility pattern. Furthermore, our scheme incurs low (computation and communication) overhead and prevents abuse of the mechanism towards diminishing the performance of the system or harming user privacy (Sec.~\ref{sec:ieee-iot-tracking-quantitative-analysis}). We also (iii) introduce a tracking algorithm towards linking vehicles before and after a cryptographically protected mix-zone (Sec.~\ref{subsec:ieee-iot-tracking-tracking-algorithm}). We leverage information in \acp{CAM} and the road layout towards linking pseudonyms syntactically and semantically, thus compromising user privacy. To mitigate such inference attacks, we introduce cooperative dissemination of decoy traffic: vehicles and \acp{RSU} emulate a non-existing vehicle by broadcasting decoy traffic in order to generate sufficiently many vehicles, thus diminishing the probability of linking two successive pseudonyms by an eavesdropper (Sec.~\ref{sec:ieee-iot-tracking-quantitative-analysis}). Our scheme efficiently and effectively enhances user privacy and it maintains strong user privacy protection for vehicles upon pseudonym change in a mix-zone in the presence of \emph{honest-but-curious} system entities (Sec.~\ref{sec:ieee-iot-tracking-security-and-privacy-analysis}).

In the rest of the paper, we survey the state-of-the-art research efforts (Sec.~\ref{sec:ieee-iot-tracking-related-work}) and describe the system model, adversarial model, and the requirements (Sec.~\ref{sec:ieee-iot-tracking-model-objectives}). We present our novel \ac{CMIX} scheme with decoy traffic (Sec.~\ref{sec:ieee-iot-tracking-cmix-with-decoy-traffic}), followed by a qualitative analysis of security and privacy (Sec.~\ref{sec:ieee-iot-tracking-security-and-privacy-analysis}). We evaluate the performance of our scheme (Sec.~\ref{sec:ieee-iot-tracking-quantitative-analysis}) before conclusion and future work (Sec.~\ref{sec:ieee-iot-tracking-conclusion}).


\section{Related Work}
\label{sec:ieee-iot-tracking-related-work}

Due to the openness of wireless transmissions and dissemination of basic safety messages in plaintext (as confidentiality is not needed in \ac{VC} systems~\cite{1609-2016, ETSI-TS-102-940, papadimitratos2006securing, US-EU-V2V-V2I-2013, ETSI-TS-103-097, ETSI-TS-102-867}), an external entity can arbitrarily eavesdrop \ac{VC} systems~\cite{Bellatti2017, Narain2016, Gao2014}. With advances in broadcast technology to extend the transmission range of \acp{OBU}~\cite{Qualcomm2016}, \ac{VANET} messages become increasingly accessible for an attacker. This information allows semantic linking attacks that rely on location and heading information of continuously broadcast \acp{CAM}~\cite{wiedersheim2010privacy}. Prior works, e.g.,~\cite{freudiger2007mix, vaas2018nowhere}, assume that the system entities that are fully trustworthy, i.e., \acp{RSU} and \ac{VPKI} entities, could link successive pseudonyms belonging to a given vehicle. However, recent revelations of mass surveillance, e.g.,~\cite{nsa, era2015cryptography}, show that assuming service providers are fully-trustworthy is no longer a viable approach. Thus, in~\cite{beresford2004mix, freudiger2007mix, cornelius2008anonysense, yu2016mixgroup}, the \ac{VPKI} entities can easily link pseudonyms issued for the vehicles, thus tracking them for the entire trip duration. Unlike the chaff-based \ac{CMIX} scheme~\cite{vaas2018nowhere} that requires vehicles provide their intended trajectory path to the \acp{RSU}, our scheme does not provide additional information and maintains strong user privacy protection upon pseudonym change in the presence of \emph{honest-but-curious} system entities.

There are different solutions for location privacy: \emph{K-anonymity}~\cite{gruteser2003anonymous} and dummy-based privacy protection schemes, e.g.,~\cite{liu2017spatiotemporal}, ensure that a target node is not distinguishable from at least \emph{K-1} nodes within an \emph{anonymity set} with respect to the information each node disseminates. However, safety applications require precise information to operate correctly, e.g., \emph{intersection collision warning}~\cite{etsi_102_637_2}. Alternatively, one can rely on group signature schemes, e.g.,~\cite{boneh2004short, calandriello2007efficient, papadimitratos2008impact, calandriello2011performance, khodaei2017RHyTHM}, to enhance user privacy. However, the performance of safety-related applications could be degraded. For example, leveraging such anonymous authentication schemes by the majority of vehicles results in a 30\% increase in cryptographic processing overhead~\cite{khodaei2017RHyTHM}. Moreover, with all vehicle-sensitive information in \acp{CAM} and \acp{DENM}, e.g., location, velocity, and acceleration, a targeted node could be unique, or one of few, and thus, successive messages could be linked sequentially by an external observer.

Different pseudonym transition strategies, to prevent an attacker from inferring such information, have been proposed. To evade correlation attacks, each vehicle could turn its wireless transmitter off for a randomly chosen interval and change pseudonym within that silent period~\cite{Huang2005, sampigethaya2005caravan, sampigethaya2007amoeba}. Based on the quality of service required for each application, this interval of being silent or being active can be dynamically adjusted~\cite{huang2006silent}. Even though such schemes could improve user privacy, they impose a performance penalty on safety applications~\cite{Corser2016}, thus jeopardizing human safety. To mitigate such a problem, vehicles could become silent and change their pseudonyms when their speed drops below 30 km/h since the risk of a fatal accident at a slow speed is expected to be low~\cite{buttyan2009slow}. Alternatively, vehicles could change their pseudonyms when refilling fuel at a gas station~\cite{boualouache2019privanet}, or each vehicle changes its pseudonyms randomly, e.g., every 5 minutes or every 1-3 KM~\cite{tijinkPositionPaperC2CCC}. However, an adversary can still conduct syntactic linking attacks due to a lack of synchronization among vehicles~\cite{khodaei2018PrivacyUniformity}, or track vehicles across pseudonym changes by predicting their trajectories~\cite{Emara2015}. In general, any individually-determined or user-defined pseudonym changing strategy could act as user fingerprints, thus enabling an adversary to track users, i.e., syntactic linking attack.

Another line of study proposes pseudonym transitions strictly within \ac{CMIX}~\cite{freudiger2007mix}, which does not impair transportation safety applications. A cryptographic mix-zone was initially proposed~\cite{freudiger2007mix} in the \ac{VC} systems to establish a cryptographically protected region at appropriate times and places, e.g., at intersections. When crossing these regions, vehicles change their pseudonyms privately while their communication is encrypted, which prevents syntactic and semantic linking attacks. However, the achieved privacy protection highly depends on the number of vehicles participating in the mix-zone, i.e., user privacy is degraded under low traffic density, e.g., in a highway scenario~\cite{Forster2017}. Moreover, an attacker could compromise unlinkability within a mix-zone based on the traffic mobility pattern and vehicle speed~\cite{tomandl2012simulation}. To counter this, vehicles could randomly switch lanes and speed prior to entering and/or crossing the mix-zones to confuse an adversary~\cite{Wasef2010, ravi2019enhancing}. However, such schemes would not be practical as they could seriously jeopardize human safety. Unlike such schemes, we provide privacy protection without affecting the operation of safety applications and regardless of variations in road layout, vehicle density, and mobility patterns.

Another alternative approach is to participate into a dynamic mix-zone, e.g.,~\cite{Wasef2010}: each \ac{OBU} is provided with a global symmetric key, using it to initiate a pseudonym change process. However, an internal attacker could terminate the encryption period on behalf of any vehicle; this impairs the functionality and operation of the scheme, thus eliminating user privacy protection. A dynamic cooperative location privacy protection scheme was proposed~\cite{khodaei2018PosterMixzoneEverywhere}: time-aligned pseudonyms are issued for all vehicles to facilitate synchronous pseudonym changes. Upon reaching a pseudonym transition process, a dynamic mix-zone formation is initiated by a vehicle and all \acp{CAM} within each mix-zone is encrypted using a distinct symmetric session key~\cite{khodaei2018PosterMixzoneEverywhere}. However, in a low traffic density area where there are very few vehicles to cooperatively change pseudonyms, vehicles could be semantically linkable. Unlike such schemes, our system ensures that user privacy is strongly protected even in situations with inherently low traffic density, e.g., suburban areas, and during low traffic periods.

In order to mitigate inferences by a compromised \ac{RSU}, an asymmetric group key agreement protocol by leveraging identity-based cryptography (not compatible with the standardization bodies, i.e., IEEE 1609.2 WG~\cite{1609-2016} and \acs{ETSI}~\cite{ETSI-TS-102-940}) was proposed~\cite{zhang2017otibaagka}: vehicles cooperatively conduct a group key agreement protocol to derive asymmetric keys. However, it is not clear how to determine the group size and handle dynamic changes of each group to perform group key agreement. Moreover, the pseudonyms (and their corresponding private keys) are generated by a Trusted Authority, i.e., a fully-trusted entity, and the keys are pushed to the vehicles; this raises concerns in terms of accountability and has not been adopted by standardization~\cite{1609-2016}. Furthermore, the `Trusted Authority' could trivially link all pseudonyms belonging to a vehicle, and thus the pseudonymously authenticated messages towards tracking it for the entire duration of its presence in the system. Moreover, the scheme lacks an extensive performance evaluation, notably in terms of communication and computation overhead in highly congested traffic conditions.

MobiMix~\cite{Palanisamy2015, palanisamy2011mobimix} shows that an adversary could infer user-sensitive information based on the vehicle population in a mix-zone, the statistical behavior of the population, and the geometry of a mix-zone. To mitigate such inferences, it is proposed to dynamically adjust the geometry of a mix-zone based on multiple factors, e.g., the statistical behavior and the movement patterns of the users. But, an adversary could still perform semantic linking attacks when the traffic density is sparse~\cite{guo2018independent}. Swing \& Swap~\cite{li2006swing, singh2019cpesp} and MixGroup~\cite{yu2016mixgroup} propose to construct a region in which vehicles exchange their pseudonyms (and the corresponding private keys). But, such schemes do not achieve liability attribution and non-repudiation, which are basic requirements for a secure \ac{VC} system~\cite{1609-2016, papadimitratos2008secure, papadimitratos2006securing, ETSI-102-731}.


\section{Model and Objectives}
\label{sec:ieee-iot-tracking-model-objectives}

\subsection{System Model and Assumptions}
\label{subsec:ieee-iot-tracking-system-model-assumptions}

\begin{figure}[t]
	\vspace{-0.25em}
	\centering
	\includegraphics[trim=0cm 0cm 0cm 0cm, clip=true, totalheight=0.26\textheight,angle=0,keepaspectratio]{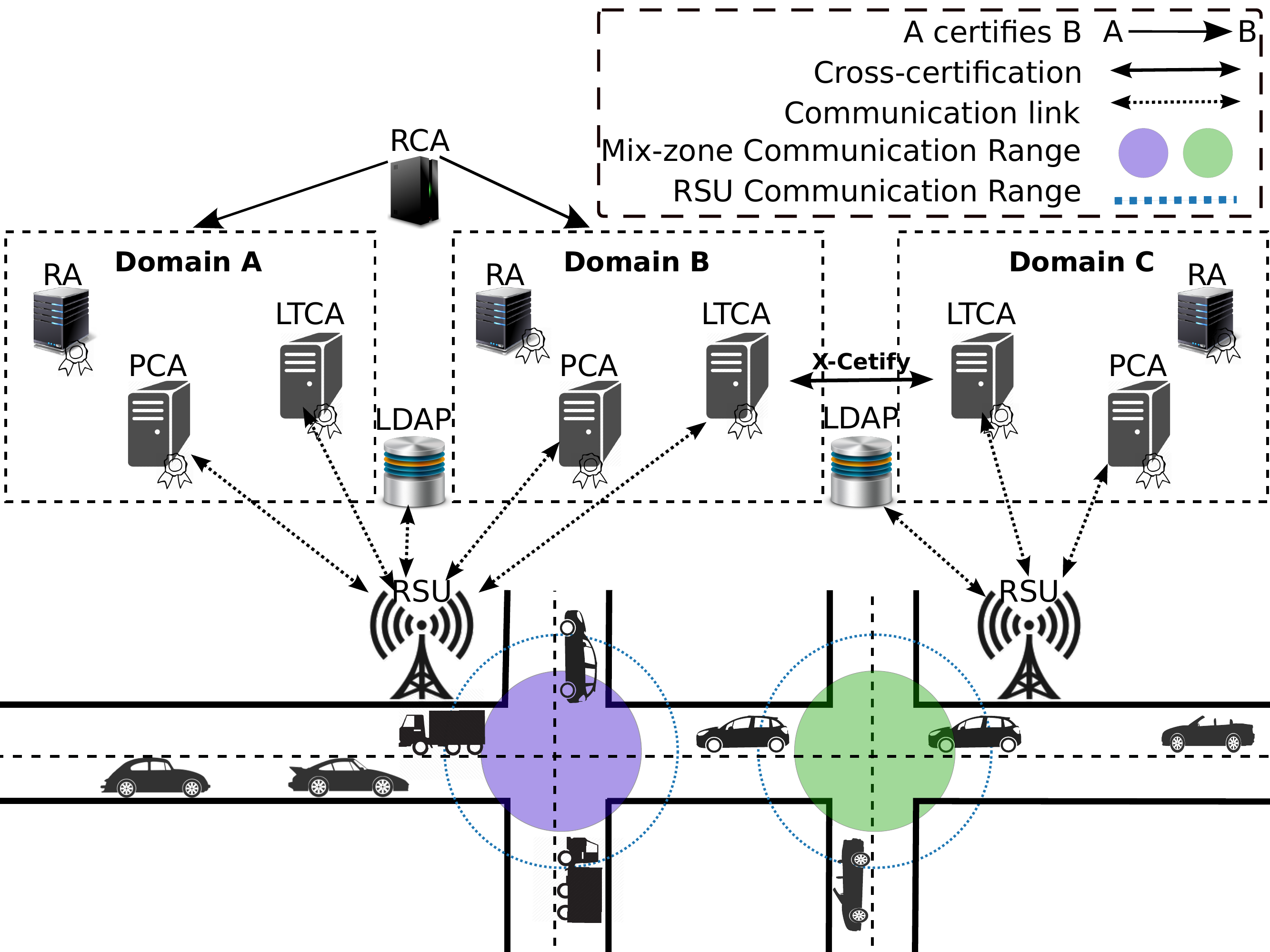}
	\vspace{-0.5em}
	\caption{An overview of secure and privacy-protecting \ac{V2X} communication.}
	\label{fig:ieee-iot-tracking-overview-secure-privacy-protecting-v2x-communication}
	\vspace{-1.5em}
\end{figure}

Fig.~\ref{fig:ieee-iot-tracking-overview-secure-privacy-protecting-v2x-communication} shows an overview of secure and privacy-preserving \ac{VC} systems. We assume a \ac{VPKI}, shown on top of Fig.~\ref{fig:ieee-iot-tracking-overview-secure-privacy-protecting-v2x-communication}, with distinct entities and roles that registers vehicles in a domain~\cite{khodaei2015VTMagazine} and issues pseudonyms, e.g.,~\cite{whyte2013security, khodaei2018Secmace}. The \ac{RCA}, the highest-level authority, certifies other lower-level authorities; the \ac{LTCA} provides registered vehicles (and \acp{RSU}) with a \acf{LTC}, used to authorize the acquisition of pseudonyms from a \ac{PCA}~\cite{khodaei2018Secmace}. To facilitate the overall intra-domain and multi-domain operations, a vehicle first finds such information from a \ac{LDAP}~\cite{sermersheim2006lightweight} server. This is carried out without disclosing the real identity of the vehicle. Any \ac{CMIX}-based scheme requires vehicles change their pseudonyms whenever crossing a mix-zone. As a result, vehicles need several pseudonyms with overlapping lifetimes, compatible with the proposals of standardization bodies, i.e., IEEE 1609.2 WG~\cite{1609-2016} and \acs{ETSI}~\cite{ETSI-TS-102-940}. This allows pseudonym changes, without pseudonym reuse, to be straightforward, i.e., without extensive prior knowledge on \ac{CMIX} placement, trip details, pseudonym lifetimes, etc.

Each vehicle triggers pseudonym acquisition process based on various factors~\cite{khodaei2016evaluating}. Our scheme requires sparse connectivity to the \ac{VPKI}, allowing an \ac{OBU} to be \emph{preloaded} with numerous pseudonyms proactively, covering a longer period, e.g., a week or a month (should the connectivity be relatively scarce). We assume that a state-of-the-art \acp{VPKI}, e.g.,~\cite{whyte2013security, khodaei2018Secmace, khodaei2019Scaling}, can provide pseudonyms in a timely manner. Moreover, we assume the \ac{VPKI} pseudonym lifetime policy is the same for all registered vehicles, so that timing information does not harm user privacy~\cite{khodaei2018Secmace, khodaei2014ScalableRobustVPKI}.

\acp{OBU} and \acp{RSU} are equipped with \acp{HSM}. Private keys stored in an \ac{HSM} cannot be extracted and only one pseudonym can be active at any time. In case of any deviation, pseudonymously authenticated messages can be used by a \ac{RA} to retrieve the long-term identity of the vehicle~\cite{khodaei2018Secmace}. The misbehaving entities should be evicted and revocation information be distributed~\cite{khodaei2018VehicleCentric, khodaei2019TMCVehicleCentricCRL}. The certificates of higher-level authorities are installed on the \acp{OBU} and their clocks are (loosely) synchronized with the \ac{VPKI} servers through \ac{GNSS} or other means, e.g., \ac{NTP} servers over the Internet.

We assume that each vehicle and \ac{RSU} have their own location information; \acp{RSU} could communicate with \ac{VPKI} entities and they are aware of the road layout (within their communication ranges). We further assume that appropriate countermeasures are in place to prevent location spoofing, e.g.,~\cite{papadimitratos2008gnss}, towards enabling secure neighborhood discovery~\cite{papadimitratos2008secureneighborhood}, and facilitating physical position verification~\cite{FioreCCP:J:2013}. Upon a pseudonym change inside a mix-zone, vehicles change their Media Access Control (MAC) and IP addresses~\cite{papadimitratos2007architecture} to prevent their old and new pseudonyms from being (trivially) linked based on these interfaces~\cite{fonseca2007support, Petit2014}. The impact of pseudonym change on the quality of services, e.g., safety applications~\cite{eckhoff2016marrying}, or protocols, e.g., geographic routing~\cite{SchochKLSP:C:2006, festag2010design}, is orthogonal to our investigation.

The choice of \acp{RSU} to establish \acp{CMIX} depends on different factors, e.g., desired level of privacy, traffic conditions, physical constraints of the road layout, and efficiency; for example, the more mix-zones are constructed, the higher the frequency of changing pseudonyms becomes. This would result in higher number of unlinkable segments for any journey, thus, enhance user privacy. The frequent change of mix-zone location makes it even harder for an adversary to eavesdrop the communication: an adversary would need to deploy eavesdropping facilities near most, if not all, the \acp{RSU} to improve her chance, which seems to be practically infeasible. However, full deployment of mix-zones over all \acp{RSU}, i.e., mandating frequent pseudonym changes, could affect the operation of safety application and impose communication overhead. The \ac{VPKI} system chooses a fraction of \acp{RSU} to establish a cryptographically protected mix-zone; in Fig.~\ref{fig:ieee-iot-tracking-overview-secure-privacy-protecting-v2x-communication}, two mix-zones are constructed, each established by an \ac{RSU}, to facilitate private pseudonym changes. In our performance evaluation in Sec.~\ref{sec:ieee-iot-tracking-quantitative-analysis}, the mix-zones locations are fixed and known to an adversary, i.e., the best case for an adversary. Leveraging an optimal placement of mix-zones, e.g.,~\cite{Freudiger2009Optimal, humbert2009optimal}, in order to balance the achieved level of privacy and cost, is orthogonal to this investigation.

\subsection{Adversarial Model}
\label{subsec:ieee-iot-tracking-adversarial-model}

We consider the general adversary model in~\cite{papadimitratos2008secure, papadimitratos2006securing} for secure and privacy-preserving \ac{VC} systems and more specifically the adversarial model assumptions of \ac{CMIX} schemes~\cite{beresford2004mix, freudiger2007mix, vaas2018nowhere, cornelius2008anonysense, yu2016mixgroup} that consider external eavesdroppers, possibly with broad or global coverage range. Along these lines, we assume that \acp{RSU} and participating users/vehicles are honest (i.e., trustworthy entities). We consider external adversaries with wireless receivers placed near each mix-zone, to eavesdrop \ac{VC} systems to infer user-sensitive information towards harming user privacy. They passively eavesdrop communication of vehicles entering and exiting the mix-zone, covering all entry and exit points of the mix-zones, towards linking pseudonyms before and after a mix-zone. This is based on information derived from \acp{CAM}, e.g., timing, velocity, and location. We do not constrain the choice and design of the inference algorithm, i.e., a tracking algorithm to link two pseudonyms of a vehicle, prior to and after pseudonym change in a mix-zone. Rather, in order to achieve tangible results, we devise a tracking algorithm (see Sec.~\ref{subsec:ieee-iot-tracking-tracking-algorithm}), orthogonal to the defense mechanism.

In addition, we explore the consequences of strengthening the adversarial model in Sec.~\ref{subsec:ieee-iot-tracking-performance-comparison}. In particular, we consider (i) \acp{RSU} and \ac{VPKI} entities that are \emph{honest-but-curious}, i.e., entities complying with security protocols and policies, but motivated to profile users by collecting or inferring user sensitive information based on the execution of the protocols. Moreover, (ii) the collaboration (collusion) of honest-but-curious entities that share information individually inferred by each (see Sec.~\ref{sec:ieee-iot-tracking-security-and-privacy-analysis} for detailed security analysis). Finally, we consider (iii) a set of non-cooperative actions by registered vehicles that can affect the operation (or level) of protection of the scheme (and any \ac{CMIX} scheme).

\textbf{Extending the passive eavesdropper model:} In this paper, we focus on the effect and improvement of the \ac{CMIX} approach. The investigation can be extended to the entire network, considering the optimal placement of eavesdroppers, increasing their coverage, and overall pseudonym usage. The adversarial model can be further strengthened if internal adversaries, including the non-cooperative vehicles joining the mix-zone, report the symmetric keys of the mix-zones and the observed communication to an external adversary (collection point). For example, an \ac{RSU} could share a transcript of pseudonymously authenticated messages with an honest-but-curious \ac{VPKI} entity to perform syntactic and semantic linking attacks. However, this adversarial model is beyond the scope of this investigation. Moreover, non-\ac{VC} mechanisms such as traffic monitoring cameras (with object recognition techniques), Radio Frequency (RF)-based characteristics, e.g., angle-of-arrival~\cite{golle2004detecting, xiao2006detection}, physical layer device identification~\cite{rasmussen2007implications, brik2008wireless, danev2012physical}, and physical layer localization with additional equipments, e.g.,~\cite{xiao2007fingerprints, xiao2008using, van2016exploiting, vaas2018poster}, which can localize vehicles based on the physical layer attributes of transmitters or identify decoy traffic from the actual one, are out of scope and warrant a separate investigation. Further, attacks on \ac{GNSS} are orthogonal to our work. In fact, the consequence of location spoofing would be that mix-zones are not formed properly; however, the effect is more dire for the \ac{VC} systems to begin with. These extensions of the adversarial model is part of future work.

\subsection{Requirements}
\label{subsec:ieee-iot-tracking-security-privacy-requirements}

Security and privacy requirements for \ac{V2X} communications have been specified in the literature~\cite{papadimitratos2006securing}, and additional requirements for \ac{VPKI} entities in~\cite{khodaei2018Secmace}. The security and privacy requirements for the \ac{CRL} distribution are in~\cite{khodaei2018VehicleCentric, khodaei2019TMCVehicleCentricCRL}. Beyond `conventional' security requirements~\cite{papadimitratos2006securing}, it is ingrained in \ac{CMIX} schemes to establish neighbors proximity (physical or communication neighborhood~\cite{papadimitratos2008secureneighborhood, papadimitratos2008protection, brands1993distance, narain2019security, zeng2018all}). This can be done in a secure manner, e.g., with \ac{RSU}/vehicle protocols, or increased protection of the vehicle location information~\cite{festag2010design}. In the following, we describe the security and privacy, as well as functional and performance, requirements for a privacy-preserving \ac{CMIX} scheme.

\emph{R1. Privacy (anonymity and unlinkability):} Vehicles should participate in the \ac{VC} system \emph{anonymously}, i.e., vehicles should communicate with others without revealing their long-term identifiers and credentials. Anonymity is conditional in the sense that the corresponding long-term identity can be retrieved by the \ac{VPKI} entities, and accordingly, the long-term credential revoked if vehicles deviate from system policies. In order to achieve \emph{unlinkability}, we need to diminish the inference by an eavesdropper upon pseudonym change, i.e., mitigating syntactic and semantic linking attacks.

\emph{R2. Availability:} The system should ensure any legitimate vehicle is notified about \ac{CMIX} parameters, e.g., the location, geometry, and the symmetric key corresponding to an approaching mix-zone, to facilitate their participate in the mix-zone. Moreover, a small fraction of bandwidth should be used for the distribution of mix-zone related material, to not interfere with the safety- and time-critical operations.

\emph{R3. Auditability and misbehavior detection:} Auditability refers to the ability of a system to audit the processes and operations of the system entities. In case of any deviation, the system should be able to initiate a (resolution) process to identify the misbehaving entity. This essentially allows an \ac{RSU} to interact with the \ac{VPKI} system towards detecting misbehavior. Depending on the situation, appropriate actions could be initiated, e.g., de-anonymizing the misbehaving entity, and/or revoking its cryptographic materials and evicting it from further accessing the system. In the context of this work, each \ac{RSU} monitors the behavior of vehicles when entering and exiting the mix-zone; if a substantial fraction of vehicles exit the mix-zone without changing their pseudonyms, the \ac{RSU} would increase the percentage of decoy traffic in order to achieve a desired level of privacy protection.

\emph{R4. Efficiency and scalability:} All mix-zone operations should be efficient and scale with the number of vehicles. The scalability results from fast generation and lightweight dissemination of the credentials, efficient operations, and fault-tolerant design to ensure that the system remains operational in the presence of benign failures or be resilient to resource depletion attacks.


\section{\ac{CMIX} with Decoy Traffic}
\label{sec:ieee-iot-tracking-cmix-with-decoy-traffic}

\subsection{System Overview}
\label{subsec:ieee-iot-tracking-system-overview}

The \ac{VPKI} system chooses a subset of \acp{RSU}, located near intersections where vehicles physically mix~\cite{freudiger2007mix}, to establish a cryptographically protected area and construct a \ac{CMIX} for private pseudonym changes. \acp{RSU} are responsible for the initiation of the pseudonym transition process and maintaining a symmetric key to establish the encrypted region. To mitigate syntactic and semantic linking attacks, we introduce broadcasting decoy traffic at each mix-zone. Such traffic emulates vehicles that do not exist in reality. The \ac{RSU} at each mix-zone facilitates obtaining \emph{\acp{CP}} in order to generate \emph{chaff \acp{CAM}} (or \emph{chaff \acp{DENM}}). The purpose is to decrease the probability of linking two pseudonyms of a vehicle prior to and after pseudonym change. In case of sparse traffic (low vehicle density), \acp{RSU} could also emulate a chaff vehicle by periodically broadcasting chaff \acp{CAM} (signed under the private key of a chaff pseudonym). Our system can be configured so that for each vehicle, multiple \emph{seemingly identical} chaff vehicles could (potentially) appear as if they uniformly exit from different exit points of a mix-zone. As a result, it is hard for an eavesdropper to identify actual traces based on the \acp{CAM} attributes, e.g., velocity, acceleration, mix-zone geometry, and time spent in a mix-zone. Each vehicle could request multiple chaff pseudonyms (and the corresponding chaff private keys) from an \ac{RSU}. For ease of exposition, we assume each vehicle requests one chaff pseudonym in each mix-zone. Extension to multiple chaff pseudonyms and multiple \acp{PCA} operating in a domain is straightforward.

\begin{figure}[t]
	\vspace{-0.25em}
	\centering
	\includegraphics[trim=0cm 0cm 0cm 0cm, clip=true, totalheight=0.29\textheight,angle=0,keepaspectratio]{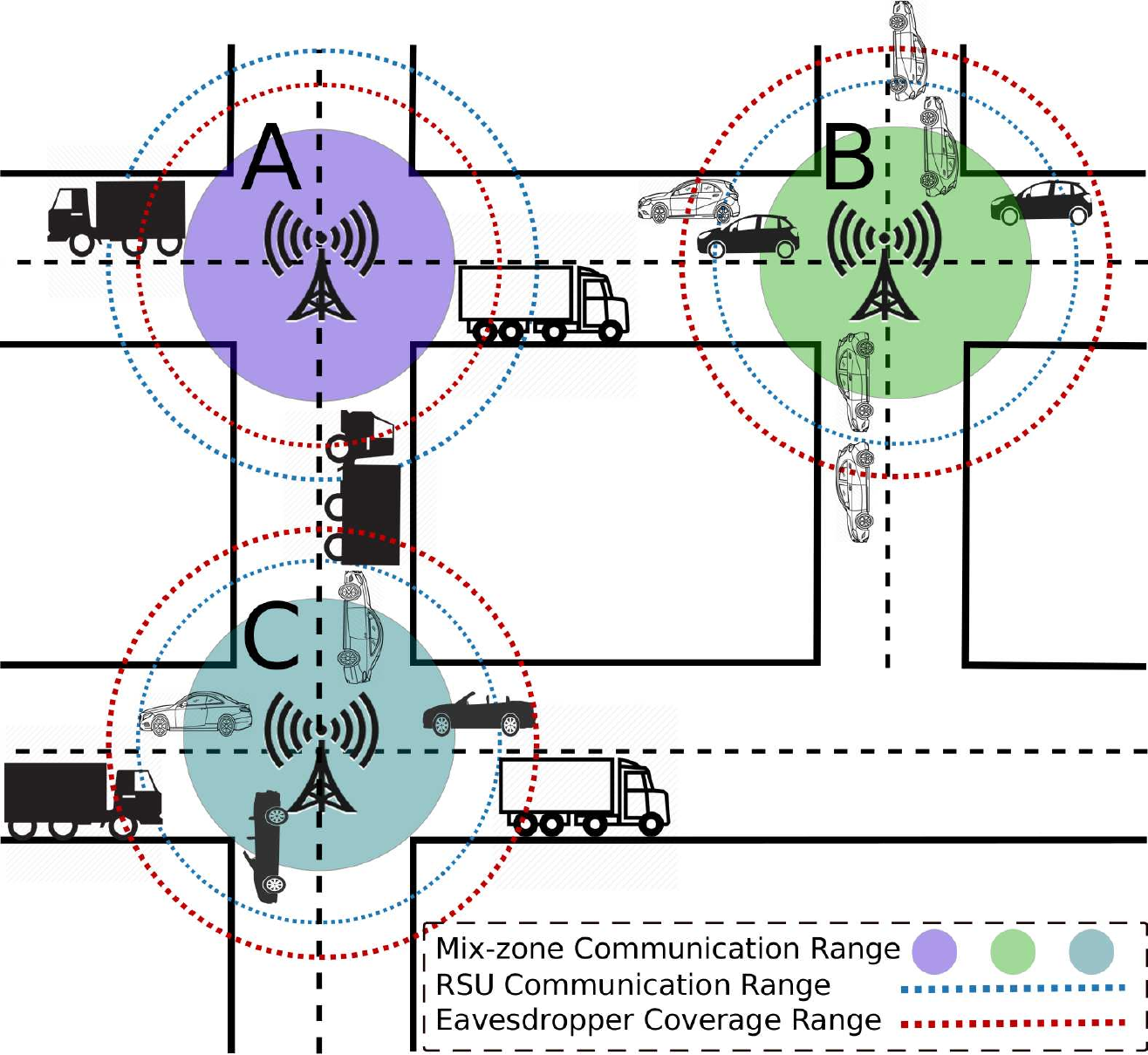}
	\vspace{-0.25em}
	\caption{Mix-zone construction with decoy traffic.}
	\label{fig:ieee-iot-tracking-mixzone-with-decoy-traffic}
	\vspace{-1em}
\end{figure}

Fig.~\ref{fig:ieee-iot-tracking-mixzone-with-decoy-traffic} shows three mix-zones: the colored disks indicate the approximate encrypted range of a mix-zone; the blue dotted circles denote the transmission range of \acp{RSU}. The coverage range of eavesdroppers denoted by red dotted circles; for mix-zones $B$ and $C$, the external adversaries eavesdrop all entry and exit points of the \acp{RSU} while for mix-zone $A$, the eavesdropper eavesdrops all entry and exit points of the mix-zone. The \ac{RSU} coverage range can be either larger or smaller than the local eavesdropper; however, the operation of our scheme does not depend on these ranges. The \ac{RSU} range needs to always exceed the mix-zone range, simply in order to allow vehicles to execute the \ac{CMIX} participation protocol, notably obtaining the mix-zone symmetric key. Black vehicles are the real ones while the white ones represent non-exiting vehicles, i.e., the decoy traffic. Once a vehicle enters a mix-zone, it requests to obtain the mix-zone symmetric key. An \ac{RSU} leverages its knowledge about the road layout and vehicles to determine how many chaff vehicles are required. In the case of sparse traffic density, an \ac{RSU} generates synthetic \acp{CAM}, resembling the traces towards an exit point of the mix-zone. The system can be configured to have \acp{RSU} provide and/or emulate one (see mix-zone $C$ in Fig.~\ref{fig:ieee-iot-tracking-mixzone-with-decoy-traffic}) or multiple (see mix-zone $B$ in Fig.~\ref{fig:ieee-iot-tracking-mixzone-with-decoy-traffic}) chaff vehicles. In our scheme, each vehicle only provides its length to the \ac{RSU}; this information is used by an \ac{RSU} to coordinate with another vehicle in the mix-zone towards disseminating decoy traffic, i.e., generating synthetic \acp{CAM} towards resembling a non-existing, \emph{but seemingly identical}, vehicle, exiting from an opposite exit point of the mix-zone.

Each \ac{PCA} pre-generates a distinct set of chaff public and private keys (chaff pseudonyms) and delivers them to an \ac{RSU}, responsible for a mix-zone construction. Each vehicle could send a request to the \ac{RSU} to obtain one chaff pseudonym. The \ac{RSU} randomly assigns chaff pseudonyms to a subset of vehicles, termed \emph{relaying vehicles}. The \ac{VPKI} system cannot correlate a vehicle and a chaff pseudonym since the \ac{RSU} randomly assigns a chaff pseudonym to a requesting vehicle. Note that accountability for chaff \acp{CAM} is not paramount as such (chaff) credentials are not valid and they cannot be used for any application. In case of deviation from system protocols, a misbehaving vehicle can still be identified (see Sec.~\ref{subsec:ieee-iot-tracking-chaff-pseudonym-resolution}).

In order to preserve the correct functionality of transportation safety applications, our scheme provides vehicles with information to identify chaff messages. Therefore, each \ac{PCA} proactively constructs a \ac{CF}~\cite{Fan2014} by including chaff pseudonyms in a probabilistic data-structure and \acp{RSU} distribute these condensed fingerprints of chaff pseudonyms among legitimate vehicles across a region. This facilitates discarding chaff pseudonyms by legitimate vehicles, thus, ensuring the correct operation of safety applications. Similarly to \ac{BF}~\cite{bloom1970space, mitzenmacher2002compressed}, \acp{CF} provide fast membership tests at the cost of a false positive rate ($\rho$), but in contrast support dynamic updates of the underlying set. This data structure includes the fingerprints of the chaff pseudonyms used to sign chaff \acp{CAM} and chaff \acp{DENM}. When receiving a \ac{CAM} or a \ac{DENM}, an \ac{OBU} could efficiently validate the attached pseudonym against the corresponding \ac{CF}; if the membership test is positive, the \ac{CAM} or the \ac{DENM} is discarded; otherwise, the signature will be verified.

Chaff \acp{CAM} are to be disseminated until a vehicle reaches another mix-zone or the end of the trip duration. When a relaying vehicle intends to stop disseminating chaff \acp{CAM}, e.g., entering another mix-zone, it queries the \ac{PCA}, signed under the private key of the chaff pseudonym, to remove that chaff pseudonym from the corresponding \ac{CF}. Further dissemination of chaff \acp{CAM} using such a chaff pseudonym is considered a misbehavior and it can be identified by a misbehavior detection system, e.g.~\cite{bissmeyer2014misbehavior}, that triggers the revocation. The \acp{CF} are frequently updated by the \acp{PCA} and pushed to the corresponding \acp{RSU}.

An \ac{RSU} operating a mix-zone cannot filter out chaff pseudonyms, originated from other mix-zones; the \ac{PCA} prepares a distinct set of chaff pseudonyms for each \ac{RSU}, operating a mix-zone. As a result, an \ac{RSU} cannot distinguish between a real pseudonym and a chaff one of another \ac{RSU}. However, a vehicle might encounter other relaying vehicles with chaff pseudonyms obtained from other mix-zones. For example, when a vehicle is crossing mix-zone $A$ and moving towards mix-zone $B$ in Fig.~\ref{fig:ieee-iot-tracking-mixzone-with-decoy-traffic}, it might encounter chaff pseudonyms originated from mix-zone $B$. Thus, it needs to request and obtain the \ac{CF} corresponding to mix-zone $B$. The vehicle could directly interact with the \ac{PCA} and request to obtain the \acp{CF}, corresponding to the nearby mix-zones. The \ac{PCA} needs to identify the physical location\footnote{Physical identification of vehicles is also a key requirement in the original mix-zone scheme~\cite{beresford2003location, beresford2004mix, freudiger2007mix}; this prevents an adversary from remotely requesting the symmetric keys of the mix-zones.}, e.g.,~\cite{papadimitratos2008secureneighborhood, papadimitratos2008protection, brands1993distance}, of requesting vehicles; in fact, requesting vehicles should be physically ``close'' to a mix-zone to obtain the corresponding \ac{CF} for. Otherwise, an external adversary could request to obtain all \acp{CF}, thus filtering out all chaff pseudonyms exiting the mix-zones.

In what follows, we describe the credential acquisition protocols in Sec.~\ref{subsec:ieee-iot-tracking-credential-acquisition-protocols}. We present updated mix-zone advertisement with chaff pseudonym acquisition protocols in Sec.~\ref{subsec:ieee-iot-tracking-mix-zone-participation-protocol}. Next, we present \ac{CF} dissemination in Sec.~\ref{subsec:ieee-iot-tracking-tracking-chaff-dissemination}. Table~\ref{table:ieee-iot-tracking-chaff-cmix-zones-protocols-notation} provides a description of the functions and notion used.

\begin{table} [!t]
	\vspace{-0em}
	\caption{Notation used in the protocols.}
	\vspace{-0.25em}
	\centering
	\resizebox{0.49\textwidth}{!}
	{
		\renewcommand{\arraystretch}{1.1}
		\begin{tabular}{l | *{1}{c} r}
			\hline \hline
			$Cert_{rsu}$ & Long-term certificate of an \acs{RSU} \\\hline 
			$CP$ & Chaff Pseudonym\\\hline 
			$CF^{i}$ & Cuckoo Filter corresponding to $PCA^{i}$ \\\hline
			$E_k(msg)$, $D_k(msg)$ & Encryption and decryption of $msg$ using key $k$\\\hline
			$(K^i_v, k^i_v)$ & \shortstack{Pseudonymous public/private key pairs} \\\hline 
			$L^{i}_{v}$ & Length of vehicle $i$ \\\hline
			$(LK_v, Lk_v)$ & Long-term public/private key pairs \\\hline
			$(msg)_{\sigma_{v}}$ & A signed message with the vehicle's private key \\\hline
			$(P^{i}_{v})_{pca}$, $P^{i}_{v}$ & A pseudonym signed by the \acs{PCA} \\\hline
			$Pos_{cmix}$, $R_{cmix}$ & The center and radius of a mix-zone \\\hline 
			$Req_{SK}/Req_{CP}/Req_{CF}$ & Requesting SK, \acs{CP}, \acs{CF} \\\hline 
			$Sign(Lk_{}, msg)$ & Signing a message with the private key ($Lk$) \\\hline 
			$SK_{cmix}^{i}$ & Symmetric Session Key inside mix-zone $i$ \\\hline 
			$t$ & Current timestamp \\\hline
			$Verify(LK_{}, msg)$ & Verifying a message with the public key \\ \hline 
			\hline
		\end{tabular}
		\renewcommand{\arraystretch}{1}
		\label{table:ieee-iot-tracking-chaff-cmix-zones-protocols-notation}
	}
	\vspace{-0em}
\end{table}

\subsection{Credentials Acquisition}
\label{subsec:ieee-iot-tracking-credential-acquisition-protocols}

A vehicle first requests an anonymous ticket~\cite{khodaei2014ScalableRobustVPKI, khodaei2016evaluating} from its \ac{LTCA}, using it to interact with the desired \ac{PCA} to obtain pseudonyms. Upon reception of a valid ticket, it generates \ac{ECDSA} public/private key pairs~\cite{1609-2016, ETSI-TS-102-940} and sends the request to the \ac{PCA}~\cite{khodaei2014ScalableRobustVPKI, khodaei2016evaluating}. Having received a request, the \ac{PCA} verifies the ticket signed by the \ac{LTCA} (assuming trust is established between the two). Then, the \ac{PCA} initiates a proof-of-possession protocol to verify the ownership of the corresponding private keys by the vehicle. Finally, the \ac{PCA} issues the pseudonyms and delivers the response. In order to achieve full unlinkability, each pseudonym should be obtained with a single ticket. For detailed security protocols and comprehensive performance evaluation, we refer interested readers to~\cite{khodaei2018Secmace, khodaei2014ScalableRobustVPKI, khodaei2018VPKIaaS, khodaei2019Scaling}.

\subsection{Cryptographic Mix-zone Participation}
\label{subsec:ieee-iot-tracking-mix-zone-participation-protocol}

\begin{figure}[t]
	\vspace{-0.75em}
	\centering
	\includegraphics[trim=8.05cm 8.5cm 1.5cm 8.5cm, clip=true, totalheight=0.32\textheight,angle=0,keepaspectratio] {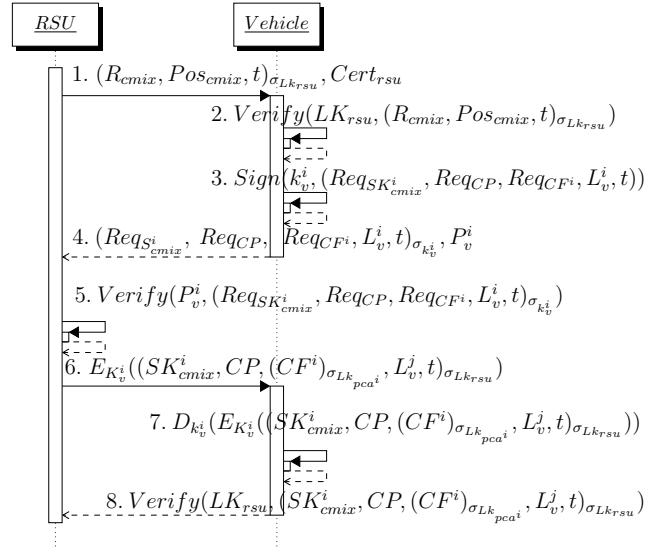}
	\vspace{-1em}
	\caption{Mix-zones advertisement \& chaff pseudonym acquisition protocols.}
	\label{fig:ieee-iot-tracking-mix-zone-participation}
	\vspace{-0.5em}
\end{figure}

Fig.~\ref{fig:ieee-iot-tracking-mix-zone-participation} shows the mix-zone advertisement and chaff pseudonym acquisition protocols. An \ac{RSU} periodically broadcasts the center of a mix-zone, $Pos_{cmix}$, its radius $R_{cmix}$, and timestamps, signed with the \ac{RSU} private key; the \ac{RSU} attaches it \ac{LTC} as well (step~\ref{fig:ieee-iot-tracking-mix-zone-participation}.1, i.e., step 1 in Fig.~\ref{fig:ieee-iot-tracking-mix-zone-participation}). To join a mix-zone, the approaching vehicle first verifies the \ac{RSU} \ac{LTC} and then the mix-zone information, by validating the signature on the message (step~\ref{fig:ieee-iot-tracking-mix-zone-participation}.2). Each vehicle needs to obtain the mix-zone symmetric session key ($SK_{cmix}^{i}$), one chaff pseudonym ($CP$), and the current \ac{CF}. It prepares a request and includes the vehicle length ($L^{i}_{v}$) and current timestamp ($t$); it signs the request with the private key corresponding to its currently valid pseudonym, and sends it to the \ac{RSU}; it attaches its pseudonyms to facilitate message validation (step~\ref{fig:ieee-iot-tracking-mix-zone-participation}.3). Upon receipt of the request (step~\ref{fig:ieee-iot-tracking-mix-zone-participation}.4), the \ac{RSU} verifies the signature (step~\ref{fig:ieee-iot-tracking-mix-zone-participation}.5) and delivers the response, first signed by the \ac{RSU} and then encrypted by the vehicle's pseudonymous public key (step~\ref{fig:ieee-iot-tracking-mix-zone-participation}.6). The response includes the mix-zone symmetric key ($SK_{cmix}^{i}$), a chaff pseudonym ($CP$), current \acp{CF} signed by the \acp{PCA} ($(CF^{i})_{\sigma_{Lk_{pca_i}}}$), and the timestamp. The \ac{RSU} also provides the length of another vehicle ($L^{j}_{v}$), which needs to be emulated, i.e., disseminating chaff \acp{CAM} for it. Finally, the vehicle decrypts the response using its private key corresponding to the currently valid pseudonym (step~\ref{fig:ieee-iot-tracking-mix-zone-participation}.7) and it verifies the signature of the message using the public key of the \ac{RSU} (step~\ref{fig:ieee-iot-tracking-mix-zone-participation}.8).

\begin{table*} [t!]
	\vspace{-0em}
	\caption{Information held by honest-but-curious system entities.}
	\vspace{-0.25em}
	\centering
	\Large
	\resizebox{1.0\textwidth}{!}
	{
		\renewcommand{\arraystretch}{3.1}
		\begin{tabular}{ | c | *{1}{c} | *{1}{c} | }
			\hline
			\Huge \textbf{Honest-but-curious (colluding) Entities} & \Huge \textbf{Information Leaked} & \Huge \textbf{Security and Privacy Implications} \\\hline
			
			\Huge \shortstack{$LTCA$} & \Huge \shortstack{\textendash\textendash\textendash\textendash\textendash\textendash\textendash} & \Huge \shortstack{\\ {} \\ An \ac{LTCA} infers no information during pseudonym changes \\ since all the communication in a mix-zone is encrypted.} \\\hline
			
			\Huge \shortstack{$PCA^{i}$} & \Huge \shortstack{$CP_{PCA^i}$} & \Huge \shortstack{\\ {} \\ A \ac{PCA} can filter out the chaff pseudonyms it issued, but it cannot link any \\ two pseudonyms upon pseudonym change or a pseudonym to a chaff one.} \\\hline 
			
			\Huge \shortstack{$RSU^{i}$ \\ {} \\ {} \\ {} \\ {}} & \Huge \shortstack{$CP^{RSU^{j}}_{PCA^{i}}, P^{i}, L^{i}_{v}$ \\ {}} & \Huge \shortstack{\\ {} An \ac{RSU} knows a \emph{distinct} set of chaff pseudonyms and the length of requesting \\ vehicle. It can link a chaff pseudonym to the pseudonym of a requesting vehicle. } \\\hline
			
			\Huge \shortstack{$LTCA$, $PCA_{H}$} & \Huge \shortstack{$CP_{PCA_H}$} & \Huge \shortstack{They can infer all chaff pseudonyms, issued by \acp{PCA}, operating in a domain.} \\\hline 
			
			\Huge \shortstack{$LTCA$, $RSU_{H}$} & \Huge \shortstack{$CP_{PCA_H}, P^{i}, L^{i}_{v}$} & \Huge \shortstack{\\ They can infer all chaff pseudonyms, the length of vehicles and their pseudonyms. } \\\hline
			
			\Huge \shortstack{$PCA_{H}$, $RSU_{H}$ \\ {} \\ {} \\ {} \\ {} \\ {} \\ {} \\ {}} & \Huge \shortstack{$CP_{PCA_H}, P_{PCA_H}, L^{i}_{v}$ \\ {} \\ {} \\ {} \\ {} \\ {} \\ {} \\ {}} & \Huge \shortstack{\\ They can infer all chaff pseudonyms issued by all the \acp{PCA} and they can \\ link a pseudonym to a chaff pseudonym. However, they cannot link \\ two successive pseudonyms as they are issued fully unlinkable.} \\\hline 
			
			\Huge \shortstack{$LTCA$, $PCA_{H}$ $RSU_{H}$} & \Huge \shortstack{$CP_{PCA_H}, P_{PCA_H}, id_{H}, L^{i}_{v}$} & \Huge 
			\makecell{\Huge \shortstack{\\ Colluding \ac{LTCA}, \acp{PCA}, and \acp{RSU} can link all successive \\ pseudonyms to their corresponding real identities.}} \\\hline 
		\end{tabular}
		\renewcommand{\arraystretch}{1}
		\label{table:ieee-iot-tracking-entity-knowledge}
	}
	\vspace{-0em}
\end{table*}

\begin{table}[!t]
	\vspace{-0.5em}
	\caption{Notation used in security \& privacy analysis.}
	\vspace{-0.25em}
	\centering
	\resizebox{0.48\textwidth}{!}
	{
		\begin{tabular}{l | *{1}{c} r}
			\hline \hline
			$LTCA^{i}$ & $LTCA^i$ operating in a domain \\\hline
			$PCA^{i}$ & $PCA^i$ operating in a domain \\\hline
			$PCA_{H}$ & A set of \acp{PCA} operating in a domain \\\hline
			$RSU^{i}$ & $RSU^{i}$ operating in a domain \\\hline
			$RSU_{H}$ & A set of \acp{RSU} operating in a domain \\\hline
			$id_{H}$ & Actual identities of the vehicles in a domain \\\hline 
			$P^{i}$ & A pseudonym issued by $\ac{PCA}^{i}$ \\\hline 
			$P_{H}$ & Pseudonyms issued by the \acp{PCA} in a domain \\\hline 
			\shortstack{\\ $CP^{RSU^{j}}_{PCA^{i}}$} & Chaff pseudonyms issued by $\ac{PCA}^{i}$ for $\ac{RSU}^{j}$ \\\hline
			$CP_{PCA_{H}}$ & Chaff pseudonyms issued by a set of \acp{PCA} in a domain \\\hline
			\hline
		\end{tabular}
		\label{table:ieee-iot-tracking-security-analysis-notations}
	}
	\vspace{-1em}
\end{table}

\subsection{\acf{CF} Acquisition}
\label{subsec:ieee-iot-tracking-tracking-chaff-dissemination}

The \acp{PCA} operating in a domain construct the \acp{CF} by inserting `enough' chaff pseudonyms. The total number of required chaff pseudonyms depends on various factors, e.g., traffic conditions and desired level of privacy protection, further evaluated in Sec.~\ref{sec:ieee-iot-tracking-quantitative-analysis}. Each \ac{PCA} pushes a (signed) distinct \ac{CF} to each \ac{RSU} operating a mix-zone. \acp{RSU} provides \acp{CF} upon request (see step 6 in Fig.~\ref{fig:ieee-iot-tracking-mix-zone-participation}). In case of being outside of an \ac{RSU} communication range, each vehicle broadcasts a signed query to its neighbors to fetch the latest \acp{CF}, e.g., similarly to~\cite{das2004spawn}. Upon receiving an authentic query for the missing \acp{CF}, each vehicle searches its local repository and randomly chooses one of the requested \acp{CF} and broadcasts it. The signed \ac{CF} is encrypted with the (pseudonymous) public key of the requesting vehicle. Upon reception, it decrypts the content using the private key of the currently valid pseudonym, it validates the signature of the \ac{CF} (signed by the \ac{PCA}), and stores them locally (evaluated in Sec.~\ref{subsec:ieee-iot-tracking-performance-evaluation}).

Each vehicle could also directly request the \ac{PCA} to obtain the \acp{CF} corresponding to the nearby mix-zones. Thus, it can filter out all chaff pseudonyms it might encounter throughout its trajectory. Upon identification of the physical location, e.g.,~\cite{papadimitratos2008secureneighborhood, papadimitratos2008protection, brands1993distance}, of the requesting vehicle, the \ac{PCA} provides the \acp{CF} corresponding to the nearby mix-zones. The vehicle-\ac{PCA} communication is over mutually authenticated \ac{TLS}~\cite{dierks2008transport} tunnels (or \ac{DTLS}~\cite{rescorla2012datagram}) leveraging the \ac{PCA}'s \ac{LTC} and the vehicle's currently valid pseudonym. Still, a vehicle might receive chaff \acp{CAM} while it has not yet received the corresponding \ac{CF} to discard them; however, it is equipped with other sensing systems, e.g., Radar and Lidar, to detect such chaff vehicles. In this case, it conducts an online pseudonym validation, e.g., \ac{OCSP}~\cite{khodaei2018Secmace, myers1999x}, to check the validity status of the pseudonym. Evaluating the impact of introducing decoy traffic on the operation of safety applications in various traffic conditions remains as one of our future work.

\subsection{Chaff Pseudonym Resolution}
\label{subsec:ieee-iot-tracking-chaff-pseudonym-resolution}

In case of a suspicious action, a report is sent to the \ac{RA}; the \ac{RA} queries the \ac{PCA} to identify the corresponding \ac{RSU}, provided the chaff pseudonym. It then sends a request to the \ac{RSU} to identify the pseudonymous identity, used to request the chaff pseudonym. Having identified the pseudonym, the \ac{RA} proceeds with the resolution process~\cite{khodaei2018Secmace}, i.e., interacting first with the \ac{PCA} and then the \ac{LTCA}. Due to lack of space,  we do not present the detailed protocol description and we refer interested readers to our earlier work~\cite{khodaei2018Secmace}.


\section{Security and Privacy Analysis}
\label{sec:ieee-iot-tracking-security-and-privacy-analysis}

We discuss how our scheme satisfies the security and privacy requirements, as well as operational requirements defined in Sec.~\ref{subsec:ieee-iot-tracking-security-privacy-requirements}. For a detailed analysis of the information held by the \ac{VPKI} entities during pseudonym provisioning, we refer to~\cite{khodaei2018Secmace, khodaei2019TMCVehicleCentricCRL}. Here, we only consider the privacy-sensitive information that can be inferred by system entities during mix-zone operations and pseudonym changes. Table~\ref{table:ieee-iot-tracking-entity-knowledge} represents the privacy implications when honest-but-curious system entities collude, based on the notation summarized in Table~\ref{table:ieee-iot-tracking-security-analysis-notations}. We do not include \ac{RCA} and \ac{RA} in our analysis; the former only authorizes the operation of other entities~\cite{khodaei2015VTMagazine}, e.g., the \ac{LTCA} and the \ac{PCA}, and the latter is involved in the process of resolution. Moreover, we do not consider the disclosure of information if vehicles collude with other system entities. The effect of colluding a set of vehicles, crossing a mix-zone operated by $\ac{RSU}^{i}$, would be equivalent to consider $\ac{RSU}^{i}$ colludes with other system entities.

All the \ac{V2X} communication in a mix-zone is encrypted and hidden from an external observer. Upon a pseudonym change in a mix-zone, an external adversary, observing the encrypted communication cannot distinguish among vehicles sets towards correlating their corresponding pseudonyms (R1). A single entity cannot fully de-anonymize a user, link two successive pseudonyms, or link a chaff pseudonym to a pseudonymous identifier of a given vehicle. An \ac{LTCA} or a \ac{PCA} can infer no information to harm user privacy during changing pseudonyms since all communication inside a mix-zone is encrypted. An external adversary observing the communication could distinguish among pseudonym and chaff pseudonym sets based on the timing information~\cite{khodaei2018Secmace}. To eliminate any distinction, the \ac{PCA} issues pseudonyms and chaff pseudonyms with fully overlapping lifetimes, thus, timing information cannot harm user privacy. Moreover, the \ac{VPKI} system issues fully unlinkable pseudonyms for all vehicles, thus, even if two pseudonyms are obtained by the same requester, they cannot be linked since each is requested using a distinct ticket~\cite{khodaei2018Secmace, khodaei2014ScalableRobustVPKI, khodaei2016evaluating}. \ac{LTCA} cannot differentiate between a chaff pseudonym and a real one. A \ac{PCA} can only differentiate chaff pseudonyms that it issued; in other words, it cannot distinguish a chaff pseudonym, issued by another \ac{PCA}, from a real one. Moreover, a \ac{PCA} cannot infer any information towards correlating a chaff pseudonym and an actual pseudonym: the \ac{RSU} randomly assigns one chaff pseudonym to a relaying vehicle.

An honest-but-curious \ac{RSU} learns the length of a requesting vehicle during mix-zone symmetric key acquisition process. However, this does not reveal additional information since the length is already included in the \acp{CAM}, frequently disseminated by the vehicle; thus, unlike the chaff-based \ac{CMIX}~\cite{vaas2018nowhere} that requires vehicles provide their intended trajectory path to the \acp{RSU}, our scheme does not provide additional information (in comparison with the \ac{CMIX} scheme~\cite{freudiger2007mix}) to the \acp{RSU}. An \ac{RSU} operating a mix-zone cannot filter out chaff pseudonyms originated from other mix-zones; this diminishes the probability of linking two successive pseudonyms belonging to the same vehicle; however, an \ac{RSU} can filter out chaff pseudonyms that it provides towards linking successive pseudonyms upon pseudonym changes in the mix-zone. We quantitatively evaluated the successful linkability in the presence of honest-but-curious \acp{RSU} in Sec.~\ref{subsec:ieee-iot-tracking-performance-comparison}. Collusion by $PCA_A$ and $PCA_B$ results in filtering out chaff pseudonyms they issued; but, they cannot observe the encrypted communication. Collusion by $RSU_{H}$ and $PCA_{H}$ enable them to decrypt the encrypted communication and filter out all chaff pseudonyms. A collusion of the \ac{LTCA}, $PCA_{H}$, and $RSU_{H}$ enable them to link all pseudonyms issued in a given domain with their real identities. As a result, they can link any pseudonym to its prior or successive pseudonyms.

Issuing chaff pseudonyms, constructing and disseminating the \acp{CF}, and validating chaff pseudonym requests are all efficient processes (see Sec.~\ref{subsec:ieee-iot-tracking-performance-evaluation}). Each \ac{RSU}, responsible for constructing a mix-zone, disseminates required information to the vehicles approaching the mix-zone, e.g., symmetric session key, mix-zone geometries, and \acp{CF}. This information is (signed by the \ac{RSU} and) encrypted using the public key of a vehicle, approaching the mix-zone. All vehicle-\ac{RSU} interactions are mutually authenticated using the currently valid vehicle's pseudonym and we leverage \acp{RSU} and car-to-car epidemic distribution to disseminate the \acp{CF} (R2). Non-cooperative vehicles could ignore changing their pseudonyms in order to degrade the anonymity set size of the mix-zone. However, as it is shown in Sec.~\ref{subsec:ieee-iot-tracking-performance-comparison}, such behavior does not degrade the user privacy protection. Vehicles could also repeatedly request to obtain multiple chaff pseudonyms from the \acp{RSU}, monopolizing a substantial portion of the chaff pseudonyms (constructed by the \ac{PCA} and pushed to the \acp{RSU}); however, each vehicle is equipped with an \ac{HSM} which guarantees all outgoing signatures are signed under the private key of a single valid pseudonym at any time. In case of deviating from the system security policy, suspicious activities or (high-rate) spurious requests are sent to the \ac{RA} to initiate a process to (possibly) resolve a pseudonym, thus identifying the long-term identity of a misbehaving vehicle, i.e., the pseudonym owner, and thus, their credentials will be revoked (R3).

The efficiency of the system stems from efficient \ac{CF} construction of chaff pseudonyms (minimal overhead on the \ac{PCA} side) and very fast validation (membership check) of chaff pseudonyms from a \ac{CF} (minimal overhead on the vehicle side) (R4). Our scheme does not introduce extra computation overhead on the \ac{RSU} side (in comparison with the \ac{CMIX} scheme~\cite{freudiger2007mix}) during mix-zone advertisement and symmetric key distribution. We allocate a small fraction of bandwidth for \acp{CF} distribution, which is sufficient to timely distribute \acp{CF} to all legitimate vehicles approaching a mix-zone (see Sec.~\ref{sec:ieee-iot-tracking-quantitative-analysis}). Our scheme introduces communication overhead to disseminate decoy traffic to enhance user privacy. In order to balance communication overhead and user privacy protection, our scheme also provides fine-grained adaptive mechanism to adjust the amount of decoy traffic in various situations, i.e., less decoy traffic during the rush-hours or more decoy traffic in sparse traffic conditions. Given a data rate of several Mbit/sec for modern IEEE 802.11p interfaces~\cite{IEEEWAVE2016}, dissemination of decoy traffic does not pose a significant communication overhead. Sec.~\ref{subsec:ieee-iot-tracking-performance-evaluation} provides a detailed quantitative analysis of our scheme on computation and communication overhead: disseminating decoy traffic for all vehicles introduces resealable computation and communication overhead.

Each \ac{CF} is signed by the corresponding \ac{PCA} which generated the chaff pseudonyms. Upon receiving a request from a vehicle, an \ac{RSU} encrypts the \ac{CF} (along with symmetric key and a chaff pseudonym) with the public key of the pseudonymous certificate of requesting vehicle. Thus, an eavesdropper cannot identify the chaff pseudonyms to filter out the decoy traffic. Moreover, an adversary cannot infer the number of active chaff pseudonyms from the size of a \ac{CF}: each \ac{PCA} overfills the \ac{CF} with extra chaff pseudonyms; thus the size of a \ac{CF} remains constant. This results in hiding the number of active chaff pseudonyms from an eavesdropper as well as diminishing the need to frequently update the \ac{CF} and re-broadcast the updated fingerprint. There is a trade-off: the higher the number of chaff pseudonyms is, the larger the \ac{CF} size becomes, thus the less frequent \ac{CF} updates and broadcast are. Obtaining a large \ac{CF} (e.g., valid for a day) could enable an adversary to filter out all chaff pseudonyms during that period. Thus, the more frequent \ac{CF} updates, the lower the vulnerability window becomes.

Changing pseudonyms require changing addresses across a vehicle's protocol stack (i.e., MAC and IP addresses) to prevent their old and new pseudonyms from being (trivially) linked based on these interfaces~\cite{papadimitratos2007architecture, fonseca2007support, hinden2003internet, eastlake2005randomness, narten2007privacy}. For the \ac{CMIX} scheme~\cite{freudiger2007mix}, each vehicle changes its addresses when changing pseudonym, i.e., once every pseudonym lifetime ($\tau_{P}$); however, by leveraging our scheme, each vehicle should change its IP and MAC addresses twice every beacon interval ($\gamma_{v}$), e.g., 20 times per second if $\gamma_{v}$ is 0.1s. To facilitate a fast handover, a vehicle could have potentially multiple virtual IP and MAC addresses at the same time, e.g.,~\cite{ishibashi2004proposal}. Thus, the relaying vehicles (and the \acp{RSU}), responsible for disseminating decoy traffic, would broadcast their actual \acp{CAM} and the chaff ones under distinct addresses. This eliminates (i) the trivial linking between the old and the new pseudonyms by the eavesdroppers, (ii) the ability of an adversary to filter out chaff \acp{CAM} based on the same interface identifier, and (iii) the need to change IP and MAC addresses twice every beacon interval.

Although communications inside the mix-zones are cryptographically protected, the physical properties of wireless radio signals, e.g., \ac{RSSI}, time of arrival, Doppler shift, etc., could be used by an adversary to localize and identify propagation path from a transmitter, e.g.,~\cite{brik2008wireless}. Tracking an object using such properties, e.g.,~\cite{vaas2018poster}, raises privacy concerns as such interfaces are uniquely associated with a single vehicle. Beyond that, by leveraging our scheme to disseminate decoy traffic, an adversary could filter out chaff \acp{CAM} from the actual ones since both are originating from the same transmitter, e.g., based on the Doppler shift and \ac{RSSI}~\cite{yao2017voiceprint, so2019physical}, or by identifying the source \ac{NIC} of an IEEE 802.11 frame~\cite{brik2008wireless}. Based on our adversarial model, an adversary cannot differentiate decoy traffic from the actual ones using the properties of physical layer device identification. Leveraging such techniques to identify vehicles based on the signal's device-of-origin and track them accordingly requires a stronger adversary with more sophisticated resources to conduct such attacks; this requires a detailed investigation and remains as our future work.

Any \ac{CMIX} scheme requires the \ac{VPKI} system issuing pseudonyms with overlapping intervals. This facilitates transition to a new pseudonym at any time, e.g., when encountering a mix-zone. All vehicles registered in the system are provided with \acp{HSM}, ensuring that private keys never leave the \acp{HSM}, thus mitigating Sybil attacks~\cite{douceur2002sybil}. Note that chaff pseudonyms and their corresponding private keys are not required to be inside the \acp{HSM}; they can be stored in the \acp{OBU}. Thus, even if a vehicle is provided with multiple chaff pseudonyms, it cannot perform Sybil-based misbehavior since such chaff pseudonyms will be ignored by other vehicles and they cannot be used for any specific application.


\section{Quantitative Analysis}
\label{sec:ieee-iot-tracking-quantitative-analysis}

In order to evaluate the performance of our scheme, we need a tracking algorithm to conduct semantic and syntactic linking attacks. We do not constrain the choice and design of the tracking algorithm and we do not dwell on its performance. There are other tracking algorithms in the literature, e.g.,~\cite{buttyan2009slow, wiedersheim2010privacy, Emara2013, vaas2018nowhere}. For example, the tracking algorithm in~\cite{vaas2018nowhere} was based on an exposure metric leveraging a vehicle's route length utilizing a pseudonym and the number of mix-zones traversed during a trip. However, in our work, we utilize a more sophisticated tracking algorithm by leveraging information in the \acp{CAM} in order to link two successive pseudonyms, thus tracking a vehicle. Due to the lack of a solid basis to compare the strength of the algorithms, we invented our own tracking algorithm, which is orthogonal to the defense mechanism.

\setlength{\textfloatsep}{0pt}
\begin{algorithm}[t!]
	\floatname{algorithm}{Algorithm}
	\caption{Syntactic and Semantic Linking Attacks}
	\label{algorithm:ieee-iot-tracking-syntactic-semantic-linking-attacks}
	\algblock{Begin}{End}
	\begin{algorithmic}[1]
		\Procedure{LinkingSuccessivePseudonymsAlgorithm}{$~$}
		\myState {Fetch eavesdropped beacon and road layout information}
		\myState {Classify eavesdropped beacons based on vehicle length}
		\myState {Create a list with the first \& last seen beacons for each identifier}
		\myState {Filter out trivially linked pseudonyms (not changing their pseudonyms)}
		\myState {$MaxTravTime$ $\gets$ Maximum time to traverse a mix-zone}
		\myState {$MinTravTime$ $\gets$ Minimum time to traverse a mix-zone}
		\For {Each $B^{i}$ in BEACON\_SET}
		\myState{$B_{f}^{i}$ is the first seen message for beacon $B^{i}$}
		\myState{$B_{l}^{i}$ is the last seen message for beacon $B^{i}$}
		\For {Each $B_{f}^{i+1}$ in BEACON\_SET}
		\myState {$B_{l}^{i}$ and $B_{f}^{i+1}$ are not correlated}
		\myState{diff $\gets$ time difference between $B_{l}^{i}$ and $B_{f}^{i+1}$}
		\If {diff $\geq$ $MinTravTime$ \&\& diff $\leq$ $MaxTravTime$}
		\If {pseudo-id for $B_{l}^{i}$ and $B_{f}^{i+1}$ not seen together}
		\If {exists a road path from $B_{l}^{i}$ to $B_{f}^{i+1}$}
		\If {$B_{f}^{i+1}$ direction is from an exit point}
		\myState {$B_{l}^{i}$ and $B_{f}^{i+1}$ are correlated}
		\myState {break}
		\EndIf
		\EndIf
		\EndIf
		\EndIf
		\EndFor
		\EndFor
		\EndProcedure
	\end{algorithmic}
\end{algorithm}

\subsection{Tracking Algorithm}
\label{subsec:ieee-iot-tracking-tracking-algorithm}

We introduce a tracking algorithm towards conducting \emph{syntactic} and \emph{semantic} linking attacks. An adversary might observe an isolated pseudonym change, and associate the old and new pseudonymous identifiers through syntactic linking. Alternatively, an adversary could leverage physical constraints of the road layout, and \acp{CAM} or \acp{DENM} payload, e.g., location, velocity, and time, to predict a vehicle's trajectory towards linking messages semantically. The goal of an adversary is to link two successive pseudonyms upon pseudonym change within a mix-zone. Algorithm~\ref{algorithm:ieee-iot-tracking-syntactic-semantic-linking-attacks} shows our tracking algorithm: it first fetches eavesdropped beacon information and the road layout information (step~\ref{algorithm:ieee-iot-tracking-syntactic-semantic-linking-attacks}.2, i.e., step 2 in Algorithm~\ref{algorithm:ieee-iot-tracking-syntactic-semantic-linking-attacks}). It then classifies beacons based on the length of the vehicles (step~\ref{algorithm:ieee-iot-tracking-syntactic-semantic-linking-attacks}.3). Next, it selects the first and the last observed beacons corresponding to each pseudonymous identifier (step~\ref{algorithm:ieee-iot-tracking-syntactic-semantic-linking-attacks}.4). It then removes the beacons that enter and exit the mix-zone with the same pseudonymous identifiers, i.e., filtering out trivially linked pseudonyms (step~\ref{algorithm:ieee-iot-tracking-syntactic-semantic-linking-attacks}.5). The minimum and maximum time to traverse a mix-zone is calculated based on the mix-zone geometry and vehicle speed limits (steps~\ref{algorithm:ieee-iot-tracking-syntactic-semantic-linking-attacks}.6~\textendash~\ref{algorithm:ieee-iot-tracking-syntactic-semantic-linking-attacks}.7). The algorithm aims at linking the last observed beacon, in fact, the one seen before entering the mix-zone, to one of the messages exiting the mix-zone. Two pseudonyms are deemed correlated (i.e., belonging to the same vehicle) if (i) the time difference between the two observed beacons is within the minimum and maximum time to traverse the mix-zone, (ii) the two pseudonyms have not been seen together (i.e., syntactic linking~\cite{khodaei2018PrivacyUniformity}), (iii) there exists a road path from the last seen beacon ($B_{l}^{i}$) to the first seen beacon ($B_{f}^{i+1}$)~\cite{wiedersheim2010privacy}, and (iv) the direction of the first seen beacon ($B_{f}^{i+1}$) is from one of the exit points of the mix-zone (steps~\ref{algorithm:ieee-iot-tracking-syntactic-semantic-linking-attacks}.8~\textendash~\ref{algorithm:ieee-iot-tracking-syntactic-semantic-linking-attacks}.25).

\begin{table}[!t]
	\vspace{-0.5em}
	\centering
	\caption{Simulation parameters for the experiments.} 
	\vspace{-0.5em}
	\label{table:ieee-iot-tracking-simulation-parameters}
	\resizebox{0.475\textwidth}{!}
	{
		\renewcommand{\arraystretch}{1.2}
		\begin{tabular}{ | c | c ||| c | c | }
			\hline 
			\textbf{Parameters} & \textbf{Value} & \textbf{Parameters} & \textbf{Value} \\\hline\hline 
			Beacon TX interval ($\gamma_{v}$) & 0.2s, 0.5s, 1s & Snapshot interval & 1s \\\hline 
			Carrier frequency & 5.89 GHz & Number of \acp{RSU} & 100 \\\hline 
			TX power & 20mW & \acp{RSU} transmission range & 600 meters \\\hline 
			Physical layer bit-rate & 18Mbps & Number of Mix-zones & 25 \\\hline 
			Sensitivity & -89dBm & Mix-zone transmission range & 100 meters \\\hline
			Thermal noise & -110dBm & MxZ Advertisement interval ($\gamma_{mz}$) & 0.5s, 1s \\\hline 
			Area size & 15KM $\times$ 15KM & Number of eavesdroppers & 25 \\\hline
			Average trip duration & 692.81s & Eavesdropping range & 250 meters \\\hline 
			Number of trips & 287,939 & Non-cooperative vehicles & 0\%-50\% \\\hline
			Number of vehicles & 138,259 & \ac{CF} distribution bandwidth ($\mathbb{B}$) & 50 KB/sec \\\hline 
			Duration of simulation & 24 hours & \ac{CF} TX interval & 1s \\\hline
			Rush-hour periods & 7-10, 12-14, 17-20 & Fraction of honest-but-curious \acp{RSU} & 0\%-100\% \\\hline
		\end{tabular}
	}
	\vspace{0.5em}
\end{table}

\subsection{Experimental Setup}
\label{subsec:ieee-iot-tracking-experimental-setup}

We use OMNET++, the PREXT project~\cite{Emara2017, prext-github}, and the Veins framework~\cite{veins-framework} to simulate a large-scale scenario using SUMO~\cite{behrisch2011sumo} with a realistic mobility trace, the \acs{LuST} dataset~\cite{codeca2015lust}. \ac{V2X} communication is over IEEE 802.11p~\cite{IEEEWAVE2016}. For \ac{CF} dissemination, we assume there is one \ac{PCA}, even though the extension of our implementation with multiple \acp{PCA} is straightforward. \acp{RSU} broadcast a \ac{CF} data structure, constructed and signed by a \ac{PCA}. For \ac{CF} operations (insertion and membership test), we used PYBLOOM~\cite{pybloom}. To effectively place the \acp{RSU} and mix-zones, we sorted the intersections with the highest numbers of vehicles passing by. We then placed 100 \acp{RSU} and selected 25 to be mix-zones based on these \emph{highly-visited} intersections with non-overlapping radio ranges. This aims at maximizing the chance for the vehicles to cross at least one mix-zone during their trajectory. We configured the transmission range of \acp{RSU} and mix-zones to be 600 and 100 meters, respectively. Near each mix-zone, we placed an eavesdropper with receiving antennas (250 meters range) capturing all broadcasted beacons. But, it cannot observe the communication within a cryptographically protected mix-zone. Vehicles are provided with pseudonyms with overlapping intervals, compatible with the proposals of standardization bodies, i.e., IEEE 1609.2 WG~\cite{1609-2016} and \acs{ETSI}~\cite{ETSI-TS-102-940}. They enter a cryptographically protected mix-zone and change their pseudonyms inside it. \acp{RSU} randomly assign a percentage of vehicles to be \emph{relaying} ones to broadcast decoy traffic. Table~\ref{table:ieee-iot-tracking-simulation-parameters} shows the simulation parameters.

\subsection{Metrics}
\label{subsec:ieee-iot-tracking-metrics}

To evaluate the performance of our scheme, we measure \emph{end-to-end delay} to obtain \acp{CF} of chaff pseudonyms, i.e., from the time a vehicle approaches a mix-zone until it successfully downloads them. The maximum allocated bandwidth for \acp{CF} distribution, i.e., system parameter $\mathbb{B}$, is chosen to be much smaller than $C$, the bandwidth the data link support. We choose a small amount of bandwidth (50 KB/s) in order not to interfere with safety-critical operations. We also evaluate additional computation and communication overhead, imposed by our scheme, on the overall \ac{VC} system components.

For privacy evaluation\footnote{There are different metrics for quantifying location privacy, e.g., anonymity set size, distortion~\cite{shokri2009distortion}, and exposure degree~\cite{Jin:J:2019a}. Here, we focus on a fundamental metric to quantify location privacy. The selection of an optimal metric for quantifying location privacy and a full-blown comparison by leveraging various metrics warrant a separate investigation.}, we consider \emph{average successful linkability}, the ratio of correctly linking two successive pseudonyms, $A$ and $B$, belonging to the same vehicle (by leveraging Algorithm~\ref{algorithm:ieee-iot-tracking-syntactic-semantic-linking-attacks}). We also consider \emph{average tracking duration}, i.e., average traversed distance by any single vehicle that the adversary (eavesdropper deployed across multiple locations) can cumulatively track. This implies the cumulative successive correct linking of pseudonyms (and thus \acp{CAM}) across multiple \acp{CMIX} (including the trajectories from one \ac{CMIX} to the next one).

\begin{figure} [!t]
	\vspace{-0em}
	\begin{center}
		\centering
		\includegraphics[trim=0cm 0cm 0cm 0cm, clip=true, width=0.42\textwidth,height=0.42\textheight,keepaspectratio]{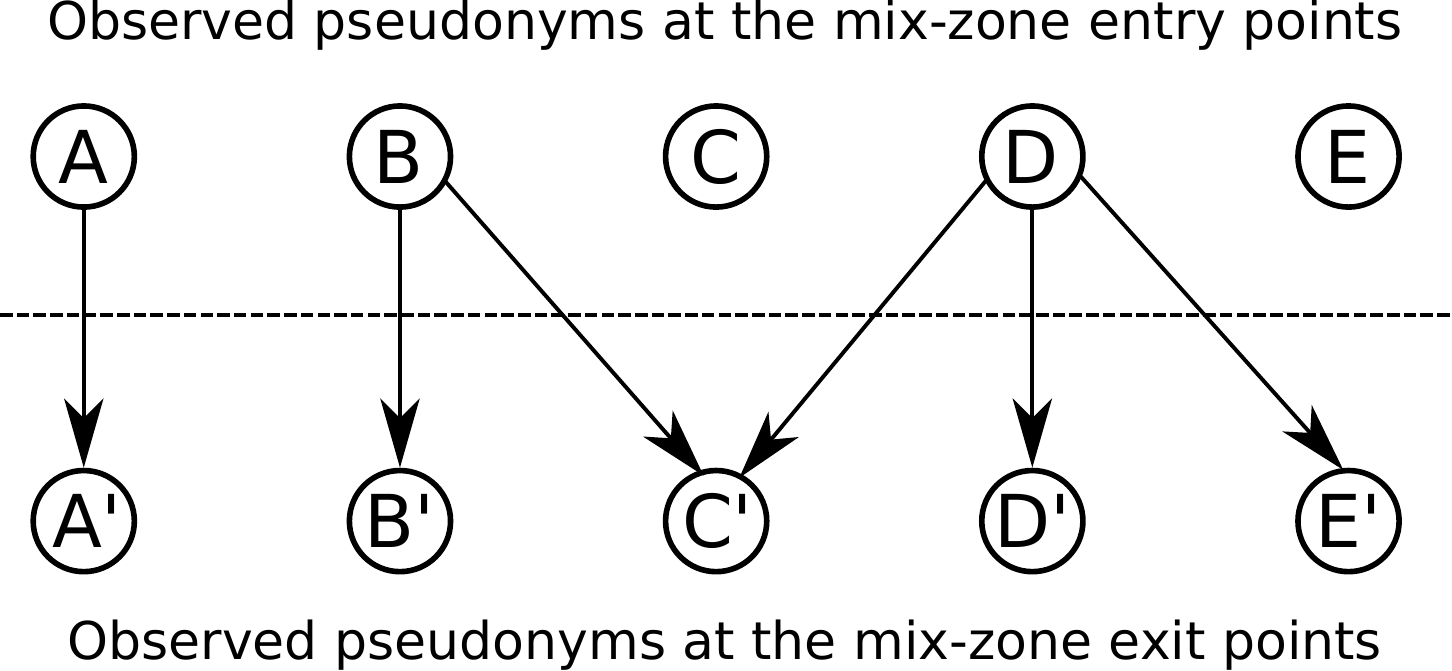}
		\vspace{-0.5em}
		\caption{Pseudonym transition from the adversary's viewpoint before entering and after exiting a mix-zone (without decoy traffic); the ground truth: $[A, A'], [B, B'], [C, C'], [D, D'], [E, E']$. Each arrow shows potential linkability between two pseudonyms based on Algorithm~\ref{algorithm:ieee-iot-tracking-syntactic-semantic-linking-attacks}.}
		\label{fig:ieee-iot-tracking-success-rate-probability}
	\end{center}
	\vspace{-0em}
\end{figure}

Fig.~\ref{fig:ieee-iot-tracking-success-rate-probability} exemplifies a snapshot of pseudonym transitions from an adversary's viewpoint before and after a mix-zone. There are 5 pseudonym transitions happened in the mix-zone. The eavesdropper leverages algorithm~\ref{algorithm:ieee-iot-tracking-syntactic-semantic-linking-attacks} towards linking successive pseudonyms. Each arrow shows potential linkability between two pseudonyms based on algorithm~\ref{algorithm:ieee-iot-tracking-syntactic-semantic-linking-attacks}.The linkability success rate is calculated as the probability of linking two pseudonyms belonging to the same vehicle. With respect to the ground truth, the probability of success rate for linking pseudonym $A$, $Pr_{A}$, is 1, $Pr_{B}$ is $\frac{1}{2}$, $Pr_{C}$ and $Pr_{E}$ are zero, and $Pr_{D}$ is $\frac{1}{3}$. Thus, the success rate for this example is {\scriptsize $\dfrac{1 + \dfrac{1}{2} + \dfrac{1}{3}}{5} = 0.36$}.

\subsection{Summary of Results}
\label{subsec:ieee-iot-tracking-summary-of-results}

\begin{figure} [!t]
	\vspace{-1.25em}
	\begin{center}
		\centering
		\subfloat[]{
			\hspace{-1.0em} 
			\includegraphics[trim=0.25cm 0.15cm 0.5cm 0.45cm, clip=true, width=0.265\textwidth,height=0.27\textheight,keepaspectratio]{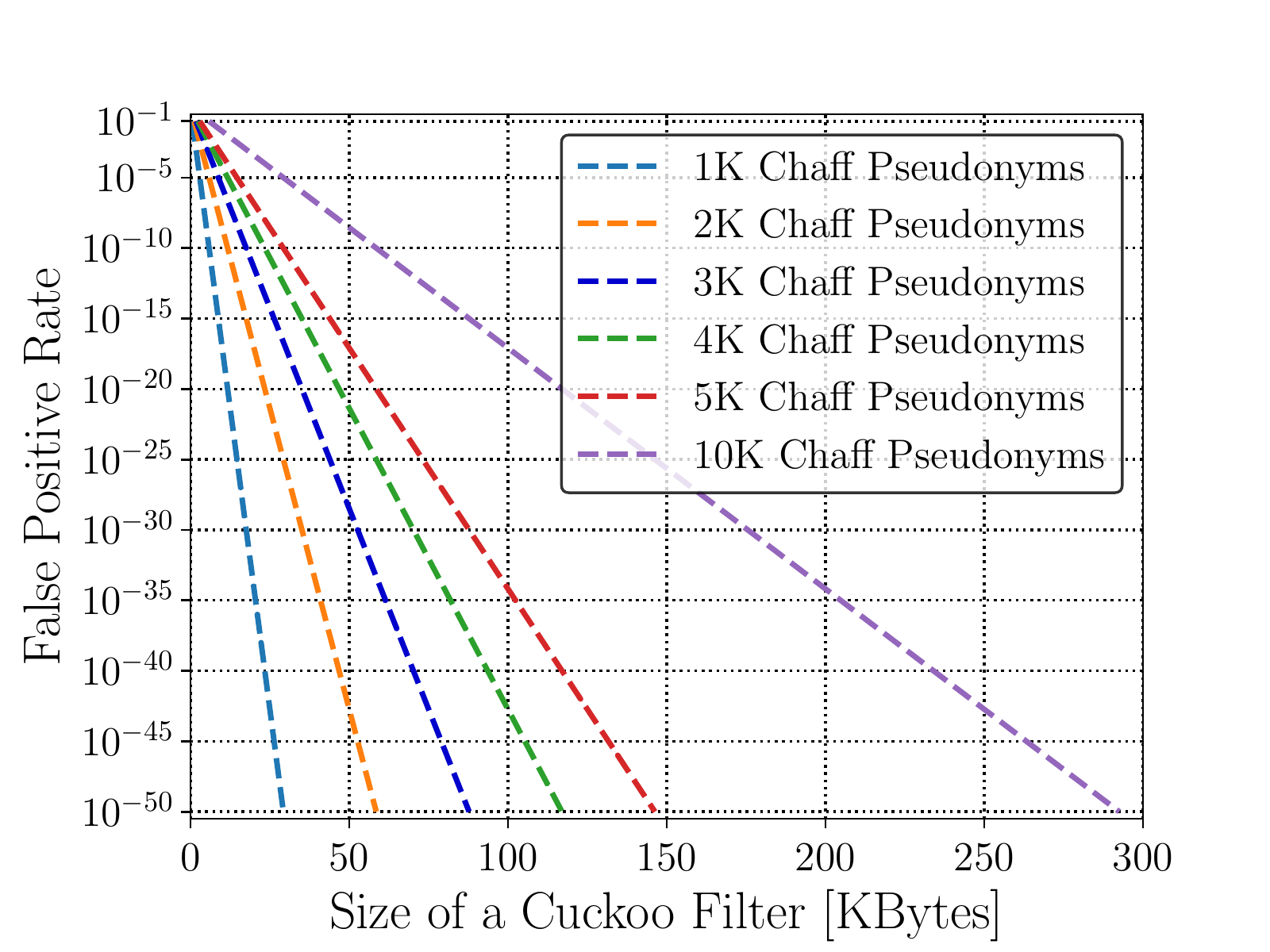}}
		\subfloat[]{
			\hspace{-1.25em} 
			\includegraphics[trim=0.25cm 0.15cm 0.25cm 0.45cm, clip=true, width=0.265\textwidth,height=0.27\textheight,keepaspectratio]{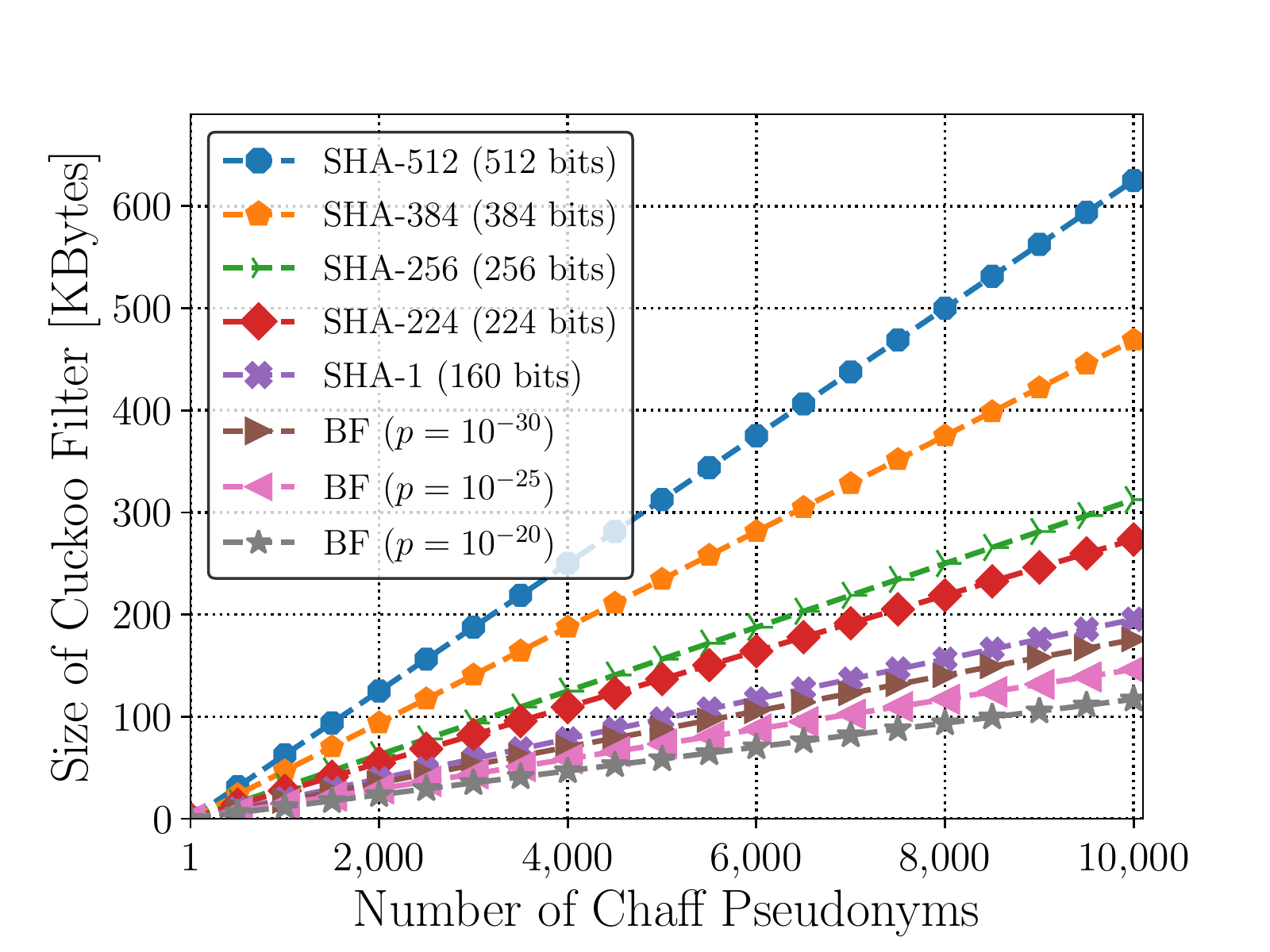}}
		\vspace{-1.15em}
		\caption{(a) The size of a \ac{CF} as a factor of false positive rate. (b) The size of a \ac{CF} as a factor of chaff pseudonyms numbers.}
		\label{fig:ieee-iot-tracking-CF-size-comparison}
	\end{center}
	\vspace{-0em}
\end{figure}

Our scheme incurs low communication and computation overhead: the size of a \ac{CF} with 1K chaff pseudonyms ($\rho=10^{-25}$) is 14.63 KB, which is sufficient to disseminate decoy traffic\footnote{The percentage of decoy traffic indicates the percentage of (relaying) vehicles disseminating chaff \acp{CAM}.} for 50\% of vehicles by all the mix-zones for an hour. Moreover, such information can be timely disseminated across a region: $F_x(t=6 \: ms)=0.99$ (Fig.~\ref{fig:ieee-iot-tracking-CF-size-comparison} and Fig.~\ref{fig:ieee-iot-tracking-e2d-delay-obtaining-CF-evaluation}(a)). Further, the additional computation overhead for a vehicle to validate a chaff pseudonym by performing a \ac{CF} membership test with 1K chaff pseudonyms ($\rho=10^{-25}$) is 3.68e-4 ms, which is highly efficient and scalable even with modest Nexcom \acp{OBU}~\cite{feiri2015} (Fig.~\ref{fig:ieee-iot-tracking-e2d-delay-obtaining-CF-evaluation}(b) and Table~\ref{table:ieee-iot-tracking-chaff-psnyms-validation-bloomfilter-membership-item}).

We compare our scheme with the \ac{CMIX}~\cite{freudiger2007mix}, namely the \emph{baseline} scheme, and the chaff-based \ac{CMIX}~\cite{vaas2018nowhere}. Enhancing user privacy, i.e., preventing linking two successive pseudonyms by disseminating decoy traffic (for all vehicles) incurs low communication and commutation overhead: in comparison with the baseline scheme~\cite{freudiger2007mix}, the average communication overhead by \acp{RSU} increases from 0.26 KB/sec to 0.88 KB/sec, the average computation overhead for an \ac{RSU} increases from 0.6 ms to 0.64 ms, and the average computation overhead on the vehicle side increases from 2.05 ms to 14 ms (Fig.~\ref{fig:ieee-iot-tracking-communication-computation-comparison}). However, even with the modest computing resources, this extra computation overhead is reasonably low (Fig.~\ref{fig:ieee-iot-tracking-communication-computation-comparison}).

\begin{figure} [!t]
	\vspace{0.25em}
	\begin{center}
		\centering
		\subfloat[]{
			\hspace{-1.4em} 
			\includegraphics[trim=0.5cm 0.15cm 0.1cm 0.65cm, clip=true, width=0.275\textwidth,height=0.275\textheight,keepaspectratio]{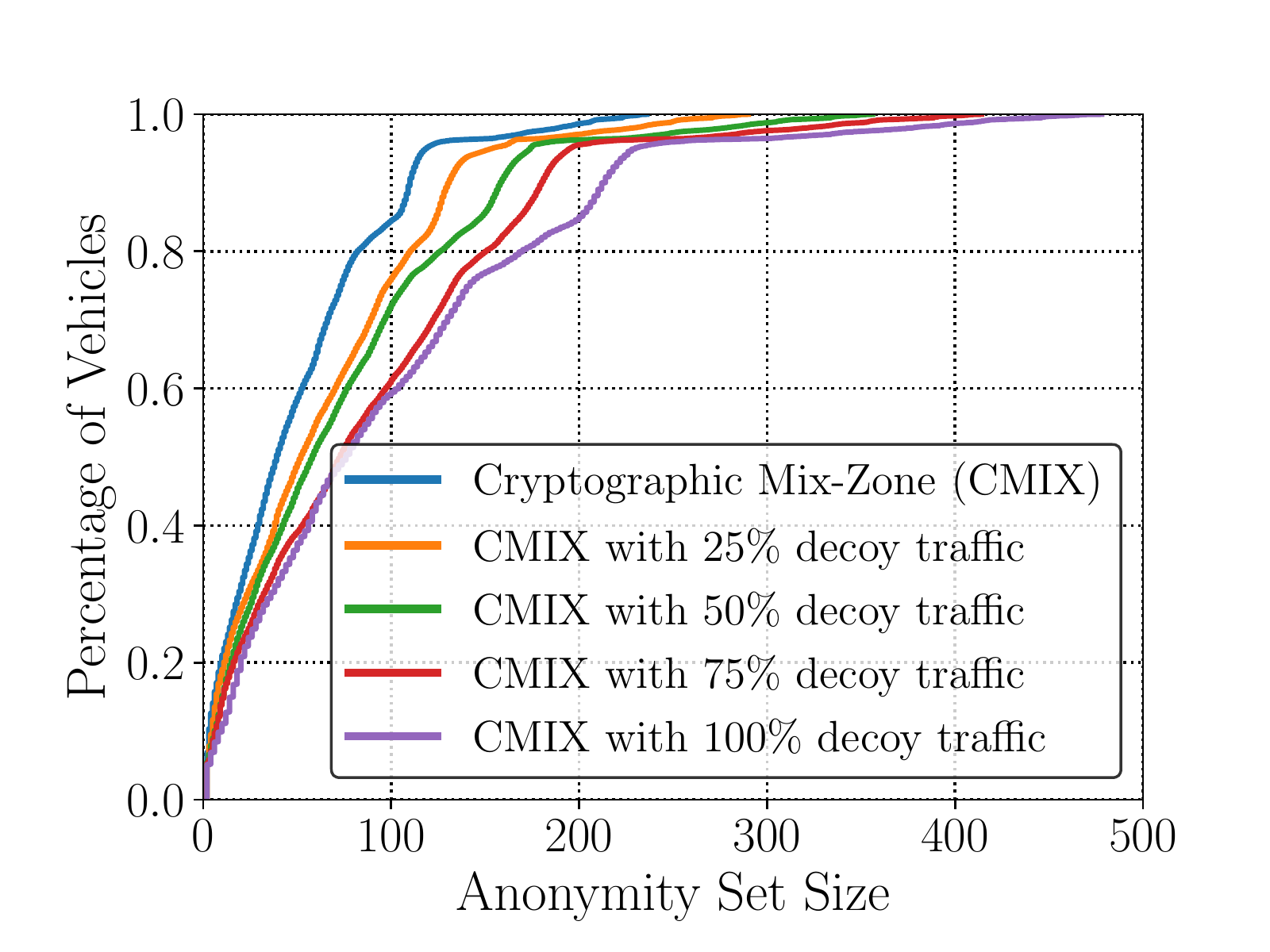}}
		\subfloat[]{
			\hspace{-1.6em} 
			\includegraphics[trim=0.5cm 0.15cm 0.1cm 0.65cm, clip=true, width=0.265\textwidth,height=0.265\textheight,keepaspectratio]{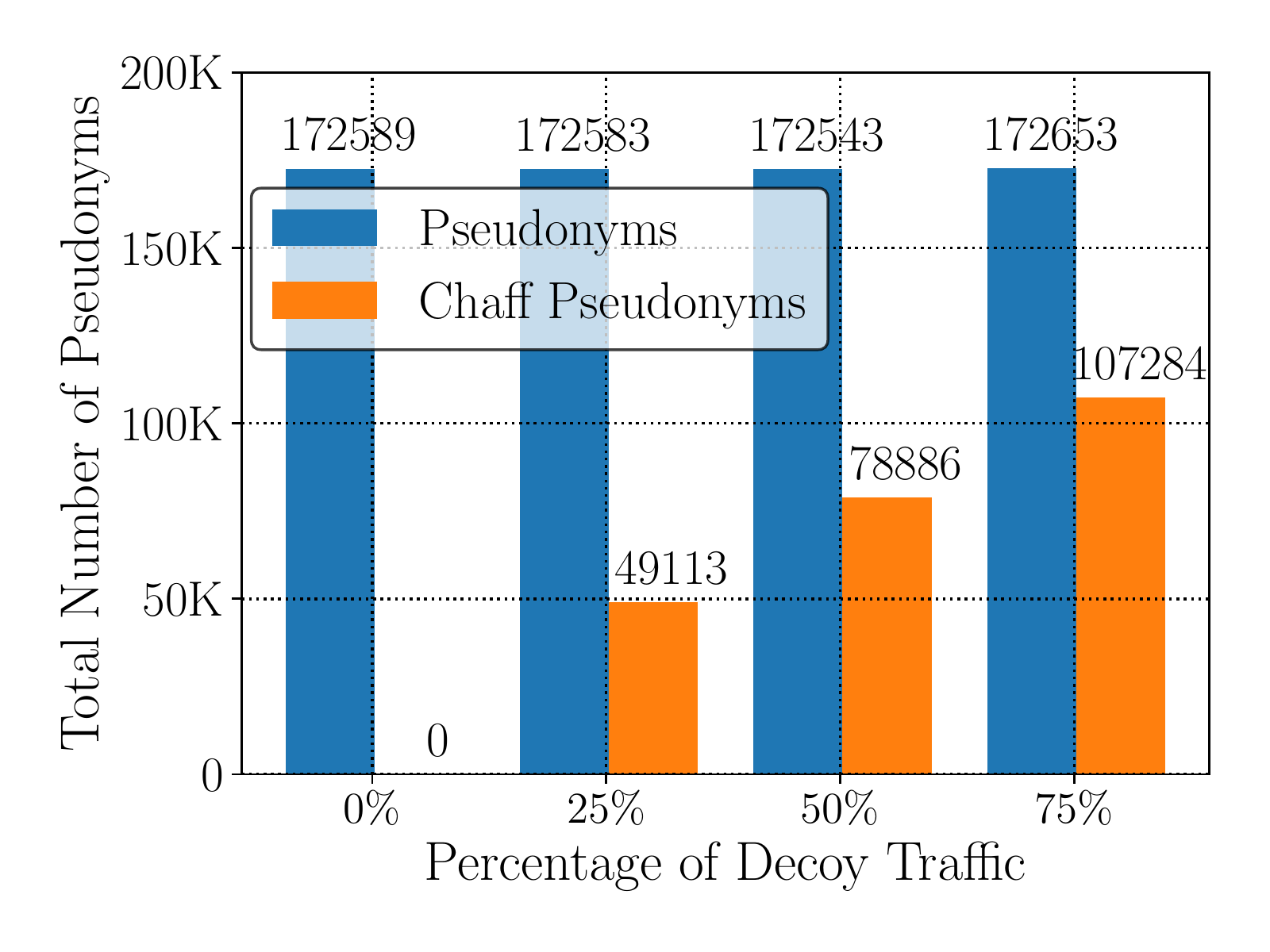}}
		\vspace{-0.75em}
		\caption{(a) \acs{CDF} of anonymity set size for \ac{CMIX} and \ac{CMIX} with decoy traffic, observed by the eavesdroppers. (b) Total number of disseminated pseudonyms and chaff pseudonyms, derived from the vehicles ($\gamma_{v}$\emph{=0.5s}).} 
		\label{fig:ieee-iot-tracking-cdf-anonymity-set-size}
	\end{center}
	\vspace{-0.25em}
\end{figure}

Our scheme outperforms prior works~\cite{freudiger2007mix, vaas2018nowhere}: for the baseline scheme~\cite{freudiger2007mix}, an eavesdropper could successfully link $\approx$68\% of pseudonyms after vehicles change their pseudonyms in a mix-zone. For chaff-based \ac{CMIX}~\cite{vaas2018nowhere} with 50\% decoy traffic, the average probability of linking pseudonyms is $\approx$50\%. In contrast, by leveraging our scheme with decoy traffic for 50\% of the vehicles, the average probability of linking pseudonyms is $\approx$19\% (Fig.~\ref{fig:ieee-iot-tracking-cdf-anonymity-set-size}, Fig.~\ref{fig:ieee-iot-tracking-success-tracking-rates-evaluation}, Fig.~\ref{fig:ieee-iot-tracking-success-tracking-rates-comparision-with-two-baseline-schemes}, Fig.~\ref{fig:ieee-iot-tracking-success-tracking-distance}, Fig.~\ref{fig:ieee-iot-tracking-histogram-psnyms-changes-percentage-of-successfully-linked-psnyms}, and Fig.~\ref{fig:ieee-iot-tracking-histogram-tracked-distance-comparison}). Even in the presence of non-cooperative vehicles, not changing their pseudonyms while crossing the mix-zones, the average successful tracking is reasonably low (Fig.~\ref{fig:ieee-iot-tracking-success-tracking-rates-with-non-cooperative-vehicles}).

\subsection{Performance Evaluation}
\label{subsec:ieee-iot-tracking-performance-evaluation}

Representing chaff pseudonyms in a space-efficient \ac{CF} trades off communication overhead for a false positive rate ($p$)~\cite{Fan2014}. Fig.~\ref{fig:ieee-iot-tracking-CF-size-comparison}(a) shows that the \ac{CF} size linearly increases as the false positive rate decreases. For example, for 1000 chaff pseudonyms with $p = 10^{-30}$ (with the optimal number of hash functions), the \ac{CF} size is 17.55 KB. This eliminates the need to validate chaff \acp{CAM}, thus enabling the correct functionality of safety applications. Note that there could be multiple \acp{CF} from different \acp{PCA} and the chaff pseudonyms are pro-actively integrated into \acp{CF} while they are updated over time, i.e., removing the expired ones and adding new ones. Given a data rate of several Mbit/sec for modern IEEE 802.11p interfaces~\cite{IEEEWAVE2016}, dissemination of \ac{CF} updates do not pose a significant communication overhead. For example, the average number of chaff pseudonyms, per hour, for all the mix-zones to disseminate 25\%, 50\%, 75\%, and 100\% of decoy traffic is 688, 1140, 1567, and 1929, respectively.

\begin{figure} [!t]
	\vspace{-0em}
	\begin{center}
		\centering
		\subfloat[Communication Latency]{
			\hspace{-1.5em} 
			\includegraphics[trim=0.5cm 0.15cm 0.5cm 0.85cm, clip=true, width=0.27\textwidth,height=0.27\textheight,keepaspectratio]{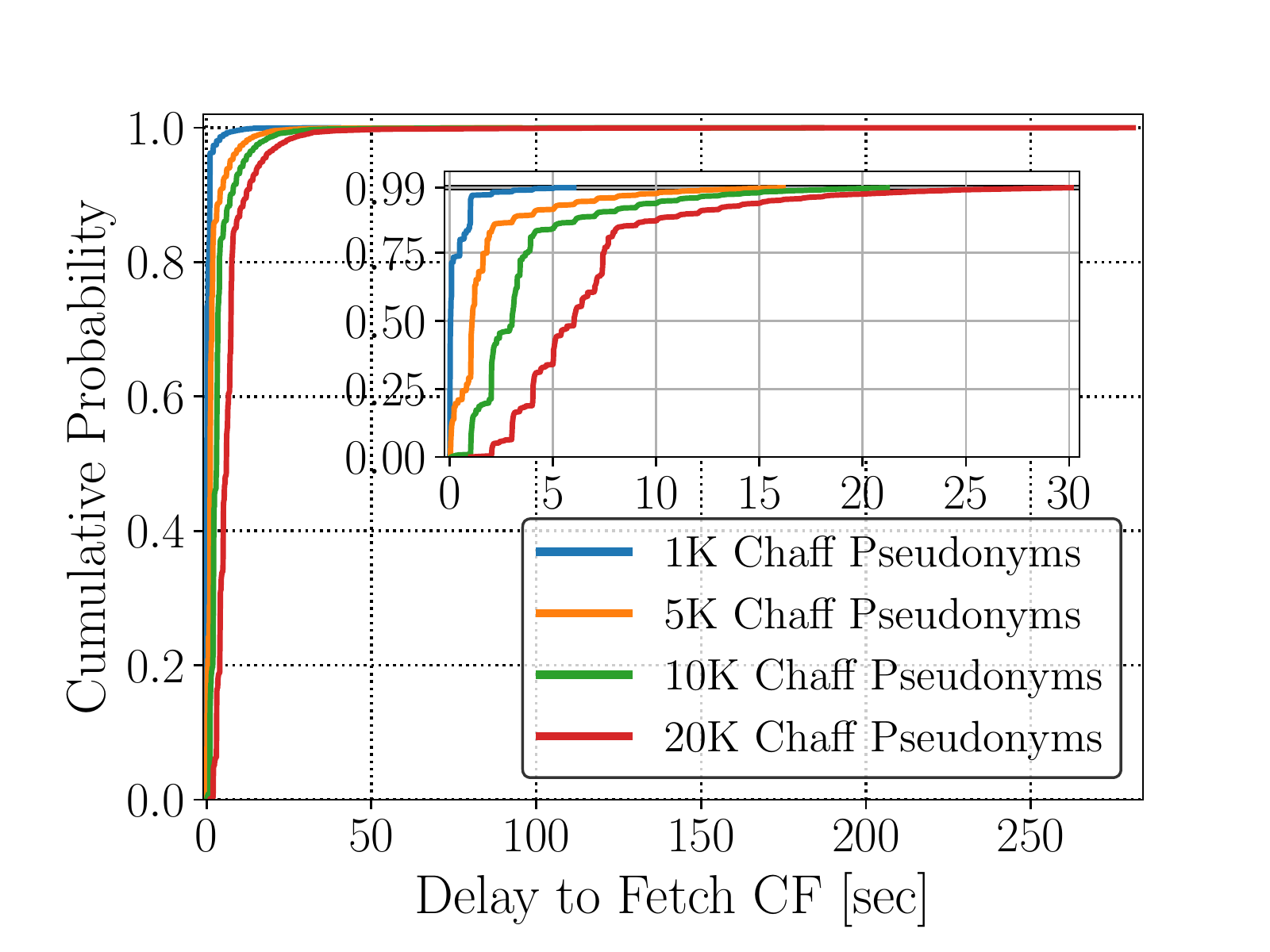}}
		\subfloat[Computation Latency]{
			\hspace{-1.6em} 
			\includegraphics[trim=0.5cm 0.15cm 0.5cm 0.85cm, clip=true, width=0.27\textwidth,height=0.27\textheight,keepaspectratio]{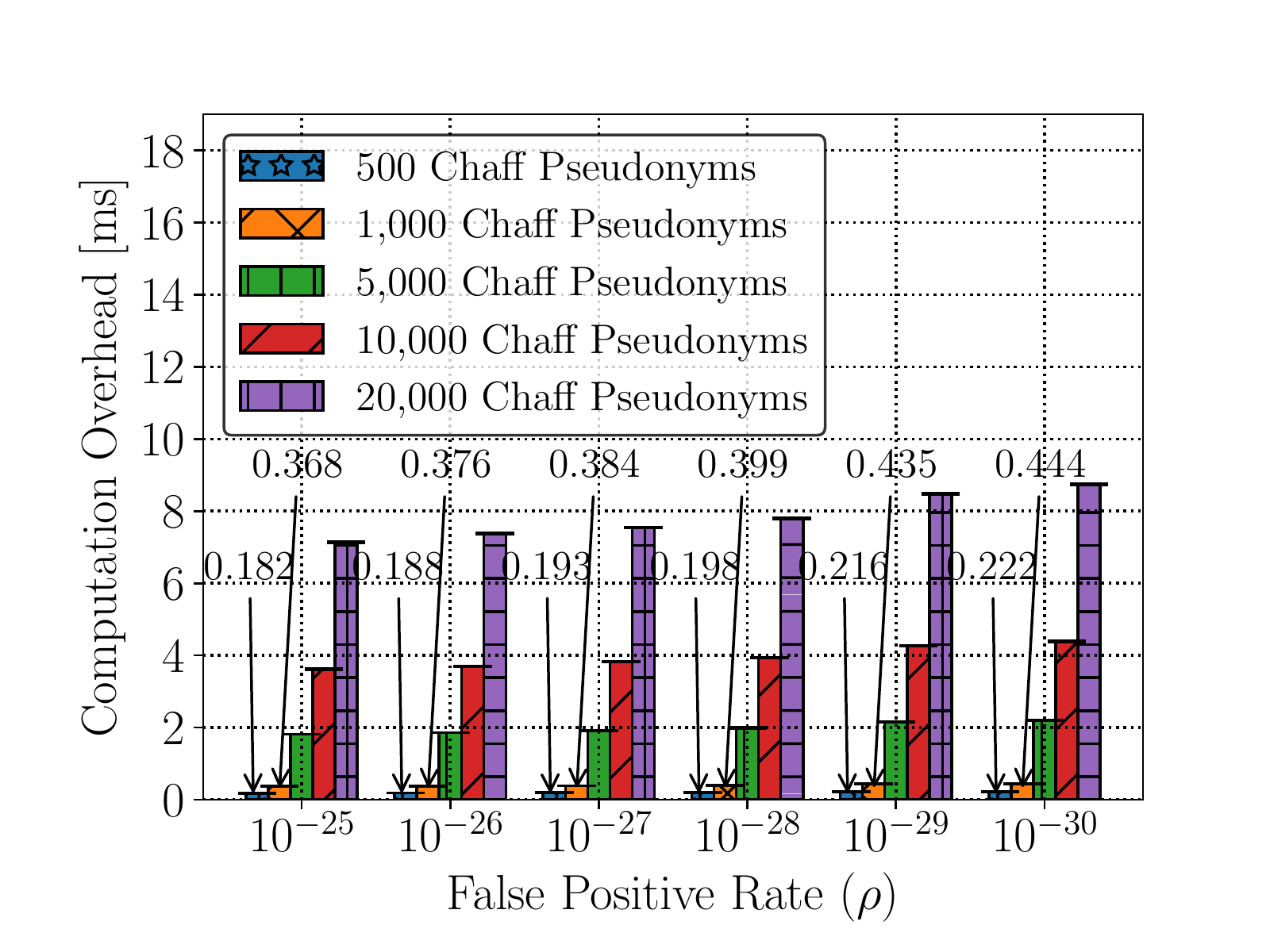}}
		\vspace{-1em}
		\caption{(a) End-to-end delay to obtain \ac{CF} of chaff pseudonyms to vehicles approaching mix-zones ($\rho=10^{-30}$, $\mathbb{B}=50KB/s$, $\gamma_{v}=0.5s$, $\gamma_{rsu}=1s$). (b) Computation overhead to validate chaff pseudonyms.}
		\label{fig:ieee-iot-tracking-e2d-delay-obtaining-CF-evaluation}
	\end{center}
	\vspace{-0em}
\end{figure}

\begin{table}[!t]
	\centering
	\vspace{-1em}
	\caption{Latency for validation of chaff pseudonyms using a \ac{CF}, executed on a Nexcom \ac{OBU}, averaged over 5000 runs.} 
	\vspace{-0.25em}
	\label{table:ieee-iot-tracking-chaff-psnyms-validation-bloomfilter-membership-item}
	\resizebox{0.5\textwidth}{!}
	{
		\renewcommand{\arraystretch}{0.99}
		\hspace{-1em}
		\begin{tabular}{|c|c|c|c|c|c|c|c|}
			\hline
			\rowcolor{LightCyan} \multirow{1}{*}{\textbf{\shortstack{chaff pseudonyms}}} & \multirow{1}{*}{\textbf{\shortstack{false positive}}} & \multicolumn{2}{c|}{\textbf{\ac{CF} size}} & \multicolumn{2}{c|}{\textbf{delay}} & \multicolumn{2}{c|}{\textbf{check/sec.}} \\
			\cline{1-8}
			\hline\hline
			\rowcolor{LightCyan} 500 & p=$10^{-25}$ & \multicolumn{2}{c|}{\shortstack{{} \\ 7.31 KBytes}} & \multicolumn{2}{c|}{0.182 ms} & \multicolumn{2}{c|}{2740} \\\hline
			\cline{2-8}
			\rowcolor{LightCyan} 1000 & p=$10^{-25}$ & \multicolumn{2}{c|}{\shortstack{{} \\ 14.63 KBytes}} & \multicolumn{2}{c|}{0.368 ms} & \multicolumn{2}{c|}{2719} \\\hline
			\cline{2-8}
			\rowcolor{LightCyan} 5000 & p=$10^{-25}$ & \multicolumn{2}{c|}{\shortstack{{} \\ 73.13 KBytes}} & \multicolumn{2}{c|}{1.814 ms} & \multicolumn{2}{c|}{2755} \\\hline
			\cline{2-8}
			\rowcolor{LightCyan} 10000 & p=$10^{-25}$ & \multicolumn{2}{c|}{\shortstack{{} \\ 146.26 KBytes}} & \multicolumn{2}{c|}{3.611 ms} & \multicolumn{2}{c|}{2769} \\\hline
			\cline{2-8}
			\rowcolor{LightCyan} 20000 & p=$10^{-25}$ & \multicolumn{2}{c|}{\shortstack{{} \\ 292.51 KBytes}} & \multicolumn{2}{c|}{7.135 ms} & \multicolumn{2}{c|}{2803} \\\hline
			\hline
			\rowcolor{LightCyan} 500 & p=$10^{-30}$ & \multicolumn{2}{c|}{\shortstack{{} \\ 8.78 KBytes}} & \multicolumn{2}{c|}{0.222 ms} & \multicolumn{2}{c|}{2254} \\\hline
			\cline{2-8}
			\rowcolor{LightCyan} 1000 & p=$10^{-30}$ & \multicolumn{2}{c|}{\shortstack{{} \\ 17.55 KBytes}} & \multicolumn{2}{c|}{0.444 ms} & \multicolumn{2}{c|}{2254} \\\hline
			\cline{2-8}
			\rowcolor{LightCyan} 5000 & p=$10^{-30}$ & \multicolumn{2}{c|}{\shortstack{{} \\ 87.75 KBytes}} & \multicolumn{2}{c|}{2.191 ms} & \multicolumn{2}{c|}{2282} \\\hline
			\cline{2-8}
			\rowcolor{LightCyan} 10000 & p=$10^{-30}$ & \multicolumn{2}{c|}{\shortstack{{} \\ 175.51 KBytes}} & \multicolumn{2}{c|}{4.387 ms} & \multicolumn{2}{c|}{2279} \\\hline
			\cline{2-8}
			\rowcolor{LightCyan} 20000 & p=$10^{-30}$ & \multicolumn{2}{c|}{\shortstack{{} \\ 351.02 KBytes}} & \multicolumn{2}{c|}{8.735 ms} & \multicolumn{2}{c|}{2289} \\\hline
		\end{tabular}
	}
	\vspace{0.5em}
\end{table}

The \ac{PCA} can concatenate hash values for chaff pseudonyms. Fig.~\ref{fig:ieee-iot-tracking-CF-size-comparison}(b) compares our \ac{CF}-based chaff pseudonyms fingerprint size with the five approved hash algorithms~\cite{dang2008recommendation}: SHA-1, SHA-224, SHA-256, SHA-384 and SHA-512, each producing hash digest size of 160, 224, 256, 384 and 512 bits, respectively. For instance, by employing SHA-256 (32 bytes output size) as the pseudonym serial number, the size of a fingerprint for 5,000 chaff pseudonyms becomes 156 KB; whereas employing our scheme results in 73.13 KB ($p=10^{-25}$) or 87.75 KB for the extremely low false positive rate ($p=10^{-30}$). Alternatively, one can utilize truncated hash digests; however, truncated message digests must be carefully used: if the message digest length is too small, computation of pre-image, second pre-image or collisions becomes feasible~\cite{gerbet2015power}. All in all, truncation will not guarantee the expected security strength of a hash digest~\cite{dang2008recommendation}. For a detailed investigation of different types of attacks on \acp{CF} (or \acp{BF}), we refer readers to~\cite{khodaei2019TMCVehicleCentricCRL, gerbet2015power}.

Fig.~\ref{fig:ieee-iot-tracking-e2d-delay-obtaining-CF-evaluation}(a) shows the \ac{CDF} of end-to-end latencies to obtain the \acp{CF} with different number of chaff pseudonyms. We consider \emph{end-to-end latency metric}, i.e., from the time a vehicle approaches a mix-zone until it successfully downloads all `pieces' of a \ac{CF}. Vehicle beacon frequency is $\gamma_{v}=0.5s$ and \acp{RSU} beacon frequency is $\gamma_{rsu}=1s$. The maximum allocated bandwidth to disseminate the \acp{CF} is $\mathbb{B}=50$ KB/sec. In general, the higher the number of chaff pseudonyms, the larger \ac{CF} size, thus the higher the latency to download \acp{CF}. For example, with 1000 chaff pseudonyms in a \ac{CF} ($\rho=10^{-30}$), 99\% of the vehicles approaching a mix-zone received \ac{CF} in less than 5s: $F_x(t=5s)=0.99$. Fig.~\ref{fig:ieee-iot-tracking-e2d-delay-obtaining-CF-evaluation}(b) compares the computation delays for validating chaff pseudonyms in a \ac{CF} with different number of inserted chaff pseudonyms. We performed our experiments on the Nexcom \ac{OBU} boxes~\cite{feiri2015} (Dual-core 1.66 GHz, 1GB memory). For example, the average latency to perform a 1000 membership check for a \ac{CF} with 1000 chaff pseudonyms ($\rho=10^{-25}$) is $\approx$0.368 ms, i.e., the average latency to validate one chaff pseudonym is 3.68e-4 ms. Table~\ref{table:ieee-iot-tracking-chaff-psnyms-validation-bloomfilter-membership-item} shows the latency for validating chaff pseudonyms with different false positive rates. For example, the latency to validate 10K chaff pseudonyms with 10K items in a \ac{CF} with $p=10^{-25}$ is $\approx$3.611 ms, i.e., 3.611e-4 ms to validate one chaff pseudonym. This is the extra overhead to filter out a chaff pseudonym, if seen, and discard all upcoming \acp{CAM}, signed under the private key of such a pseudonym. This shows that our scheme incurs minimal overhead on the vehicle side to filter out chaff pseudonyms from the real ones. We do not include latency for the \ac{PCA} to insert chaff pseudonyms into a \ac{CF}. Conducting such efficient operations on a \ac{PCA}, with stronger computational resources, does not impose significant overhead. We refer to~\cite{khodaei2019TMCVehicleCentricCRL} on the evaluation of latency to insert items into a \ac{CF}.

\subsection{Performance Comparison}
\label{subsec:ieee-iot-tracking-performance-comparison}

\begin{figure} [!t]
	\vspace{-0em}
	\begin{center}
		\centering
		\subfloat[Overhead Comparison]{
			\hspace{-1em} 
			\includegraphics[trim=0.65cm 0.9cm 0.5cm 0.75cm, clip=true, width=0.25\textwidth,height=0.25\textheight,keepaspectratio]{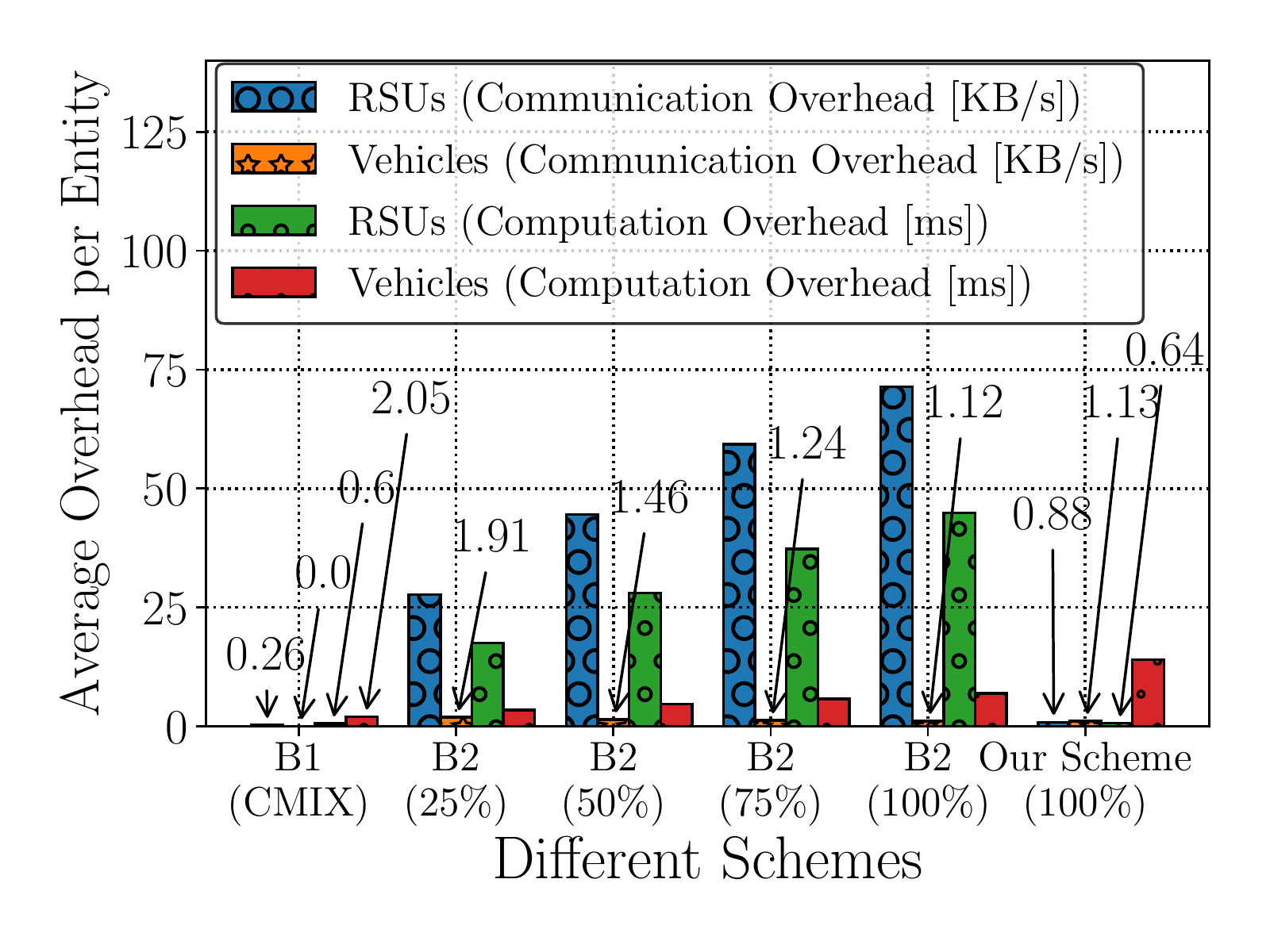}}
		\subfloat[Communication Comparison]{
			\hspace{-0.5em} 
			\includegraphics[trim=0.5cm 0cm 0.5cm 0.75cm, clip=true, width=0.25\textwidth,height=0.25\textheight,keepaspectratio]{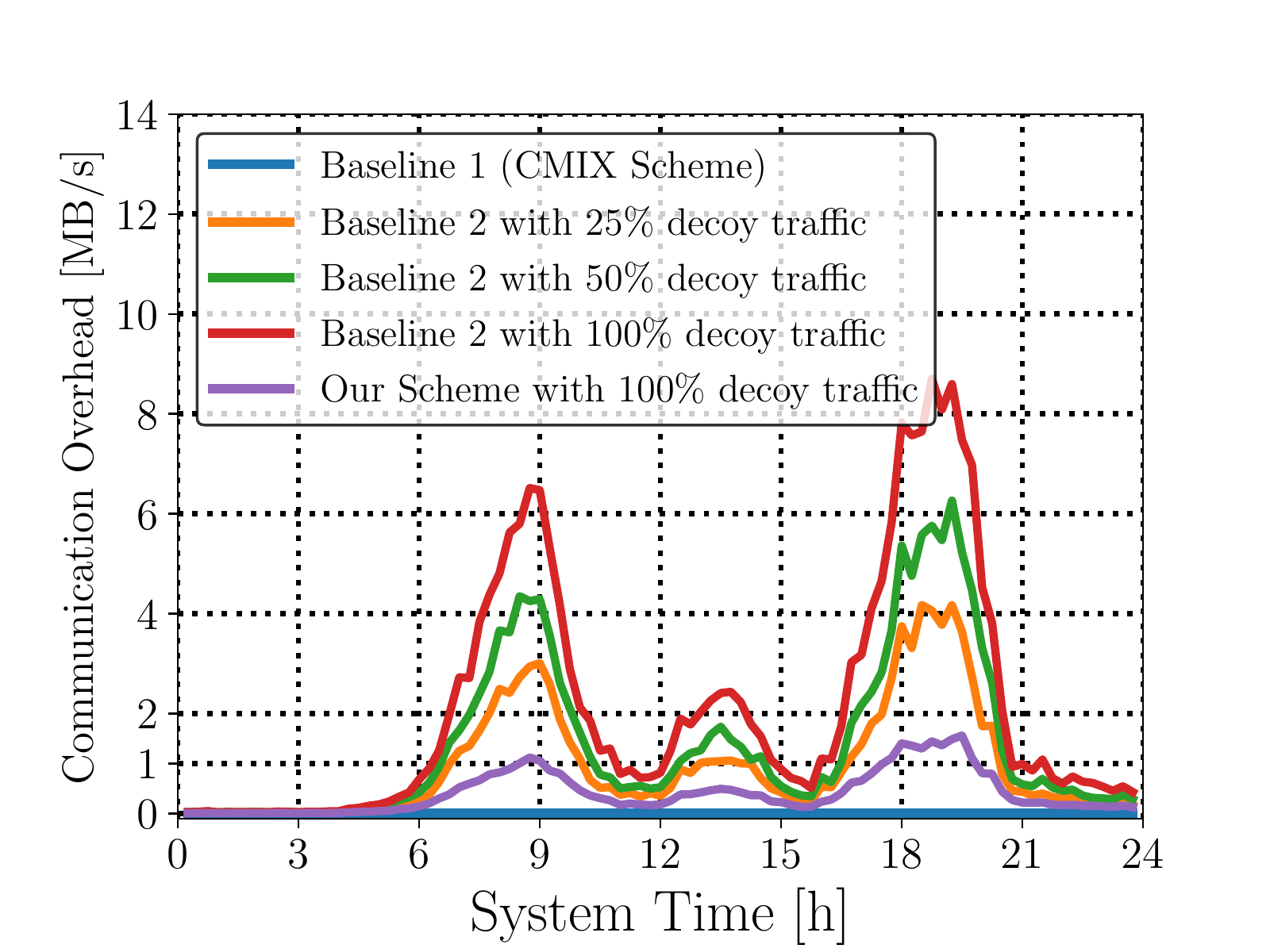}}
		\vspace{-0.75em}
		\caption{Comparison among \ac{CMIX} (B1)~\cite{freudiger2007mix}, chaff-based \ac{CMIX} (B2)~\cite{vaas2018nowhere}, and our scheme: 1K chaff pseudonyms in a \ac{CF} with $\rho=10^{-25}$; beacon frequency: $\gamma_{mz}=0.5$, $\gamma_{v}=0.2$. (a) Computation and communication overhead. (b) Communication overhead, averaged every 300s.}
		\label{fig:ieee-iot-tracking-communication-computation-comparison}
	\end{center}
	\vspace{-0em}
\end{figure}

Fig.~\ref{fig:ieee-iot-tracking-cdf-anonymity-set-size}(a) shows the \ac{CDF} of anonymity set size for the baseline and our scheme: dissemination of decoy traffic would increase the anonymity set size of the vehicles inside the mix-zones. This would diminish the power of an adversary to successfully track vehicles upon exiting the mix-zones. This obviously trades off communication overhead for a higher privacy protection level (see Fig.~\ref{fig:ieee-iot-tracking-communication-computation-comparison}(b)). Fig.~\ref{fig:ieee-iot-tracking-cdf-anonymity-set-size}(b) shows the total number of pseudonyms and chaff pseudonyms for the baseline (0\% of decoy traffic) and our scheme: the higher the percentage of decoy traffic is, the higher the number of chaff pseudonyms needed. For example, with 50\% of vehicles chosen to be relaying nodes, $\approx$78K chaff pseudonyms needed for 24 hours (for all 25 mix-zones). This is helpful to disseminate the decoy traffic via \acp{RSU} and the relaying vehicles. From our experimental results, the average number of chaff pseudonyms, per hour, required to disseminate 50\% decoy traffic for each mix-zone is 52. Thus, a \ac{PCA} (assuming there is only one) needs to construct a distinct \ac{CF} with 52 chaff pseudonyms for each mix-zone.

Fig.~\ref{fig:ieee-iot-tracking-communication-computation-comparison}(a) compares the computation and communication overhead for the \ac{CMIX}~\cite{freudiger2007mix}, chaff-based \ac{CMIX}~\cite{vaas2018nowhere}, and our scheme. For our experiments, we assumed that \acp{RSU} are configured with \emph{n1-standard-1} on the \ac{GCP}~\cite{gcp}. With this setup, the signature generation latency for \ac{ECDSA} 256 key size is $\approx$0.3 ms and verification latency is $\approx$0.4 ms. Vehicles are provided with Nexcom boxes~\cite{feiri2015}: the signature generation latency is $\approx$3 ms and the verification latency is $\approx$3.5 ms. In our experiments, the size of a pseudonym (and a chaff pseudonym) is 140 bytes and the size of a \ac{CAM} is 350 bytes~\cite{1609-2016, ETSI-302-637-2, US-EU-V2V-V2I-2013, Kenney2013}. For the \ac{CMIX} scheme~\cite{freudiger2007mix}, the computation and communication overhead is minimal: the average communication overhead for an \ac{RSU} is 0.26 KB/sec and the average computation overhead is 0.6 ms. The communication overhead on the vehicles is zero while the average computation overhead is 2.05 ms. For the chaff-based \ac{CMIX}~\cite{vaas2018nowhere}, vehicles could request to obtain pre-generated \acp{CAM} from an \ac{RSU}, operating a mix-zone, for the remaining trip duration. Our scheme outperforms the chaff-based \ac{CMIX}~\cite{vaas2018nowhere}: by disseminating 100\% decoy traffic for our scheme, the average communication overhead for an \ac{RSU} is 0.88 KB/sec while for the chaff-based \ac{CMIX}, the overhead is $\approx$72 KB/sec. The average computation overhead for an \ac{RSU} with our scheme is 0.64 ms while this is 45 ms for the chaff-based \ac{CMIX} scheme. Leveraging our scheme incurs higher computation overhead on the vehicle side in comparison with the chaff-based \ac{CMIX} scheme~\cite{vaas2018nowhere} due to generation of chaff \acp{CAM}: the computation overhead for our scheme is 14 ms while this is 7 ms for the chaff-based \ac{CMIX} scheme. However, even with the modest Nexcom box~\cite{feiri2015} computing resources, this extra computation overhead remains reasonably low.

Fig.~\ref{fig:ieee-iot-tracking-communication-computation-comparison}(b) shows the total communication overhead for the \ac{CMIX} scheme~\cite{freudiger2007mix}, chaff-based \ac{CMIX} scheme~\cite{vaas2018nowhere}, and our scheme. Mix-zones advertisement beaconing frequency is $\gamma_{mz}=0.5$ and vehicles broadcast \acp{CAM} at the frequency of $\gamma_{v}=0.2$. The size of a \ac{CF} with 1K chaff pseudonyms ($\rho=10^{-25}$) is 14.63 KB. The communication overhead is averaged every 300 seconds. As the figure shows, the total communication overhead for the \ac{CMIX} scheme~\cite{freudiger2007mix} is minimal, i.e., 6.146 KB/sec. However, the communication overhead for the chaff-based \ac{CMIX}~\cite{vaas2018nowhere} scheme reaches $\approx$8 MB/sec when broadcasting chaff \acp{CAM} for all the vehicles. This is mainly due to the pre-generation of chaff \acp{CAM} by the \acp{RSU}. More precisely, \acp{RSU} pre-generate chaff \acp{CAM} and delivers the relaying vehicles. Thus, it has significant communication overhead during chaff \acp{CAM} acquisition process. In contrast, our scheme imposes reasonable communication overhead to the system even with the dissemination of decoy traffic for all the vehicles: the total communication overhead during the rush-hours reaches $\approx$1-1.5 MB/sec. This is due to the fact the each relaying vehicle would only receive a chaff pseudonym (along with the private key) from an \ac{RSU}; thus, it has minimal communication overhead during chaff pseudonym acquisition.

\begin{figure} [!t]
	\vspace{0.75em}
	\begin{center}
		\centering
		\hspace{-2.5em} \includegraphics[trim=3.25cm 0cm 0.5cm 0.5cm, clip=true, width=0.54\textwidth,height=0.54\textheight,keepaspectratio]{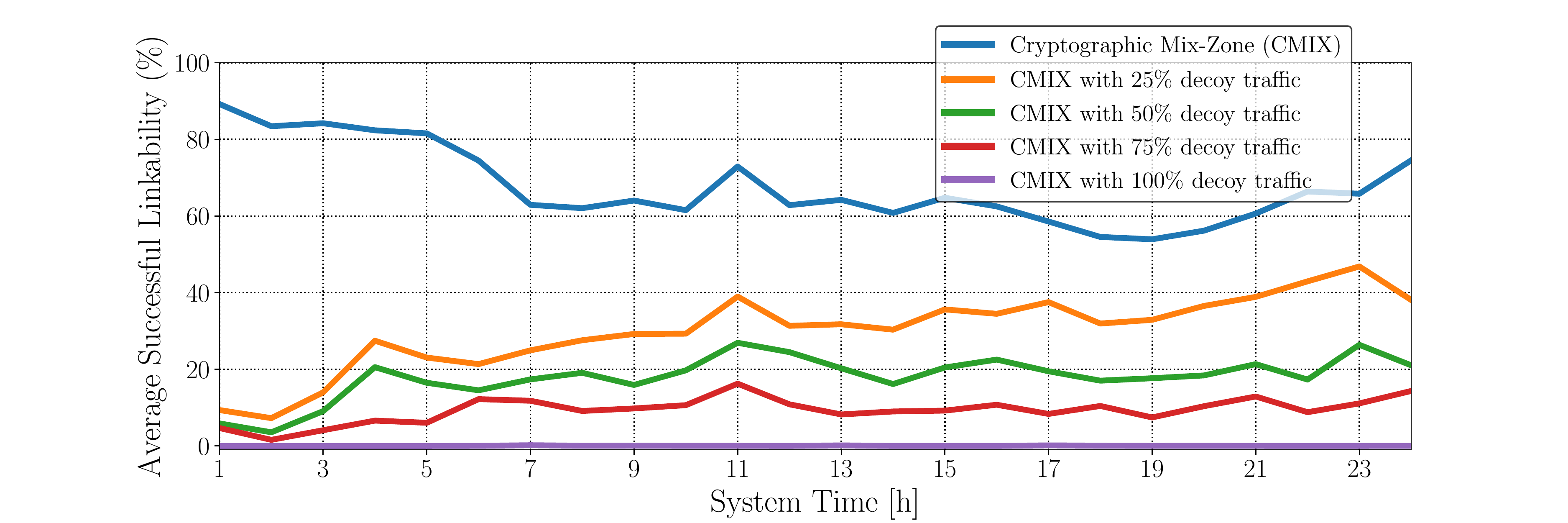}
		\vspace{-1em}
		\caption{Average successful linkability comparison with the \ac{CMIX} scheme~\cite{freudiger2007mix} through conducting syntactic and semantic linking attacks.}
		\label{fig:ieee-iot-tracking-success-tracking-rates-evaluation}
	\end{center}
	\vspace{-0em}
\end{figure}

Based on the ground truth (included in the simulation results) and leveraging our novel tracking algorithm, we compute the \emph{average successful linkability metric} towards linking pseudonyms before and after a cryptographically protected mix-zone. Fig.~\ref{fig:ieee-iot-tracking-success-tracking-rates-evaluation} shows the average pseudonym linkability by the eavesdroppers for a full-day realistic mobility pattern in the city of Luxembourg~\cite{codeca2015lust}. As we can see, the tracking algorithm could link pseudonyms for the \ac{CMIX} scheme with high probability success rate during the non-rush hours period (until system time 6). The probability of linking two successive pseudonyms decreases when the traffic density increases; but still, it can successfully link the pseudonyms with $\approx$63\% success rate at system time 7. By introducing decoy traffic for a fraction of vehicles, one can reduce the linkability: with 50\% of vehicles to be the relaying vehicles, broadcasting decoy traffic, the probability of linking drops from $\approx$63\% to $\approx$17\% at system time 7. More so, one can eliminate (syntactic and semantic) pseudonym linking attacks by disseminating decoy traffic for all vehicles.

If the number of vehicles in a mix-zone is less than a predefined (system parameter) threshold, the \ac{RSU} generates decoy traffic for all those vehicles. This stems from the results of tracking algorithm: if there are few vehicles inside a mix-zone, an adversary could easily track all those vehicles. In our simulation, we defined this threshold to be two, i.e., if there are one or two vehicles in a mix-zone, the \ac{RSU} disseminates decoy traffic for all vehicles. This is also visible in Fig.~\ref{fig:ieee-iot-tracking-success-tracking-rates-evaluation}: during very sparse traffic conditions (at system time 1), the average successful tracking is $\approx$7\%-9\%. Intuitively, the rate of decoy traffic should be inversely proportional to the traffic density, i.e., the higher the number of vehicles inside a mix-zone, the lower the probability of linking becomes, thus the less the number of chaff vehicles needed. Still, one needs to disseminate decoy traffic during the rush-hour periods: the probability of linking two successive pseudonyms during rush-hours, e.g., 7:00-10:00, is $\approx$ 62\%. This trades off pseudonyms unlinkability for (communication and computation overhead) cost, which is important for balancing the effects of chaff messages on communication overhead in dense traffic scenarios.

\begin{figure} [!t]
	\vspace{-0em}
	\begin{center}
		\centering
		\subfloat[During Non-rush Hours]{
			\hspace{-0.75em} 
			\includegraphics[trim=0.65cm 0.1cm 0.5cm 1.25cm, clip=true, width=0.25\textwidth,height=0.25\textheight,keepaspectratio]{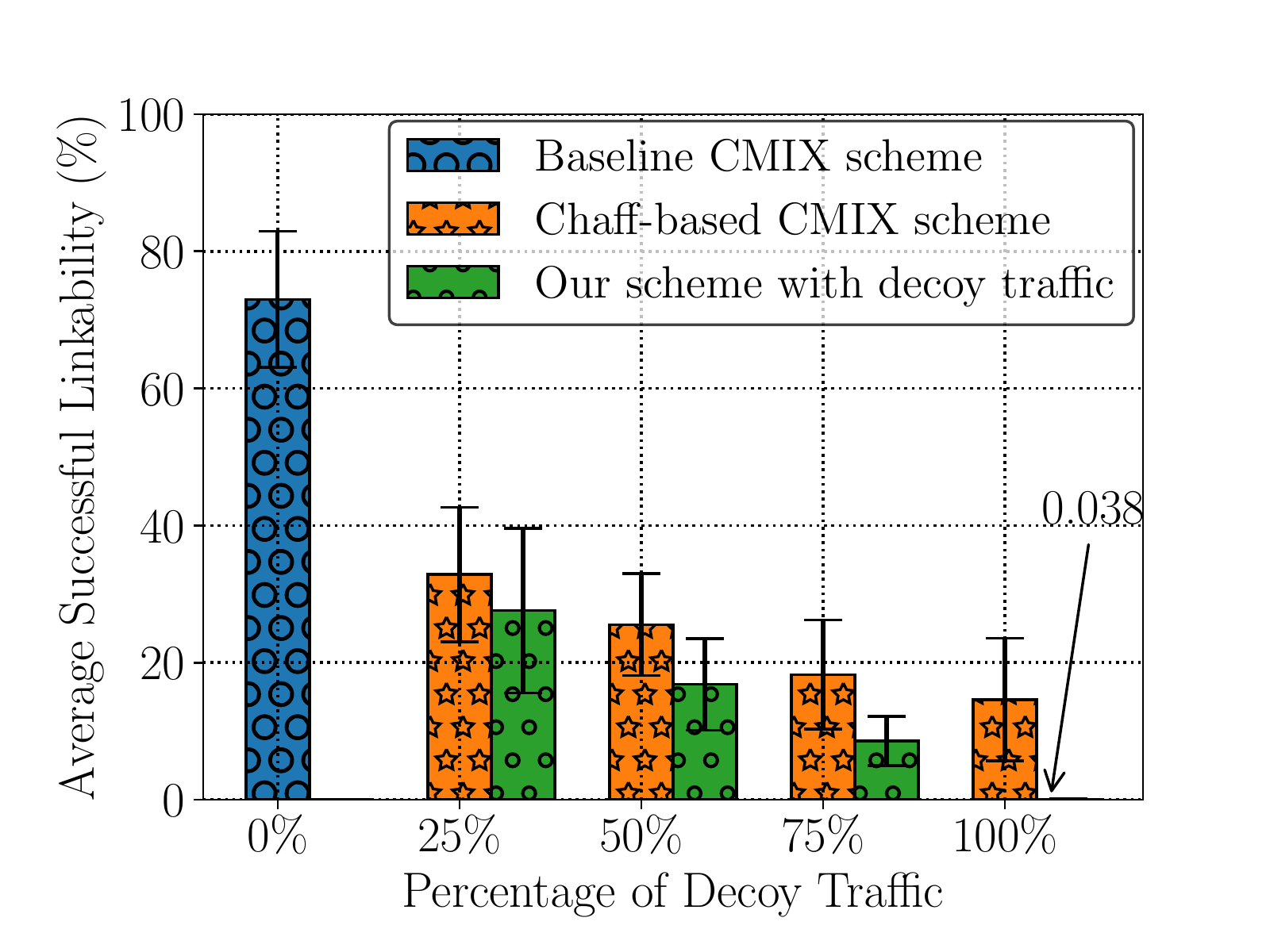}}
		\subfloat[During Rush Hours]{
			\hspace{-0.5em} 
			\includegraphics[trim=0.65cm 0.1cm 0.5cm 1.25cm, clip=true, width=0.25\textwidth,height=0.25\textheight,keepaspectratio]{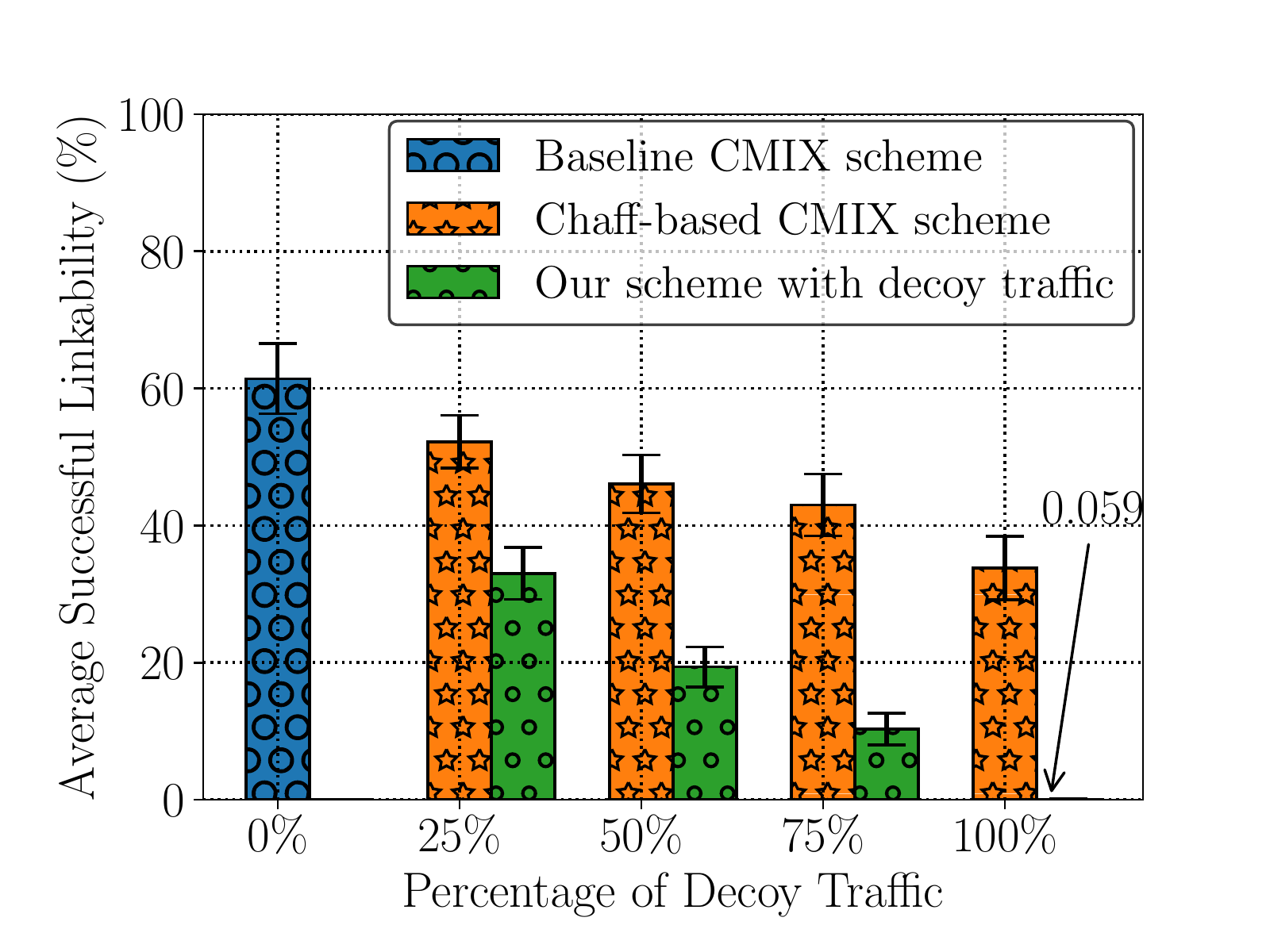}}
		\vspace{-0.75em}
		\caption{Average successful linkability comparison with the \ac{CMIX}~\cite{freudiger2007mix} and the chaff-based \ac{CMIX}~\cite{vaas2018nowhere} schemes.}
		\label{fig:ieee-iot-tracking-success-tracking-rates-comparision-with-two-baseline-schemes}
	\end{center}
	\vspace{-0em}
\end{figure}

Fig.~\ref{fig:ieee-iot-tracking-success-tracking-rates-comparision-with-two-baseline-schemes} compares the average successful linkability for the \ac{CMIX} scheme~\cite{freudiger2007mix}, chaff-based \ac{CMIX} scheme~\cite{vaas2018nowhere}, and our scheme. The average successful linkability for the \ac{CMIX} scheme~\cite{freudiger2007mix} during non-rush hours (Fig.~\ref{fig:ieee-iot-tracking-success-tracking-rates-comparision-with-two-baseline-schemes}(a)) is $\approx$73\%; during rush-hours (Fig.~\ref{fig:ieee-iot-tracking-success-tracking-rates-comparision-with-two-baseline-schemes}(b)), it is $\approx$62\%. With the chaff-based \ac{CMIX} scheme~\cite{vaas2018nowhere}, an adversary could filter out chaff \acp{CAM} from the real ones since the chaff \acp{CAM} are pre-generated by the \acp{RSU} (without considering the vehicles mobility pattern): if the distance of two \acp{CAM}, signed under two distinct pseudonyms, is less than the length of a vehicle, a chaff vehicle would stand out. The higher the percentage of decoy traffic, the higher the probability of filtering out chaff \acp{CAM}, thus the higher the average successful linkability. In contrast, with our schemes, vehicles disseminate chaff \acp{CAM} according to the traffic conditions. For the chaff-based \ac{CMIX} scheme~\cite{vaas2018nowhere} with 50\% decoy traffic, the average successful linkability, during the rush hours (Fig.~\ref{fig:ieee-iot-tracking-success-tracking-rates-comparision-with-two-baseline-schemes}(b)), is $\approx$46\% while with the same set up for our scheme, the average successful linkability, during the rush hours, is $\approx$19\%.

\begin{figure} [!t]
	\vspace{-0em}
	\begin{center}
		\centering
		\subfloat[]{
			\hspace{-0.75em} 
			\includegraphics[trim=0.65cm 0.9cm 0.5cm 0.5cm, clip=true, width=0.25\textwidth,height=0.25\textheight,keepaspectratio]{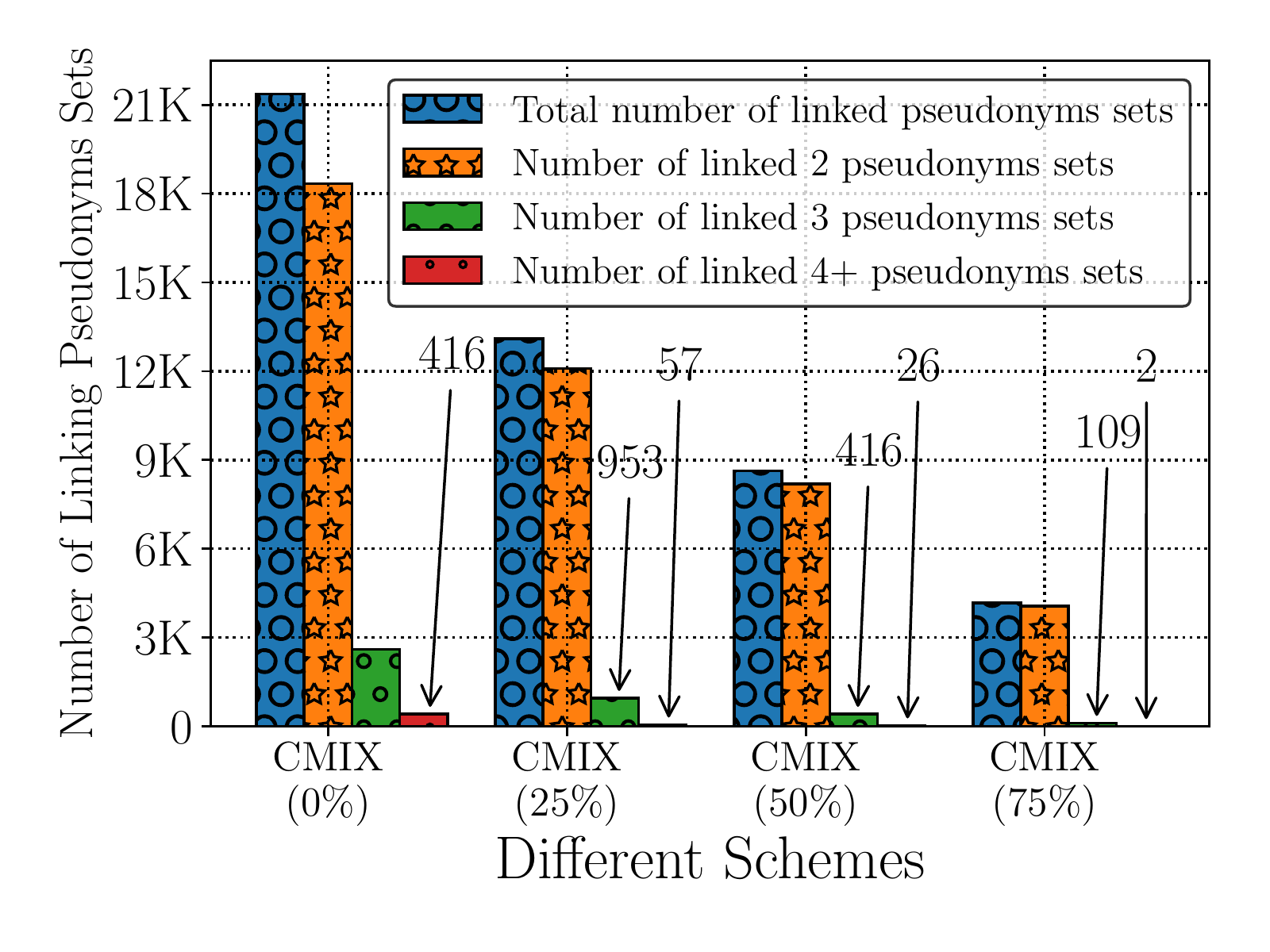}}
		\subfloat[]{
			\hspace{-0.5em} 
			\includegraphics[trim=0.65cm 0.9cm 0.5cm 0.5cm, clip=true, width=0.25\textwidth,height=0.25\textheight,keepaspectratio]{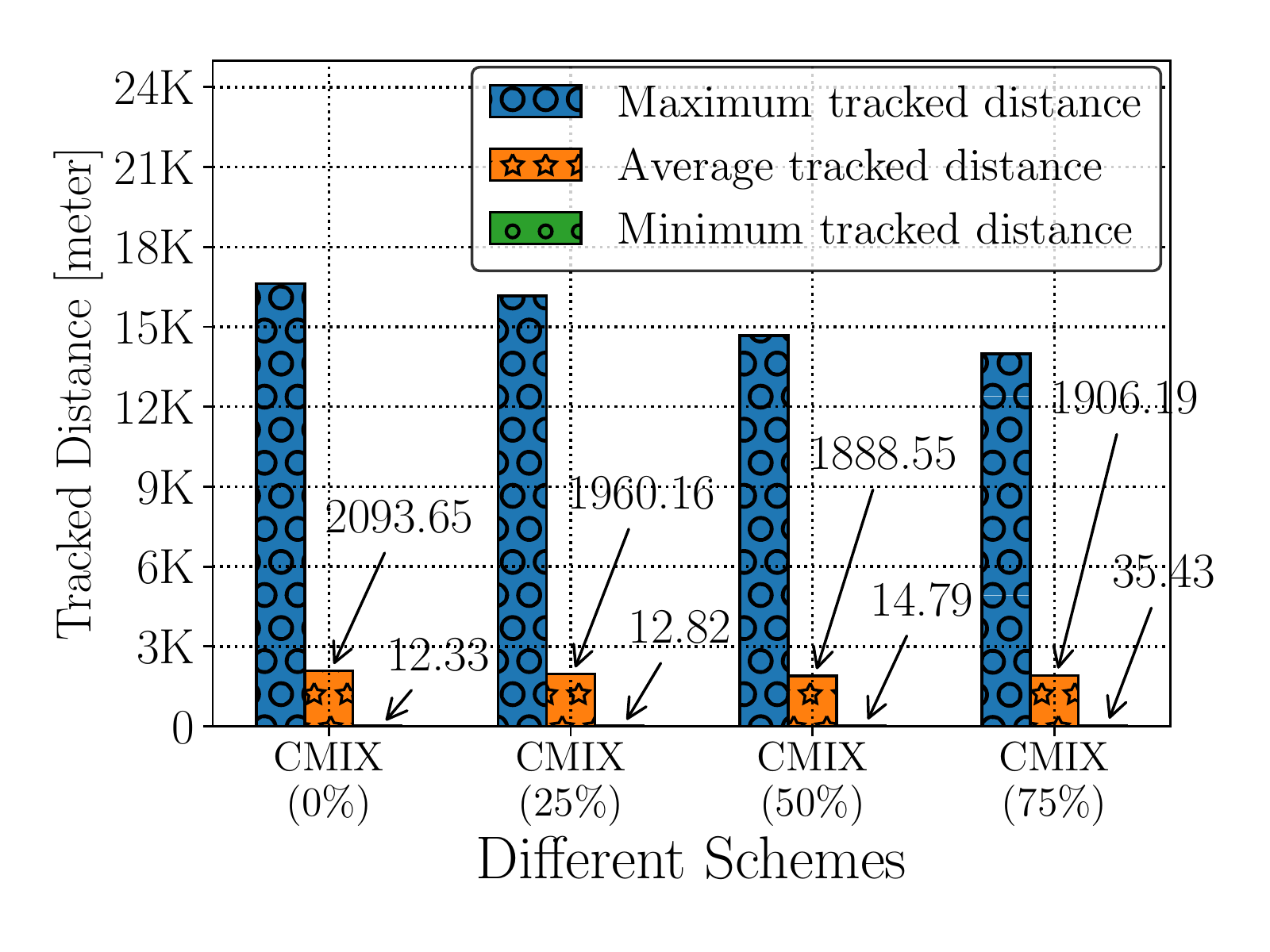}}
		\vspace{-0.75em}
		\caption{(a) Linking pseudonym sets for the baseline and our scheme. (b) Successful tracked distance for the baseline and our scheme.}
		\label{fig:ieee-iot-tracking-success-tracking-distance}
	\end{center}
	\vspace{-0em}
\end{figure}

Fig.~\ref{fig:ieee-iot-tracking-success-tracking-distance}(a) considers the \emph{average successful linkability metric} and compares the number of successfully linked pseudonyms sets for the baseline and our scheme. We refer to a successfully linked pseudonyms set as the number of pseudonyms, linked by the eavesdroppers, corresponding to the same vehicle. The figure shows the number of linked two-pseudonyms sets, three-pseudonyms sets, and four(+)-pseudonyms sets. For the baseline scheme, the total number of linked pseudonyms sets is 21367, i.e., 21367 sets of pseudonyms, each corresponding to the same vehicle, were successfully linked by the eavesdroppers. The total number of vehicles with two-pseudonyms sets linked is 18343, and the total number of vehicles with three-pseudonyms sets is 2608. Our scheme reduces the number of linked pseudonyms sets: the higher the percentage of decoy traffic is, the lower the number of linked pseudonyms sets becomes. With 75\% of decoy traffic, the total number of linked pseudonyms sets is 4168, the total number of vehicles, linked with two-pseudonyms sets is 4057, and the total number of vehicles, linked with three-pseudonyms sets is 109. In Fig.~\ref{fig:ieee-iot-tracking-success-tracking-distance}(b), we consider the \emph{tracking duration metric}, i.e., the total distance that was successfully tracked by the eavesdroppers. The average tracked distance diminishes by increasing the percentage of decoy traffic. These numbers were calculated based on the total number of linked pseudonyms sets and the distances observed by the eavesdroppers. More precisely, if eavesdroppers could link multiple-pseudonyms set corresponding to the same vehicle, then they could accumulate the total distance observed for all the \acp{CAM}, signed under the linked pseudonyms sets. For example, the average tracked distance for the baseline scheme~\cite{freudiger2007mix} is 2093 meters; with 50\% of vehicles disseminating decoy traffic, the average tracked distance becomes 1960 meters.

\begin{figure*} [!t] 
	\vspace{-1em}
	\centering
	\subfloat[{Pseudonyms changes}]{
		\hspace{-1.35em}
		\includegraphics[trim=0.8cm 0.4cm 1cm 0.75cm, clip=true, totalheight=0.205\textheight, width=0.205\textwidth, angle=0, keepaspectratio] {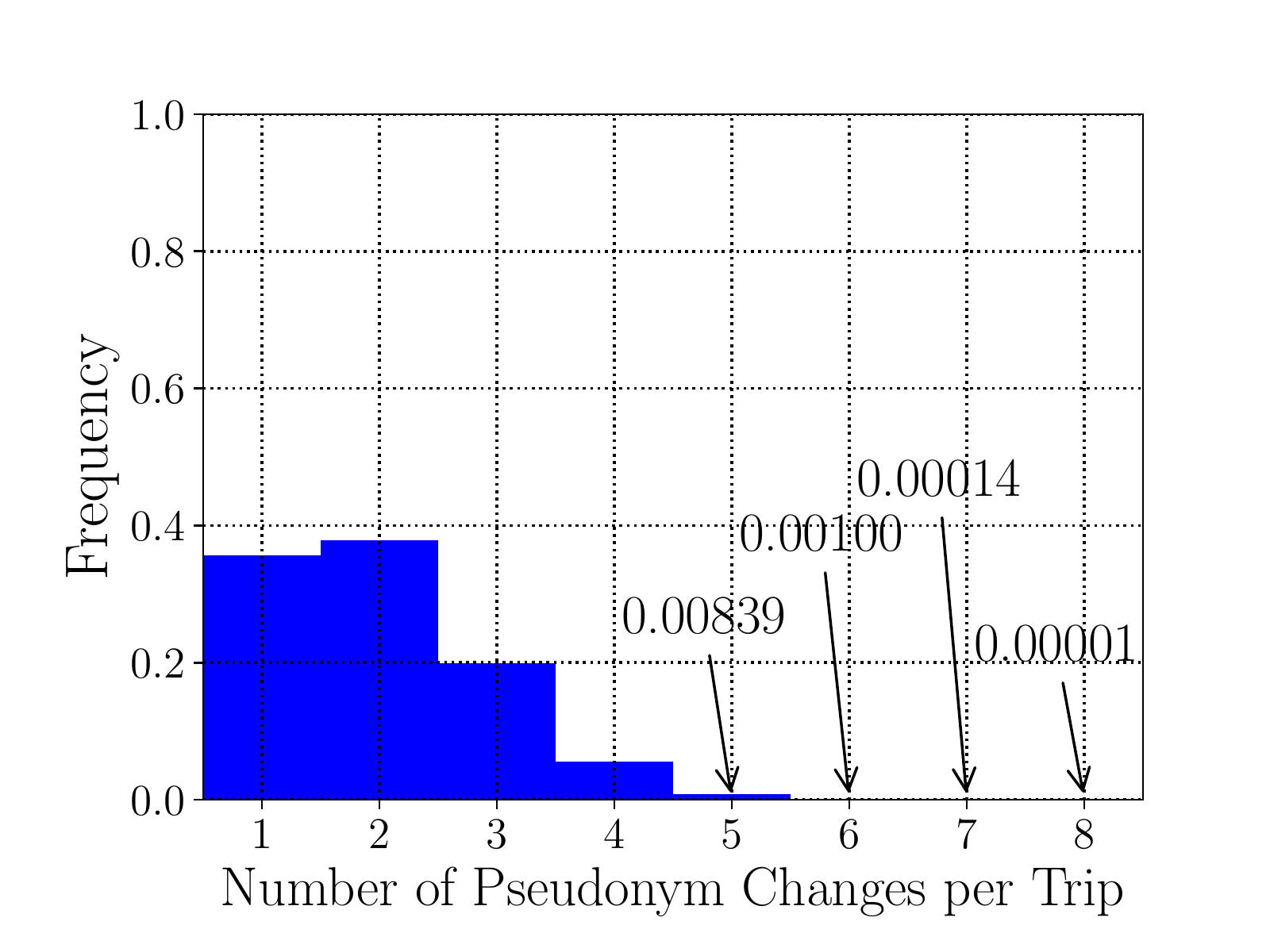}}
	\subfloat[{\ac{CMIX}: 0\% decoy traffic}]{
		\hspace{-0.75em}
		\includegraphics[trim=0.8cm 0.4cm 1cm 0.75cm, clip=true, totalheight=0.205\textheight, width=0.205\textwidth, angle=0, keepaspectratio] {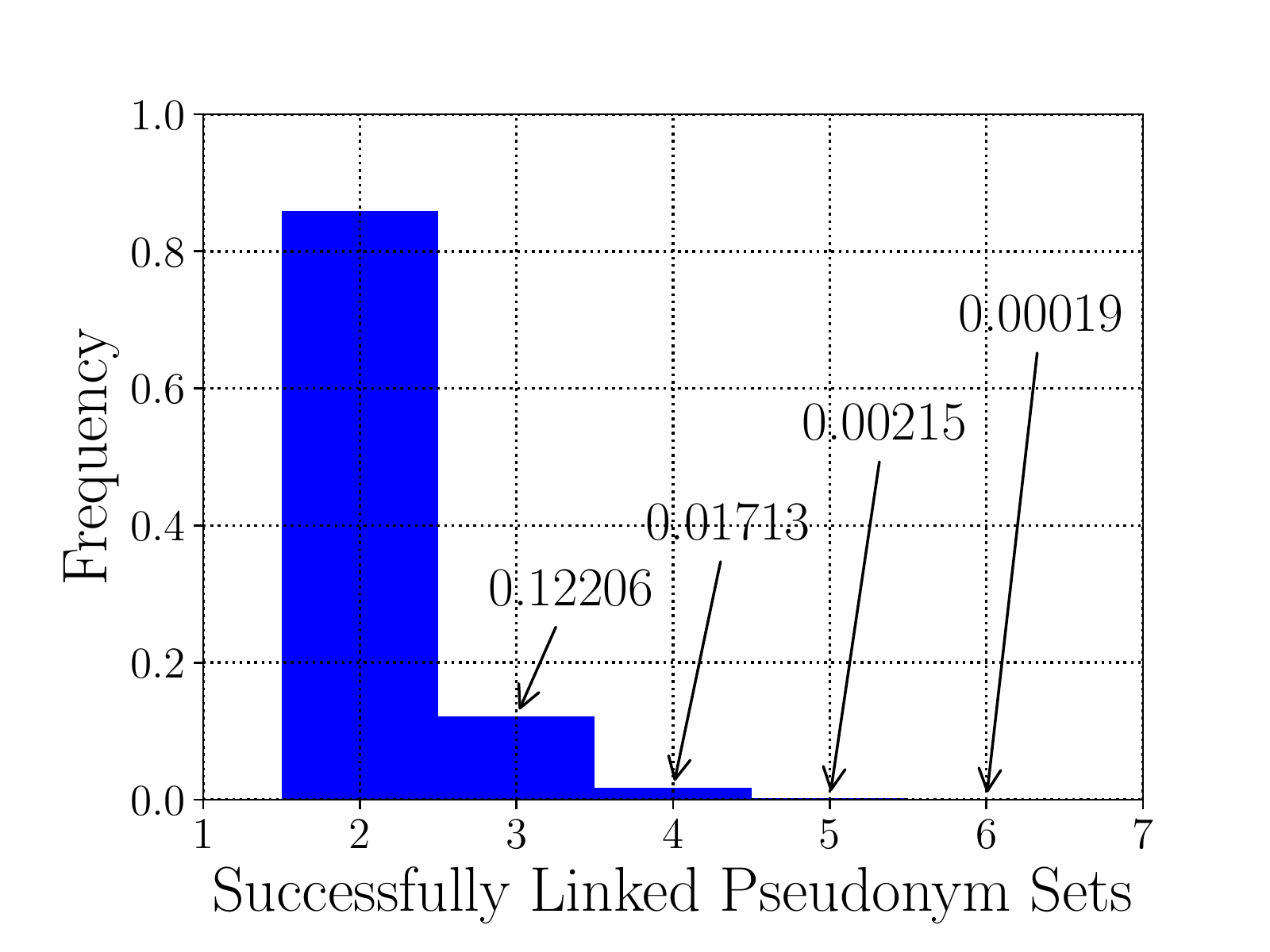}}
	\subfloat[{\ac{CMIX}: 25\% decoy traffic}]{
		\hspace{-0.75em}
		\includegraphics[trim=0.8cm 0.4cm 1cm 0.75cm, clip=true, totalheight=0.205\textheight, width=0.205\textwidth, angle=0, keepaspectratio] {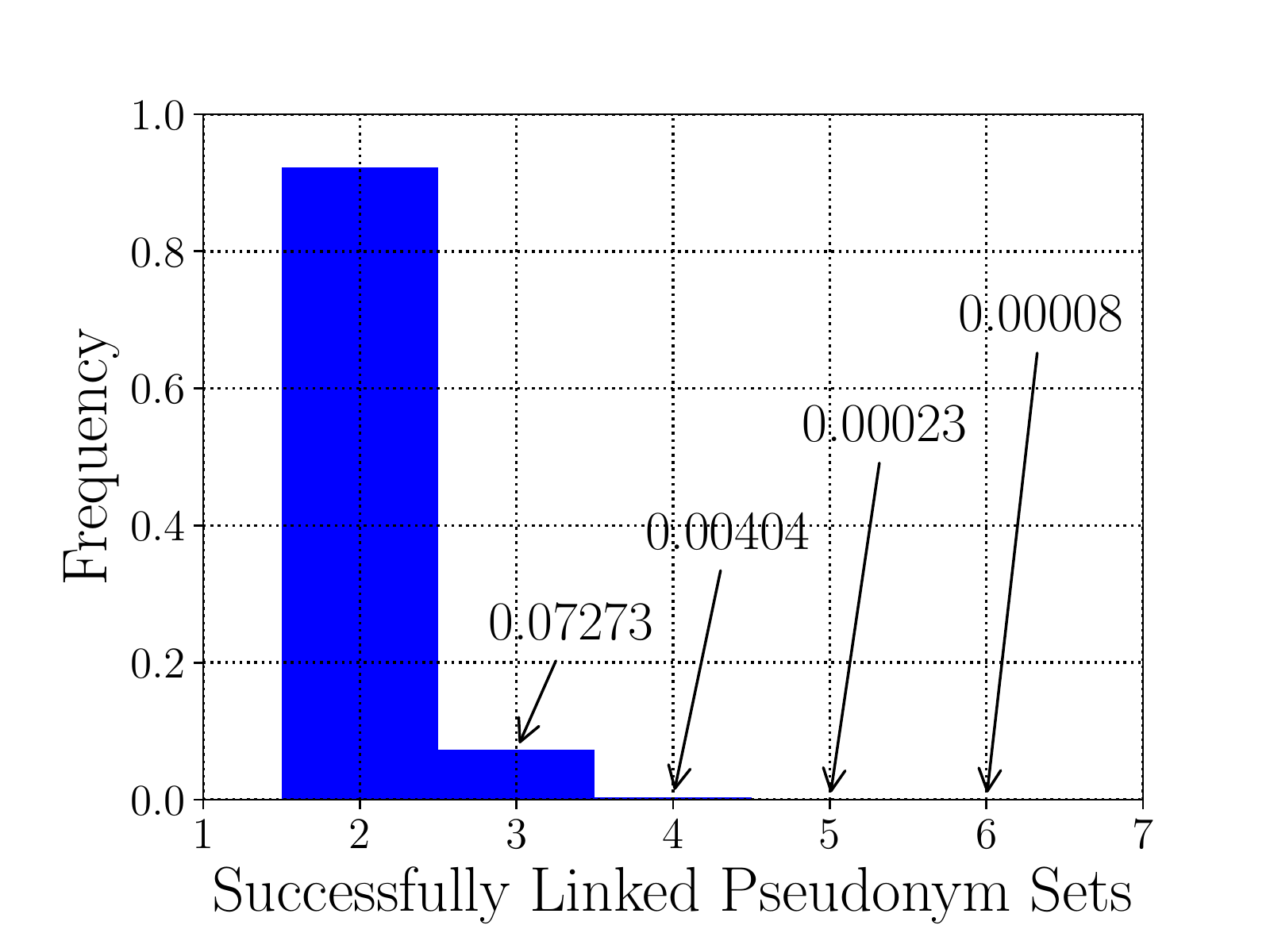}}
	\subfloat[{\ac{CMIX}: 50\% decoy traffic}]{
		\hspace{-0.75em}
		\includegraphics[trim=0.8cm 0.4cm 1cm 0.75cm, clip=true, totalheight=0.205\textheight, width=0.205\textwidth, angle=0, keepaspectratio] {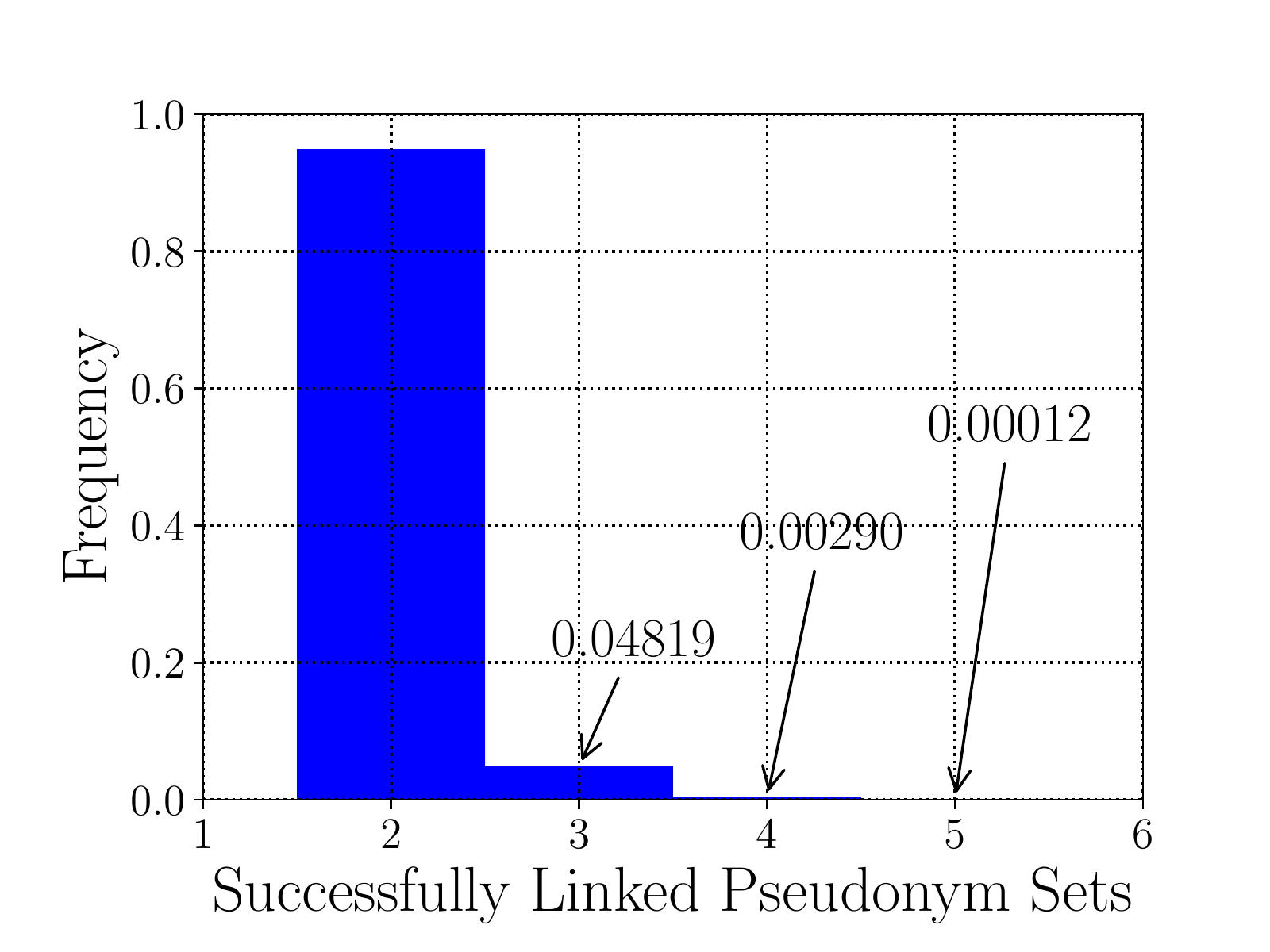}}
	\subfloat[{\ac{CMIX}: 75\% decoy traffic}]{
		\hspace{-0.75em}
		\includegraphics[trim=0.8cm 0.4cm 1cm 0.75cm, clip=true, totalheight=0.205\textheight, width=0.205\textwidth, angle=0, keepaspectratio] {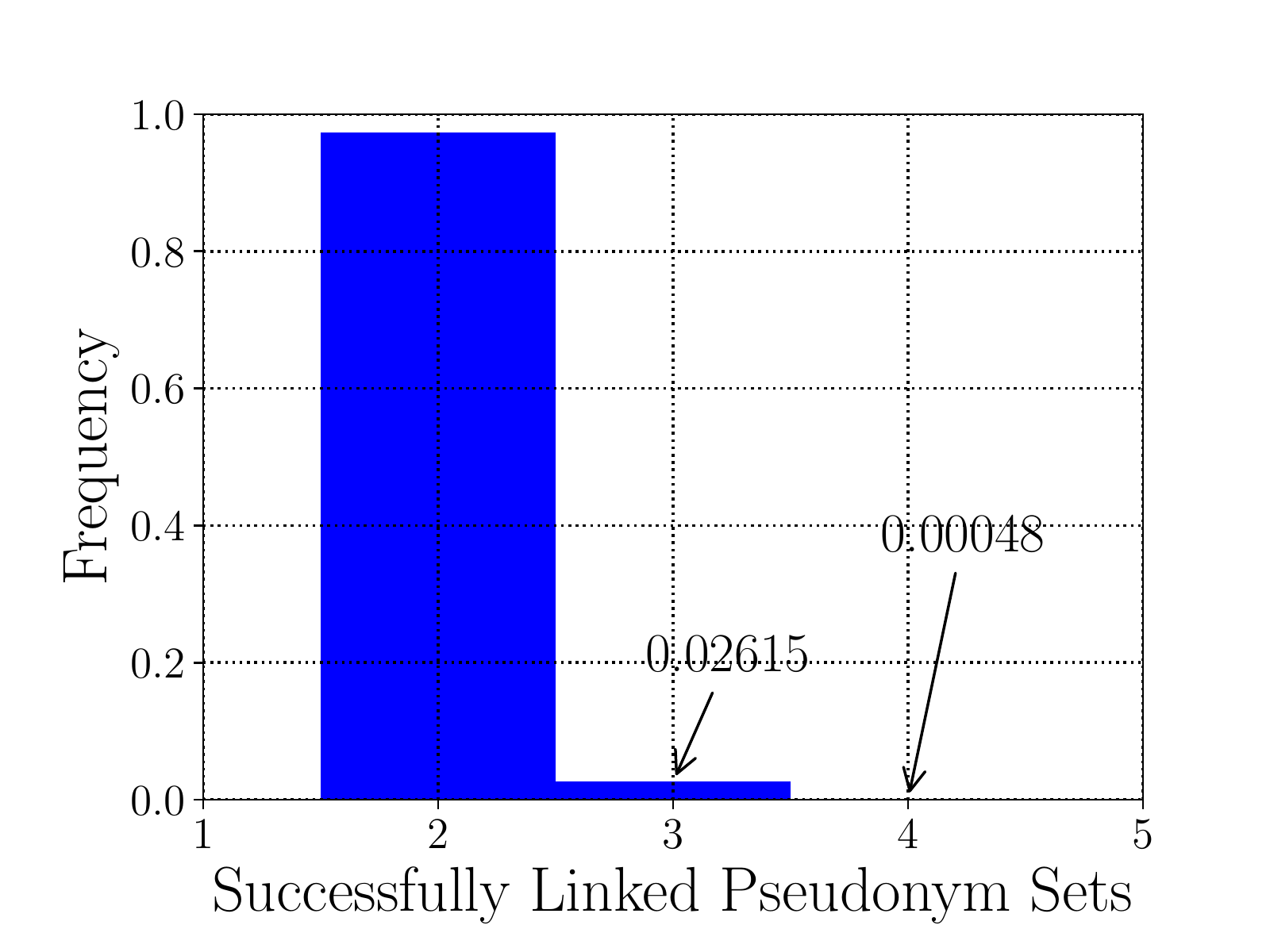}}
	\vspace{-0.5em}
	\caption{(a) Histogram of pseudonyms changes. (b) Histogram of successfully linked pseudonym sets for (b) the baseline scheme (\ac{CMIX}), (c) our scheme (\ac{CMIX} with 25\% decoy traffic), (d) our scheme (\ac{CMIX} with 50\% decoy traffic), (e) our scheme (\ac{CMIX} with 75\% decoy traffic).}
	\label{fig:ieee-iot-tracking-histogram-psnyms-changes-percentage-of-successfully-linked-psnyms}
	\vspace{-0em}
\end{figure*}

Fig.~\ref{fig:ieee-iot-tracking-histogram-psnyms-changes-percentage-of-successfully-linked-psnyms}(a) shows the histogram of the number of pseudonym changes per trip. Vehicles only change their pseudonyms when they cross a mix-zone during their trip duration: 36\% of the vehicles changed their pseudonyms only once, 38\% changed twice, and 20\% of them changed three times. There are also few vehicles changing their pseudonyms more than five times. The more mix-zones vehicles encounter, the higher the frequency of changing pseudonyms becomes; this would result in the higher number of unlinkable segments for any journey, thus enhancing user privacy protection. Fig.~\ref{fig:ieee-iot-tracking-histogram-psnyms-changes-percentage-of-successfully-linked-psnyms}(b) - Fig.~\ref{fig:ieee-iot-tracking-histogram-psnyms-changes-percentage-of-successfully-linked-psnyms}(e) show the histogram of successfully linked pseudonyms sets by the eavesdroppers for the baseline and our scheme. With the baseline scheme (Fig.~\ref{fig:ieee-iot-tracking-histogram-psnyms-changes-percentage-of-successfully-linked-psnyms}(b)), the eavesdroppers could link 86\% of two-pseudonyms sets while there are successfully linked sets with three-, four-, and five-pseudonyms. By disseminating decoy traffic, the percentage of linking pseudonyms sets decreases: with 75\% of decoy traffic (Fig.~\ref{fig:ieee-iot-tracking-histogram-psnyms-changes-percentage-of-successfully-linked-psnyms}(e)), the eavesdroppers link 97\% of two-pseudonyms sets while there are very few three- or four-pseudonyms set, linked by the eavesdroppers (and no five- or six-pseudonyms sets). The higher the percentage of decoy traffic, the lower the probability of linking pseudonyms by the eavesdroppers, thus the smaller the number of linked pseudonyms corresponding to the same vehicle. This results in a smaller percentage of the trips which could be linked by the eavesdroppers to harm user privacy.

\begin{figure*} [!t] 
	\vspace{-1em}
	\centering
	\subfloat[{\ac{CMIX}: 0\% decoy traffic}]{
		\hspace{-1.25em}
		\includegraphics[trim=0.9cm 0.4cm 0.5cm 1.25cm, clip=true, totalheight=0.263\textheight, width=0.263\textwidth, angle=0, keepaspectratio] {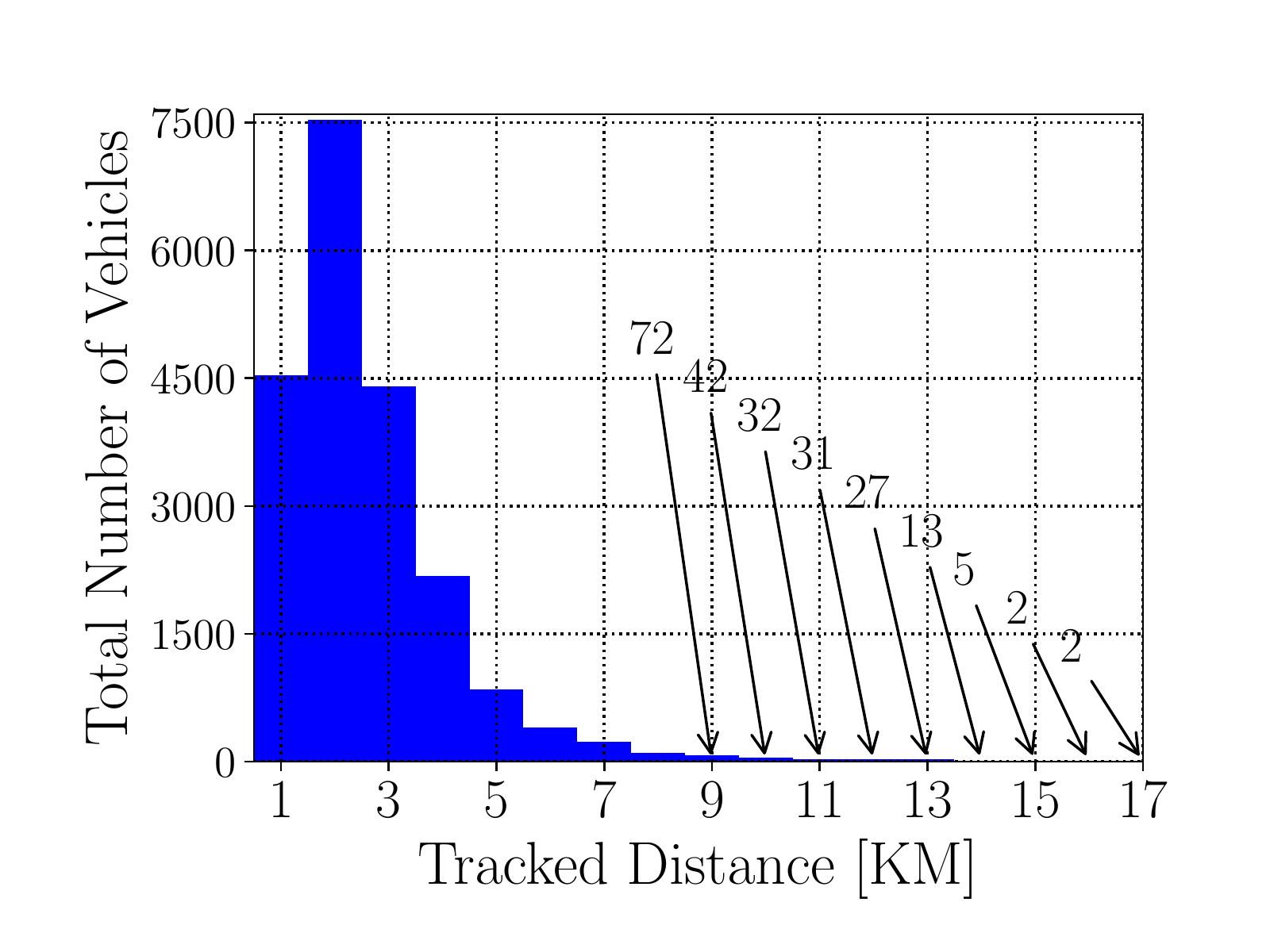}}
	\subfloat[{\ac{CMIX}: 25\% decoy traffic}]{
		\hspace{-1.15em}
		\includegraphics[trim=0.9cm 0.4cm 0.5cm 1.25cm, clip=true, totalheight=0.263\textheight, width=0.263\textwidth, angle=0, keepaspectratio] {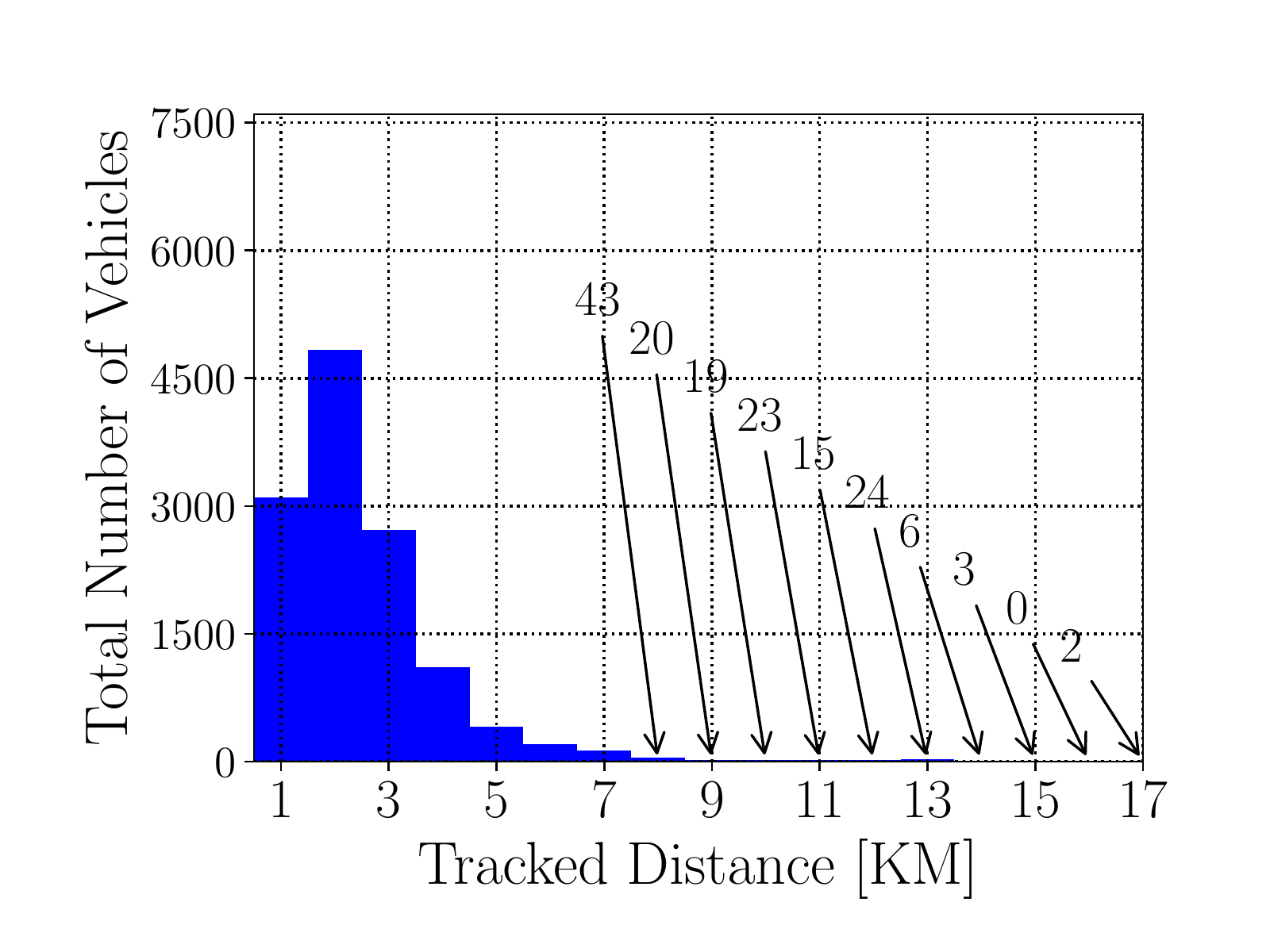}}
	\subfloat[{\ac{CMIX}: 50\% decoy traffic}]{
		\hspace{-1.15em}
		\includegraphics[trim=0.9cm 0.4cm 0.5cm 1.25cm, clip=true, totalheight=0.263\textheight, width=0.263\textwidth, angle=0, keepaspectratio] {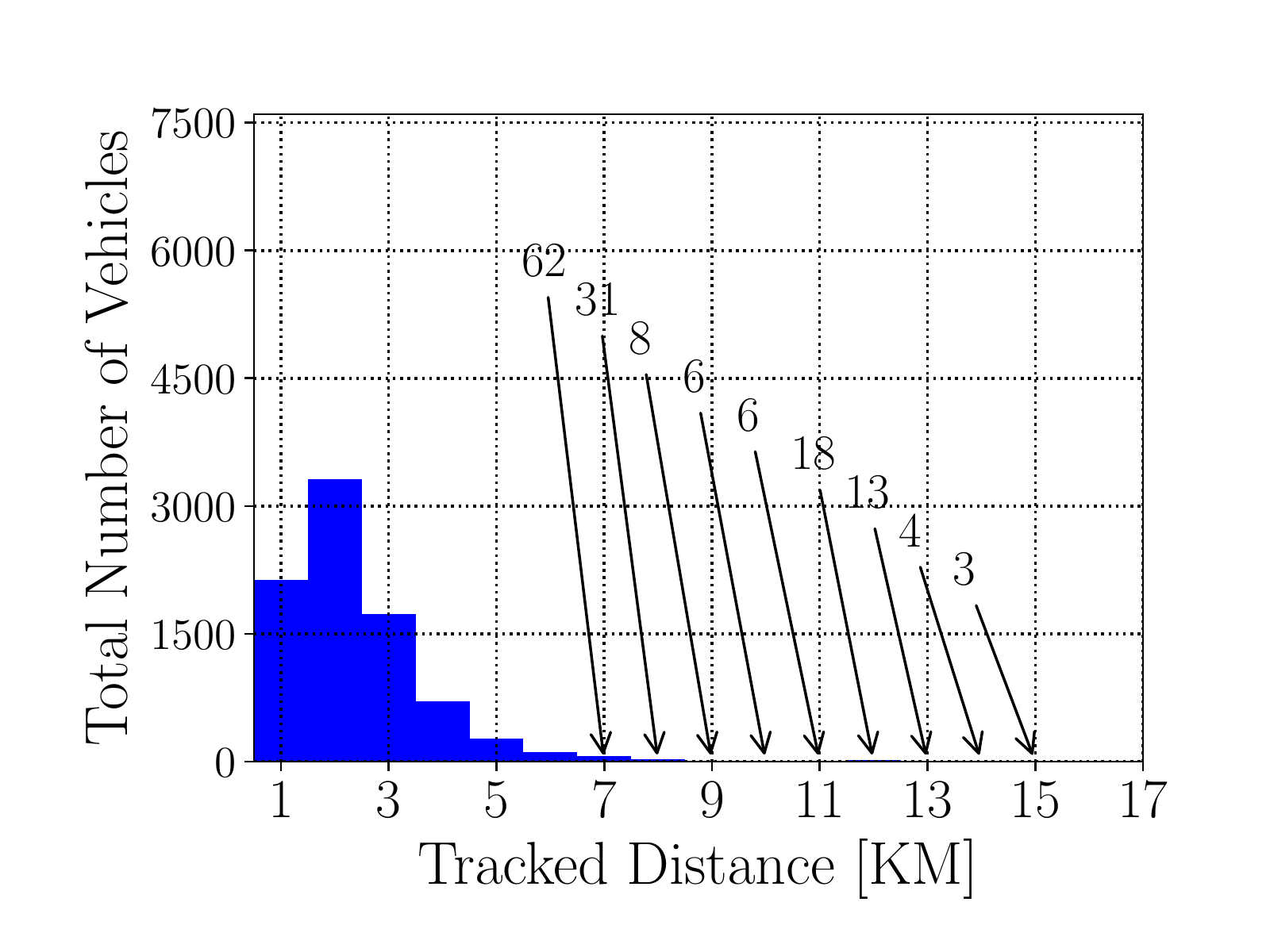}}
	\subfloat[{\ac{CMIX}: 75\% decoy traffic}]{
		\hspace{-1.15em}
		\includegraphics[trim=0.9cm 0.4cm 0.5cm 1.25cm, clip=true, totalheight=0.263\textheight, width=0.263\textwidth, angle=0, keepaspectratio] {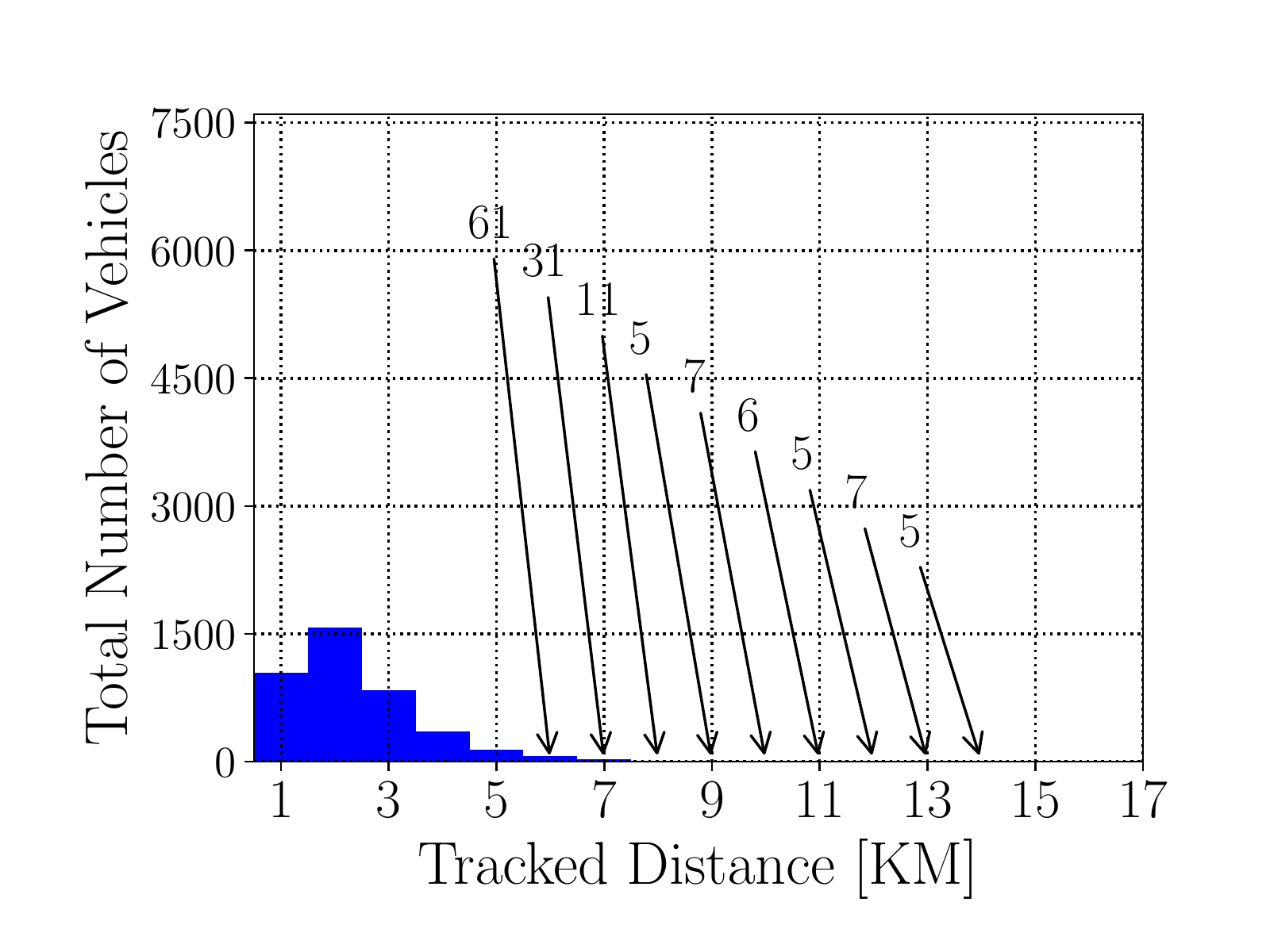}}
	\vspace{-0.5em}
	\caption{Histogram of tracked distances by eavesdroppers based on the linked pseudonyms sets for the baseline scheme (\ac{CMIX}) and our scheme.}
	\label{fig:ieee-iot-tracking-histogram-tracked-distance-comparison}
	\vspace{-0em}
\end{figure*}

\begin{figure*} [!t] 
	\vspace{-1em}
	\centering
	\subfloat[{During Rush Hours}]{
		\hspace{-0.75em}
		\includegraphics[trim=0.5cm 0.3cm 0.75cm 0.95cm, clip=true, totalheight=0.33\textheight, width=0.33\textwidth, angle=0, keepaspectratio] {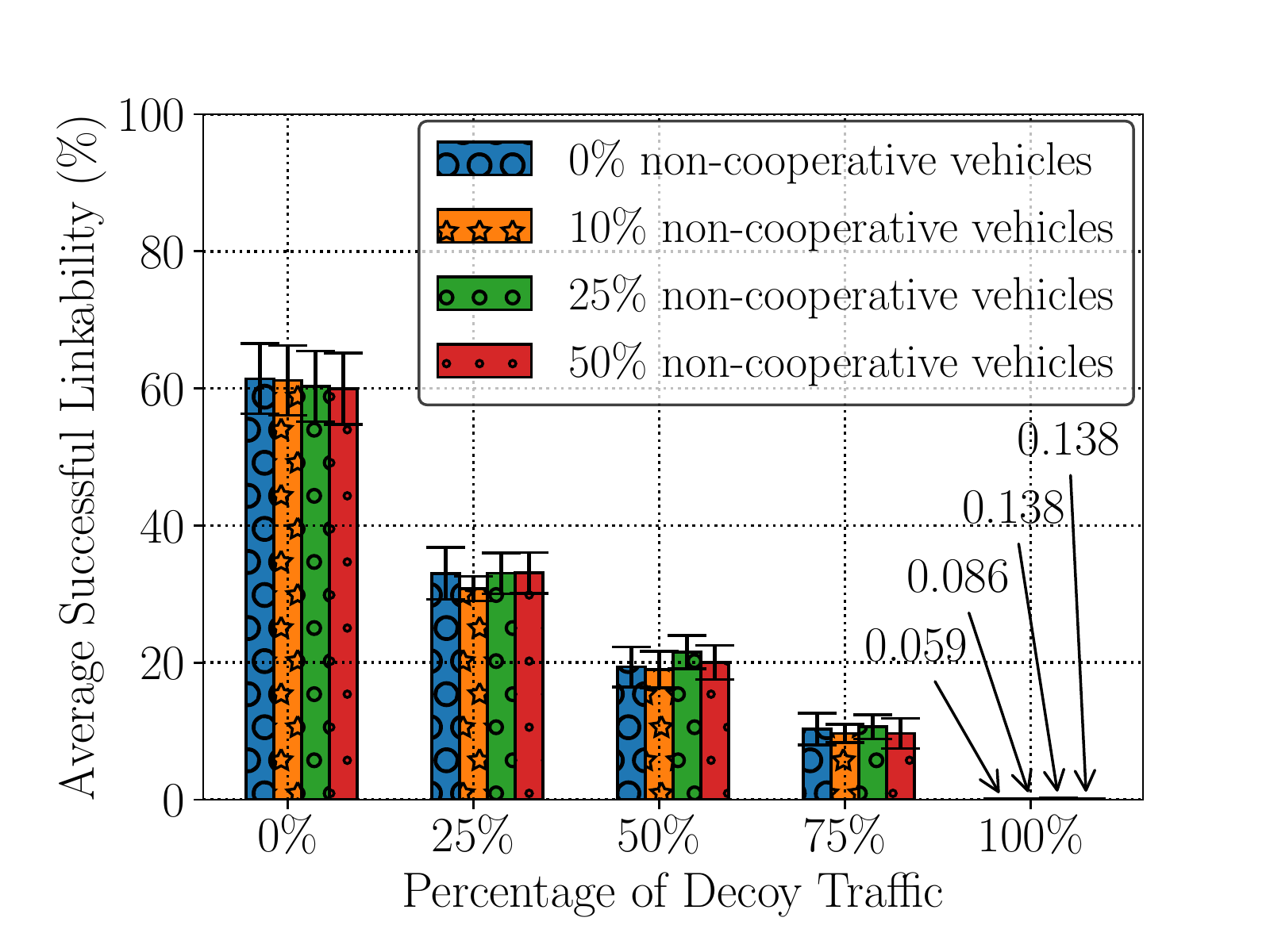}}
	\subfloat[{During Non-rush Hours}]{
		\includegraphics[trim=0.5cm 0.3cm 0.75cm 0.95cm, clip=true, totalheight=0.33\textheight, width=0.33\textwidth, angle=0, keepaspectratio] {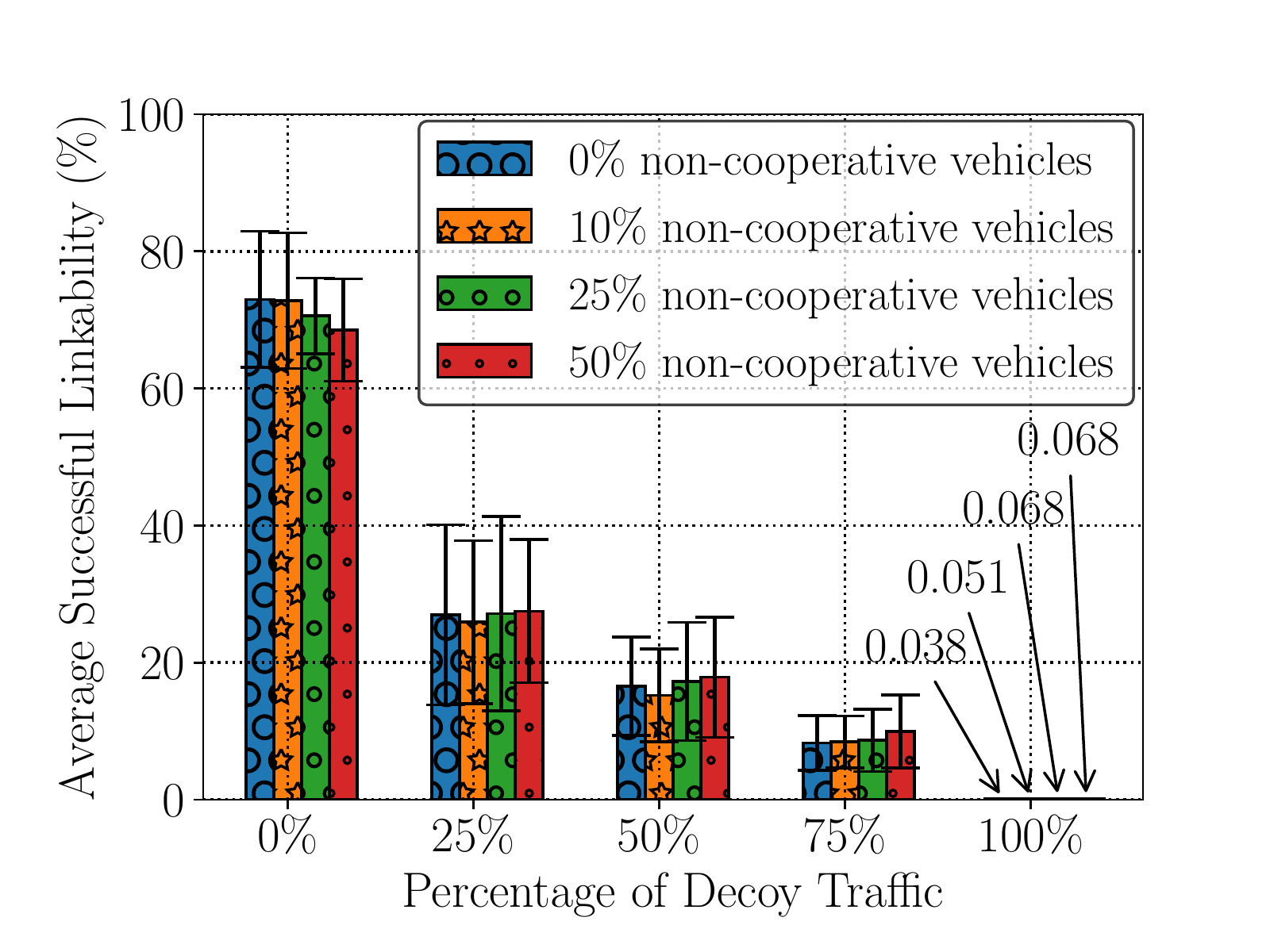}}
	\subfloat[{During 24 Hours}]{
		\includegraphics[trim=0.5cm 0.3cm 0.75cm 0.95cm, clip=true, totalheight=0.33\textheight, width=0.33\textwidth, angle=0, keepaspectratio] {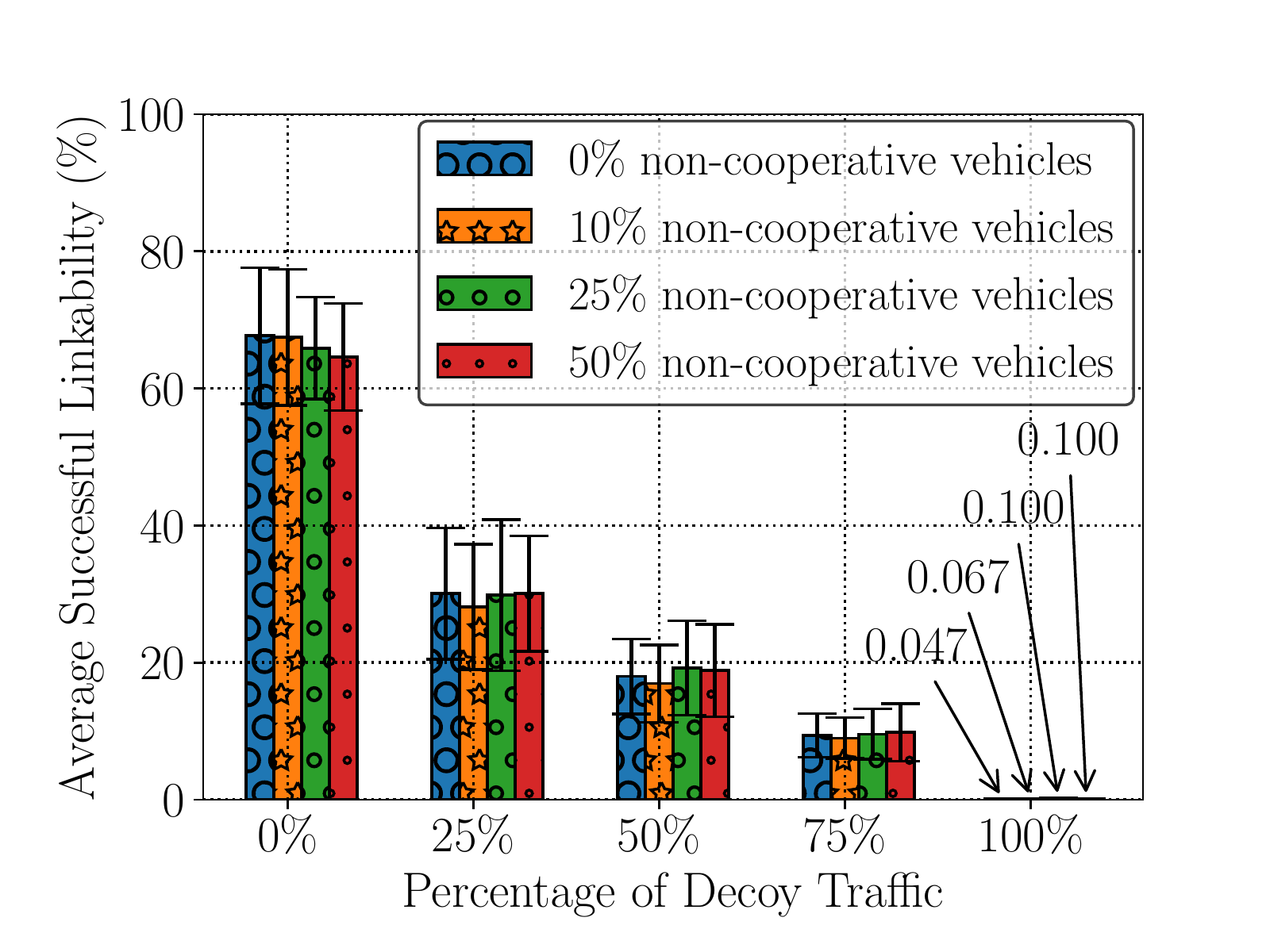}}
	\vspace{-0.5em}
	\caption{Average successful linkability in the presence of non-cooperative vehicles, not changing their pseudonyms while crossing the mix-zones.}
	\label{fig:ieee-iot-tracking-success-tracking-rates-with-non-cooperative-vehicles}
	\vspace{-0em}
\end{figure*}

Fig.~\ref{fig:ieee-iot-tracking-histogram-tracked-distance-comparison} shows the histogram of the number of vehicles, tracked by the eavesdroppers, based on the linked pseudonyms sets. With the baseline scheme (Fig.~\ref{fig:ieee-iot-tracking-histogram-tracked-distance-comparison}(a)), the eavesdroppers could link 4536 vehicles for 1 KM, 7532 vehicles for 2 KMs, and 4409 vehicles for 3 KMs. In contrast, by introducing decoy traffic for vehicles exiting the mix-zones, the total number of vehicles, tracked by the eavesdroppers, drastically decreases: with 75\% of decoy traffic (Fig.~\ref{fig:ieee-iot-tracking-histogram-tracked-distance-comparison}(d)), the eavesdroppers could only link 1044 vehicles for 1 KM, 1576 vehicles for 2 KMs, and 837 vehicles for 3 KMs. Note that by disseminating 100\% decoy traffic, the probability of linking two successive pseudonyms by the eavesdroppers is very low, thus such tracking becomes ineffective (see Fig.~\ref{fig:ieee-iot-tracking-success-tracking-rates-evaluation} and Fig.~\ref{fig:ieee-iot-tracking-success-tracking-rates-with-non-cooperative-vehicles}).

Fig.~\ref{fig:ieee-iot-tracking-success-tracking-rates-with-non-cooperative-vehicles} shows the average success rates in the presence of non-cooperative vehicles that try to diminish the anonymity set size of a mix-zone. Such vehicles exit the mix-zone without changing their pseudonyms; also, if chosen to be relaying vehicles, they do not disseminate decoy traffic. The tracking algorithm (step 4 in Algorithm~\ref{algorithm:ieee-iot-tracking-syntactic-semantic-linking-attacks}) filters out these trivially linked pseudonyms, i.e., \acp{CAM} of vehicles that enter and exit the mix-zone with the same pseudonym. Fig.~\ref{fig:ieee-iot-tracking-success-tracking-rates-with-non-cooperative-vehicles}(a) shows the average successful tracking during the rush hours. The average successful tracking in the presence of non-cooperative vehicles for the \ac{CMIX} scheme slightly decreases: the eavesdroppers filter out transcript of pseudonymously authenticated messages with the same pseudonym. Thus, non-cooperative vehicles, not changing their pseudonyms, do not help eavesdroppers link successive pseudonyms with higher percentage of successful tracking. During the non-rush hour periods (Fig.~\ref{fig:ieee-iot-tracking-success-tracking-rates-with-non-cooperative-vehicles}(b)), the average successful tracking for the \ac{CMIX} scheme is higher than the one during the rush-hour periods: due to lower number of vehicles in a mix-zone, the probability of linking increases; still, non-cooperative vehicles that do not change their pseudonyms, when crossing a mix-zone, do not highly affect the anonymity set size. Fig.~\ref{fig:ieee-iot-tracking-success-tracking-rates-with-non-cooperative-vehicles}(c) shows the average successful tracking for the entire intervals: eavesdroppers could successfully link 68\% of successive pseudonyms before and after pseudonym changes in the mix-zones.

The average successful tracking for our scheme is not considerably affected in the presence of non-cooperative vehicles thanks to dissemination of decoy traffic. Note that selection of non-cooperative vehicles is independent of selection of relaying vehicles, i.e., in each scenario, different sets of vehicles are selected to be non-cooperative. Thus, a direct comparison of the scenarios with different percentage of non-cooperative vehicles is not straightforward. In order to mitigate the effect of non-cooperative vehicles, an \ac{RSU} could monitor the behavior of vehicles when entering and exiting the mix-zone; if a substantial fraction of vehicles exit the mix-zone without changing their pseudonyms, the \ac{RSU} can increase the percentage of decoy traffic. Further investigation is one of our future work.

\begin{figure} [!t]
	\vspace{-0em}
	\begin{center}
		\centering
		\hspace{0em} \includegraphics[trim=0.7cm 0.35cm 1.5cm 0.25cm, clip=true, width=0.4\textwidth,height=0.4\textheight,keepaspectratio]{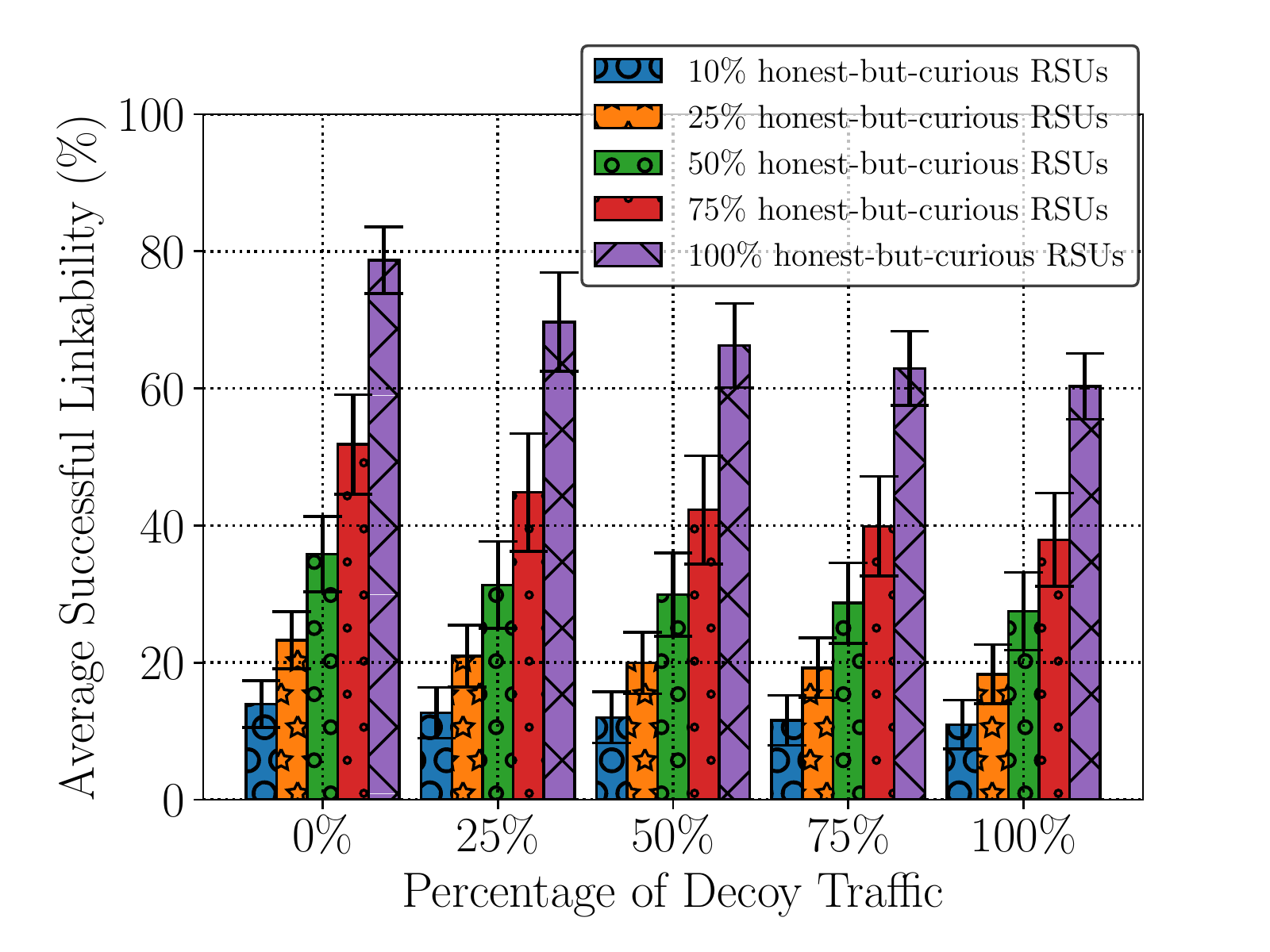}
		\vspace{-0.5em}
		\caption{Average successful linkability in the presence of a fraction of honest-but-curious \acp{RSU}, operating the mix-zones.}
		\label{fig:ieee-iot-tracking-success-tracking-rates-by-fraction-of-honest-but-curious-rsus}
	\end{center}
	\vspace{-0em}
\end{figure}

Fig.~\ref{fig:ieee-iot-tracking-success-tracking-rates-by-fraction-of-honest-but-curious-rsus} shows the average successful linkability among pseudonyms sets by a fraction of honest-but-curious \acp{RSU}. Such entities have broader communication coverage and they can observe the communication inside the encrypted area. However, each \ac{RSU} only knows a distinct set of \ac{CF}, provided by the \ac{PCA} and it cannot filter out chaff pseudonyms originated from other mix-zones. For the baseline scheme, the honest-but-curious \acp{RSU} could link the successive pseudonyms with higher probability in comparison with our scheme. For example, for the baseline scheme with 50\% of \acp{RSU} to be honest-but-curious, the average successful pseudonym linkability is $\approx$36\%. However, by introducing 100\% decoy traffic, such linkability drops to $\approx$27\%. Note that introducing chaff \acp{CAM} does not fully diminish the pseudonyms linkability in the presence of honest-but-curious \acp{RSU}. That requires introducing chaff \acp{CAM} combined with other techniques, e.g., simultaneously changing pseudonyms by all the vehicles inside a mix-zone, to fully diminish the syntactic and semantic linking attacks. This requires further investigation and remains as our future work.

\section{Conclusion and Future Work}
\label{sec:ieee-iot-tracking-conclusion}

We proposed a novel scheme to protect user privacy regardless of the geometry of the mix-zones, mobility patterns, vehicle density, and arrival rates. Our system enhances user privacy protection at the cost of low computation and communication overhead while it ensures that the operation of the safety applications remains unaffected by the dissemination of decoy traffic. Our results show that cooperative dissemination of decoy traffic, by relaying vehicles exiting a mix-zone, can significantly diminish syntactic and semantic pseudonym linking attacks. Moreover, our experiments show that the deployment of mix-zones can be cost-effective. As future work, we plan to expand our adversarial model and investigate the resiliency of our scheme against a fraction of malicious vehicles or compromised \acp{RSU} that covertly send the \ac{CMIX} symmetric key or the \acp{CF} to other (internal or external) adversaries. Moreover, we plan to investigate the effect of mix-zone transmission range on the overall communication and computation overhead of the \ac{VC} system. Further, we intend to improve our tracking algorithm towards tracking vehicles based on predicting vehicles trajectories using Kalman Filter and physical properties of the wireless radio signals. Moreover, we plan to investigate various metrics for quantifying location privacy and conduct a full-blown comparison of our scheme by leveraging different metrics. Finally, we intend to evaluate the impact of decoy traffic on the operation of safety applications in various traffic conditions.


\section*{Acknowledgements}
\label{ieee-iot-tracking-acknowledgements}

Work supported by the Swedish Foundation for Strategic Research (SSF) SURPRISE project and the KAW Academy Fellowship Trustworthy IoT project.



\bibliographystyle{IEEEtran}
\bibliography{IEEEabrv,references}

\ifCLASSOPTIONcaptionsoff
  \newpage
\fi



%

\vfill

\begin{IEEEbiography}[{\includegraphics[trim=10cm 10cm 20cm 0cm, clip=true, width=1\textwidth,height=1\textheight,keepaspectratio]{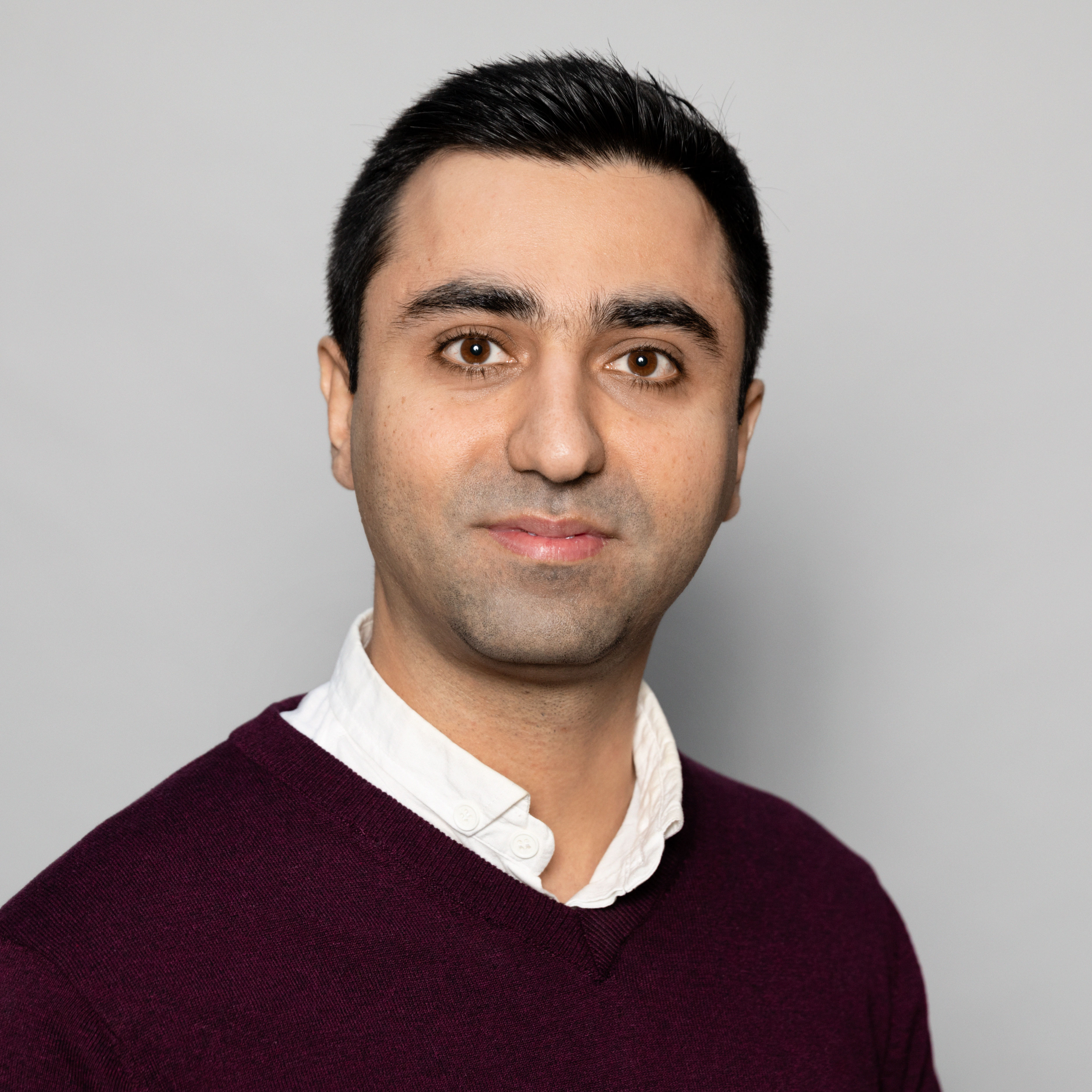}}]{Mohammad Khodaei} earned his Ph.D degree from KTH Royal Institute of Technology, Stockholm, Sweden, in 2020. He is currently a postdoctoral researcher at the Networked Systems Security Group, KTH, under the supervision of Prof. Panos Papadimitratos. His research interests include security and privacy in smart cities, the Internet of Things, distributed systems, and cloud computing. 
\end{IEEEbiography}

\vfill
\begin{IEEEbiography}[{\includegraphics[width=1in,height=1.25in,clip,keepaspectratio]{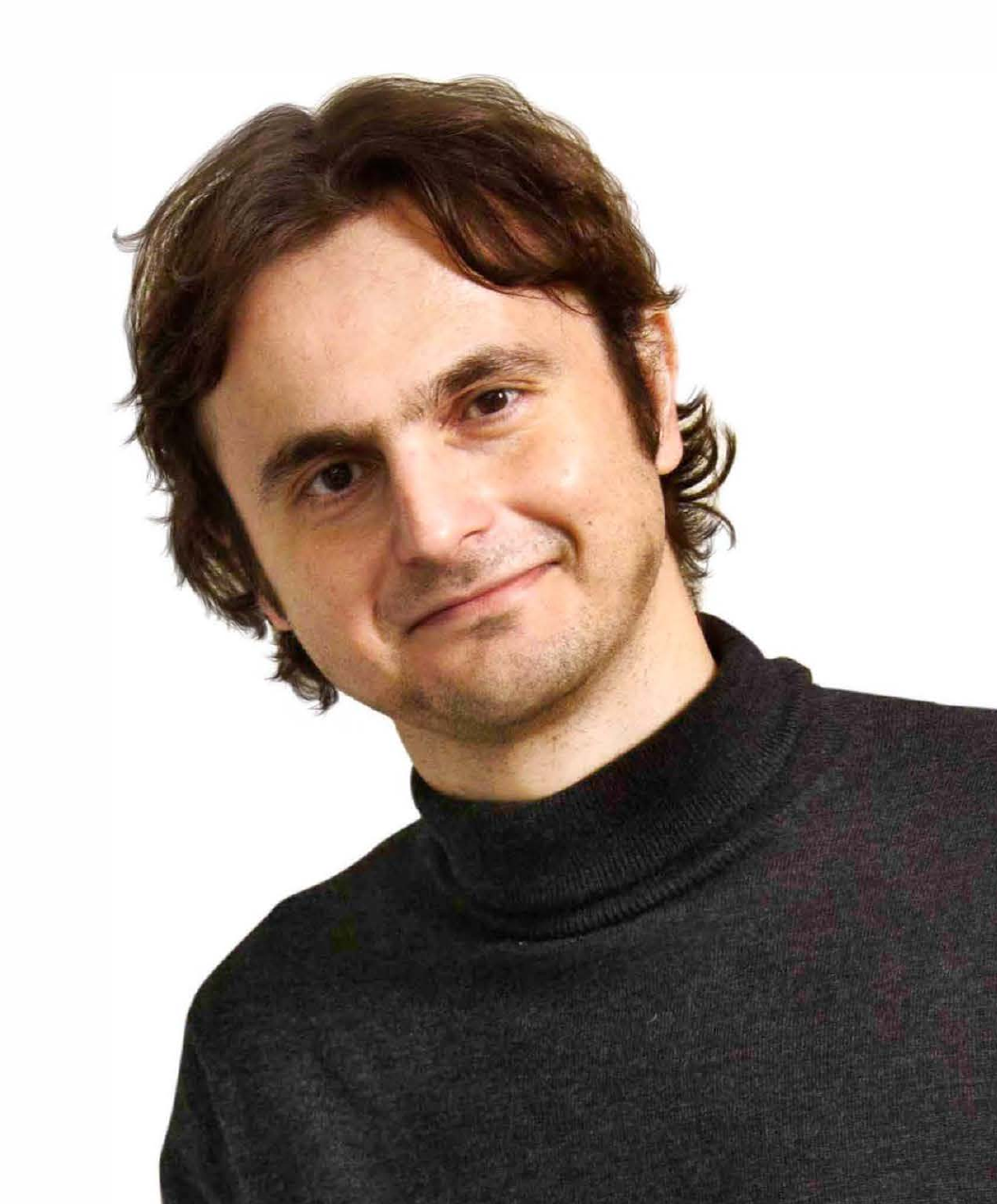}}]{Panos Papadimitratos} earned his Ph.D. degree from Cornell University, Ithaca, NY. At KTH, Stockholm, Sweden, he leads the Networked Systems Security lab, and he is a member of the steering committee of the Security Link center. He has delivered numerous invited talks, keynotes, panel addresses, and tutorials in flagship conferences. He serves or served as: Associate Editor of the IEEE TMC, the ACM/IEEE ToN and the IET IFS journals. He is a member of the PETS Editorial and Advisory Boards, and the ACM WiSec and CANS conference steering committees. He was a program chair for the ACM WiSec'16, TRUST'16 and CANS'18 conferences; a general chair for ACM WISec'18, PETS'19, and IEEE EuroS\&P'19. He is a Fellow of the Young Academy of Europe, a Knut and Alice Wallenberg Academy Fellow, and an IEEE Fellow. His group web-page is: \url{https://www.eecs.kth.se/nss}.
\end{IEEEbiography}

\vfill

\end{document}